%% file: main.tex
\documentclass[a4paper,11pt]{article}
\pdfoutput=1 % if you are submitting a pdflatex (i.e. if you have
\usepackage{amsmath}
\usepackage{preamble_ohad_copy}
\usepackage[T1]{fontenc}
\usepackage[sorting=none,maxcitenames=50,maxnames=50]{biblatex}
\title{A String Theory for Two Dimensional Yang-Mills Theory I}
\author{Ofer Aharony$^{1,2}$\footnote{ofer.aharony@weizmann.ac.il}, Suman Kundu$^1$\footnote{suman.kundu@weizmann.ac.il}, and Tal Sheaffer$^1$\footnote{sheaffertal@gmail.com}\\
\\
\footnotesize{\it ${}^1$ Department of Particle Physics and Astrophysics, Weizmann Institute of Science, Rehovot, Israel.}\\
\footnotesize{\it ${}^2$ School of Natural Sciences, Institute for Advanced Study, Princeton, NJ 08540, USA.}}
\date{}

\begin{document}

\maketitle

\begin{abstract}
Two dimensional gauge theories with charged matter fields are useful toy models for studying gauge theory dynamics, and in particular for studying the duality of large $N$ gauge theories to perturbative string theories. A useful starting point for such studies is the pure Yang-Mills theory, which is exactly solvable. Its $1/N$ expansion was interpreted as a string theory by Gross and Taylor 30 years ago, but they did not provide a worldsheet action for this string theory, and such an action is useful for coupling it to matter fields. The chiral sector of the Yang-Mills theory can be written as a sum over holomorphic maps and has useful worldsheet descriptions, but the full theory includes more general extremal-area maps; a formal worldsheet action including all these maps in a ``topological rigid string theory'' was written by \horava\ many years ago, but various subtleties arise when trying to use it for computations. In this paper we suggest a Polyakov-like generalization of \horava's worldsheet action which is well-defined, and we show how it reproduces the free limit of the Yang-Mills theory, both by formal arguments and by explicitly computing its partition function in several cases. In the future we plan to generalize this string theory to the finite-coupling gauge theory, and to analyze it with boundaries, corresponding either to Wilson loops or to dynamical matter fields in the fundamental representation.
\end{abstract}

\newpage
\tableofcontents{}

\newpage
\input{introduction}

\input{YM_review}

\input{horava_model/Horava_model}
\input{Polyakov_formalism}

\input{various_amplitudes}

\section*{Acknowledgments}
We would like to thank Netanel Barel, Matthias Gaberdiel, David Gross, Shota Komatsu, David Kutasov, Ohad Mamroud, Joe Minahan, Erez Urbach, and Edward Witten for useful discussions.
This work was supported in part by an Israel Science Foundation (ISF) center for excellence grant (grant number 2289/18), by ISF grant no. 2159/22, by Simons Foundation grant 994296 (Simons
Collaboration on Confinement and QCD Strings), by grant no. 2018068 from the United States-Israel Binational Science Foundation (BSF), by the Minerva foundation with funding from the Federal German Ministry for Education and Research, by the German Research Foundation through a German-Israeli Project Cooperation (DIP) grant ``Holography and the Swampland'', and by a research grant from Martin Eisenstein. OA is the Samuel Sebba Professorial Chair of Pure and Applied Physics. 

\appendix

\input{appendices}

\printbibliography
\end{document}

%% file: introduction.tex
\section{Introduction and Summary}

Fifty years ago this month, 't Hooft argued \cite{tHooft:1973alw} that the large $N$ limit of $SU(N)$ gauge theories, keeping fixed the 't Hooft coupling $\lambda = g_{YM}^2 N$, could be equivalent to a string theory with string coupling $g_{str} \simeq 1/N$, such that the string theory becomes classical in the large $N$ limit.
In some low dimensional examples (for instance, the double-scaling limit of matrix quantum mechanics) it has been possible to understand this duality in detail. In higher dimensions, many examples were found following the AdS/CFT correspondence \cite{Maldacena:1997re}; however, in these cases there is not yet an explicit derivation of the duality (except in topological cases \cite{Gopakumar:1998ki,Ooguri:2002gx}) nor a prescription to find the string dual for a given gauge theory.

Two dimensional gauge theories give an interesting arena where it may be possible to make further progress in deriving the string duals for a given gauge theory. The $2d$ pure Yang-Mills (YM) theory has no dynamical degrees of freedom and is exactly solvable. Gross and Taylor \cite{Gross:1993cw,Gross:1993hu,Gross:1993yt} analyzed the $1/N$ expansion of this theory and argued that it could be recast as a sum of mappings with specific properties from string worldsheets to the $2d$ space-time that the gauge theory lives in. If one could find a local worldsheet theory that reproduces this sum over mappings, then one could use this as a starting point for finding the string duals of dynamical gauge theories. Adding matter fields in the fundamental representation (scalar or fermion, massless or massive) should correspond to adding specific boundary conditions for the string worldsheet, so given such a worldsheet theory, finding the string dual of (say) the 't Hooft model \cite{tHooft:1974pnl} should be straightforward.\footnote{In fact, given a string theory for pure $2d$ YM, the appropriate boundary conditions and boundary degrees of freedom for the string to reproduce the theory with a fundamental field could be found algorithmically using the worldline formalism \cite{Strominger:1980xa,Strassler:1992zr}, which allows us to recast the matter fields as worldlines that interact with the gauge fields as line operators (Wilson lines).} Adding matter fields in 2-index representations of the gauge group (such that there is a good large $N$ limit) should also be possible, since when these fields are very massive (compared to the scale set by the gauge coupling) they are weakly coupled, and one could hope to first construct the dual worldsheet theory in this limit (where these fields should lead to weakly coupled particles that could sit at folds of the worldsheet \cite{Ganor:1994rm}), and then to follow it (analytically or numerically) as the mass is decreased. 

Unfortunately, finding a local string worldsheet that reproduces the Gross-Taylor expansion is challenging. A subsector of the $2d$ YM theory -- the so-called ``chiral sector'' (defined in the next section) -- corresponds (in an ``orthogonal gauge'' for the worldsheet diffeomorphisms) to holomorphic mappings from the worldsheet to space-time, and several suggestions have been made \cite{Cordes:1994sd,Cordes:1994fc,Vafa:2004qa,KOMATSU_forthcoming,GABERDIEL_forthcoming,Rudd:1994ta} for worldsheets that reproduce the sum over these holomorphic mappings. However, the full space of mappings includes also non-holomorphic mappings, which are singular in the orthogonal gauge, and it is less clear how to correctly include them (some suggestions were made in \cite{Cordes:1994sd,Cordes:1994fc,Kimura:2008gs}). As far as we know, the only concrete suggestion for this was made by \horava\ in \cite{Horava:1993aq,Horava:1995ic}, and we will review it in section \ref{sec: horavasec} below. This suggestion was formally argued to reproduce the full $1/N$ expansion of the YM theory, but (as we review below) there are many subtleties in making this precise, since many expressions appearing there vanish (leading to $0/0$ factors) and it is not clear if the worldsheet action correctly deals with the various singularities in the moduli spaces of mappings.

In this paper, we suggest a revised version of \horava's string theory, in which we add an additional worldsheet metric (as in the Polyakov formulation of critical string theory). We analyze this formalism in detail for the case of the closed strings dual to the YM theory with a vanishing gauge coupling, which is a non-trivial $BF$-type topological theory. The generalization to finite gauge coupling (corresponding to a finite string tension) or to generalized YM theories, and the addition of boundaries to the worldsheet (corresponding to Wilson lines, or to dynamical matter fields in the fundamental representation) is postponed to future work.

We begin in section \ref{sec: 2d YM review}, by reviewing two dimensional Yang-Mills theory, its exact solution, and Gross and Taylor's large $N$ analysis. In section \ref{sec: horavasec}, we review \horava's topological rigid string theory and its limitations.

In section \ref{sec: Polyakov formalism} we present our reformulation of the theory and discuss the general method of computation in it. 
Our reformulation treats certain pathologies that appear in \horava's theory, and formally reproduces the $1/N$ expansion of the YM theory.
However, in many cases the mappings from the worldsheet to the target space are singular, and there are also singularities in the moduli space of mappings (which is integrated over), and we have to verify that these singularities are correctly handled in our formalism.
%and naturally regulates the highly singular non-chiral maps that are needed to account for the various terms in the YM partition function. 

 In section \ref{sec: various amplitudes} we go over a number of concrete computations, and compare the results to those expected from the Gross-Taylor expansion. 
 We begin with two non-singular examples, the sphere covering the sphere once and the torus covering the torus any number of times, confirming that these give the correct answer. We then move on to examples illustrating how our formalism deals with different types of singularities.
 
 When the mapping involves the worldsheet covering the space-time with different orientations (``non-chiral mappings''), the connected mappings are singular because they always involve zero-size orientation-reversing tubes connecting the different coverings. When the worldsheet has a moduli space of complex structure deformations, the singular mappings appear at the boundary of this moduli space (which should not be confused with the moduli space of mappings), and (at least in some cases) become regular in its interior, so that their contributions can be controlled. We illustrate this in section \ref{subsec: torus on sphere} in the example of a torus mapping to a sphere. In other cases, and in particular when there is no moduli space of complex structure deformations, we need to find a different method to regularize the singular mappings. We suggest doing this by adding ``constrained instanton'' type terms, that artificially introduce extra moduli into the path integral, such that for a finite value of these moduli the mapping is regular, while the path integral is dominated by limiting values of these moduli where the mapping becomes singular. We illustrate this procedure in section \ref{subsec: sphere on sphere w/ tube} for the example of a sphere covering a space-time sphere several times with different orientations. 
 In these cases with zero-size folds there are some subtle issues with the computation of the one-loop determinant, that it would be nice to understand better.
 We expect that similar methods can be used to show that the contributions from all singular mappings of this type correctly reproduce the expected ones, but we do not analyze the general case here.
 
 In our final example, we discuss a singularity in the moduli space of mappings; in general, this moduli space has singularities when different special points (such as branch points) come together. This happens already for moduli spaces of holomorphic maps (which cover the target space with a specific orientation), and a specific way to deal with these singularities (which in many cases are orbifold-like) that reproduces the correct answer was suggested in \cite{Cordes:1994fc,Horava:1995ic}. In section \ref{subsec: sphere on sphere twice} we discuss the simplest case where this happens, when a worldsheet sphere covers a target space sphere twice. Such a connected mapping has two branch-points in the target space, and the moduli space (comprising of the positions of these two branch points) is singular when they come together. We argue that our procedure correctly deals with these singularities.

We hope that these examples provide convincing evidence that our formalism correctly reproduces the full partition function of the free Yang-Mills (or $BF$) theory. It would be interesting to confirm this by analyzing the general case and verifying that all of its singularities are correctly dealt with in our formalism.

%% file: YM_review.tex
\section{A review of \texorpdfstring{$2d$}{2d} Yang-Mills theory and its large \texorpdfstring{$N$}{N} expansion}\label{sec: 2d YM review}

Gauge fields in $d$ space-time dimensions have $(d-2)$ physical degrees of freedom, and in particular for $d=2$ gauge fields have no local degrees of freedom. This allows the $2d$ Yang-Mills theory
\begin{equation} \label{eq: sym}
    S = \frac{1}{4 g_{YM}^2} \int d^2x \sqrt{\det(g_{\mu \nu})} {\rm tr} (F_{\mu \nu} F^{\mu \nu})
\end{equation}
to be exactly solvable, on any Riemann manifold with metric $g_{\mu \nu}$ and for any gauge group $G_{YM}$. If we consider the Yang-Mills theory on a spatial circle of circumference $L$, the only degree of freedom is the holonomy of the Wilson line around the circle, and quantizing it leads to one state for every irreducible representation $R$ of the gauge group, with energy
\begin{equation}
    E = \frac{g_{YM}^2 L}{2} C_2(R),
\end{equation}
where $C_2(R)$ is the second Casimir of the representation $R$. By gluing together such cylinders one can obtain any Riemann surface, and show that the partition function of the theory on that surface only depends on its genus $G$ and area $A$, and is given by a sum over irreducible representations:\footnote{This decomposition into sectors has been related to the presence of non-invertible 1-form symmetries in the YM theory -- whose presence helps explain the persistence of Casimir-scaling \cite{Nguyen_unsal:2021naa} -- and further interpreted as a sum over Dijkgraaf-Witten correlators in \cite{Pantev_Sharpe:2023dim}. Most of these target-space 1-form symmetries do not have a simple manifestation in the stringy formulation. The invertible 1-form center symmetry of the YM theory (which is $U(1)$ for $U(N)$ gauge theories, and $\mathbb{Z}_N$ for $SU(N)$) is realized as in all oriented string theories as counting the number of worldsheets with one orientation minus the number of oppositely-oriented ones. In finite $N$ $SU(N)$ theories it is broken non-perturbatively to $\mathbb{Z}_N$ by the appearance of a ``baryon vertex'', but we will not discuss this here.}
%becomes a $U(1)$ symmetry at $N\to\infty$, should survive the introduction of adjoint matter into the theory and manifests in the conservation of the $n-\tilde{n}$ (see \eqref{eq: zymnew} and the discussion below it) in a given topological sector of maps. Its finite $N$ reduction to $\mathbb{Z}_N$ is, from the stringy theoretic point of view, a highly nonperturbative phenomenon that will require further investigation. }
\cite{Migdal:1975zg,Rusakov:1990rs,Fine:1990zz,Witten:1992xu,Blau:1991mp}
\begin{equation} \label{eq: zym}
    Z_{YM} = \sum_R ({\rm dim}(R))^{2-2G} e^{-\frac{1}{2} g_{YM}^2 A C_2(R)}.
\end{equation}
Similar exact expressions can be written also for the expectation values of Wilson loops in this theory.

The $g_{YM}\rightarrow 0$ limit of \eqref{eq: zym} depends only on the topology of space-time, and in fact in this limit the theory \eqref{eq: sym} is purely topological, as may be seen by rewriting it in terms of an additional scalar field $B$ in the adjoint representation as
\begin{equation} \label{eq: sgym}
    S = \int d^2 x \sqrt{\det(g_{\mu \nu})} {\rm tr} (\frac{i}{2} B \epsilon_{\mu \nu} F^{\mu \nu} + \frac{1}{2} g_{YM}^2 B^2).
\end{equation}
The $g_{YM}\to 0$ limit is then a topological $BF$-theory. In this language, it is clear that many generalizations of the Yang-Mills theory are also well-defined in $2d$, given by generic polynomials in $B$ that can be added to \eqref{eq: sgym}. The partition function and string theory dual to these generalized Yang-Mills theories were discussed in \cite{Ganor:1994bq}, but we will not discuss them here.

\begin{figure}[ht]
    \centering
    \begin{minipage}{0.45\columnwidth}
    \includegraphics[width=\linewidth]{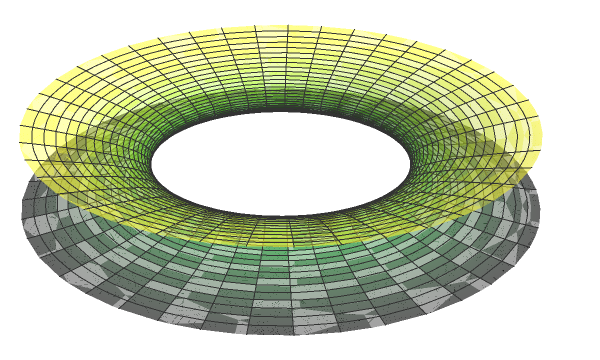}
    \caption{A simple orientation reversing tube (the fictitious height dimension is for visualization only). The extremal area case occurs when the size of the tube shrinks to 0.}
    \label{fig:simple ort}
    \end{minipage}\hfill
    \begin{minipage}{0.45\columnwidth}
    \includegraphics[width=\linewidth]{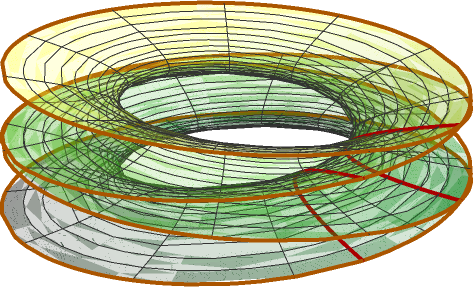}
    \caption{A branched orientation reversing tube (the fictitious height dimension is for visualization only). Circumnavigation of the tube's position permutes two of the sheets of the cover (see orange lines). The red lines connect sheets of opposite orientations. The extremal area case occurs when the size of the tube shrinks to 0.}
    \label{fig:branched ort}
    \end{minipage}
\end{figure}

As for any large $N$ gauge theory, it is natural to expect $2d$ Yang-Mills theory with $G_{YM} = SU(N)$ to have a string theory dual description with a string coupling $g_{str} \simeq 1/N$ \cite{tHooft:1973alw}. The large $N$ expansion of \eqref{eq: zym} was derived by Gross and Taylor in \cite{Gross:1993hu}. They showed that in the large $N$ limit only representations whose Young tableaux include $n$ boxes (in columns remaining at finite size for large $N$) and ${\tilde n}$ ``anti-boxes'' (in columns whose length is almost $N$, and that can be completed to length $N$ by adding ${\tilde n}$ boxes) contribute. They interpreted the contributions of these representations as coming from worldsheets that cover the space-time (which we will call the ``target space'') $n$ times with one orientation and ${\tilde n}$ times with the other orientation, and that are allowed to have specific types of singular points, at which the different sheets may be connected (with a possible permutation in $S_n \times S_{\tilde n}$ when going around the singular point). Denoting the 't Hooft coupling $\lambda = g_{YM}^2 N$, the explicit formula found in \cite{Gross:1993hu} may be rewritten \cite{Cordes:1994sd,Cordes:1994fc,Horava:1995ic} in the following way :
\begin{align} \label{eq: zymnew}
    Z_{YM} = \sum_{n,{\tilde n}=0}^{\infty} & \frac{1}{n! {\tilde n}!} e^{-(n+{\tilde n}) \lambda A / 2} \sum_{s,t=0}^{\infty} (-1)^s \frac{(\lambda A)^{s+t}}{s!t!} N^{(n+{\tilde n})(2-2G)-s-2t} \frac{(n-{\tilde n})^{2t}}{2^t} \times \\
    & \sum_{k=0}^{\infty} (-1)^k \binom{2G+k-3}{k} \sum_{p_1,\cdots,p_s} \sum_{a_1,\cdots,a_G,b_1,\cdots,b_G} \delta(p_1 \cdots p_s {\tilde \Omega}_{n,\tilde n}^k \prod_{j=1}^G (a_j b_j a_j^{-1} b_j^{-1})). \nonumber 
\end{align}
Here,
\begin{enumerate}
    \item $s$ is the number of ``simple branch points'', such that two sheets of the same orientation are permuted (with a permutation $p_i \in S_n \times S_{\tilde n}$ which has a single 2-cycle) when going around them (this requires that either $n \geq 2$ or ${\tilde n} \geq 2$);
    \item $t$ is the total number of other types of singularities allowed in the mapping: zero-size tubes (see figure \ref{fig:simple ort}) connecting two sheets of the same orientation or of opposite orientation, and zero-size collapsed handles;
    \item $k$ is the number of points which we shall call ``\ocmr-points''. Some of these points were called $\Omega$-points in \cite{Gross:1993cw}, while other points here come from expanding what \cite{Gross:1993cw} called $\Omega^{-1}$-points, following
    %in reference to CMR's 
    \cite{Cordes:1994fc,Cordes:1994sd}. %reinterpretation of what Gross and Taylor \cite{Gross:1993cw} called \ocmr-points. 
    Once the expression hidden inside $\tilde \Omega$ (see \eqref{eq: ocmr point expansion}) is unpacked, it becomes evident that an \ocmr-point can be a branch point, a collapsed tube, or any of a set of other possiblities elaborated in section \ref{subsec: stringy interpretation of the ocmr points};
    \item finally, $a_j,b_j \in S_n \times S_{\tilde n}$ are the permutations between the sheets when going around the $2G$ non-trivial cycles of the space-time manifold.
\end{enumerate}  
The delta function (in $S_n \times S_{\tilde n}$) is equal to the coefficient of the identity permutation in its argument, ensuring that the various permutations are consistent with each other. This is known as the ``holonomy constraint''. The factor $1/(n!\tilde n!)$, after partially canceling with the sum over permutations, gives a symmetry factor related to the automorphism group of the map.
%\footnote{Note that the absence of constant maps in \eqref{eq: zymnew} can be interpreted as due to a division by the volume of an infinite dimensional automorphism group acting on such maps.}
For $U(N)$ gauge theories, there is a similar expansion, but without the sum over $t$.

The expression ${\tilde \Omega}_{n,{\tilde n}}$ is given by the sum over all permutations $\sigma \in S_n$ and $\tau \in S_{\tilde n}$ with the following weight for each permutation :
\begin{equation}\label{eq: ocmr point expansion}
    1+{\tilde \Omega}_{n, \tilde n} = \sum_{\sigma,\tau} (\sigma \otimes \tau) N^{-n-{\tilde n}+K_{\sigma}+K_{\tau}} \prod_{\ell=1}^{{\rm min}(n,\tilde n)} \left( \sum_{v_{\ell}=0}^{{\rm min}(\sigma_{(\ell)},\tau_{(\ell)})}(-1)^{v_{\ell}} \ell^{v_{\ell}} v_{\ell}! \binom{\sigma_{(\ell)}}{v_{\ell}} \binom{\tau_{(\ell)}}{v_{\ell}} N^{-2v_{\ell}} \right).
\end{equation}
Here $K_{\sigma}$ is the number of cycles of the permutation $\sigma$, $\sigma_{(\ell)}$ is the number of cycles it has of length $\ell$, and similarly for $\tau$; a term with a specific value of $v_{\ell}$ corresponds to having $v_{\ell}$ zero-size orientation-reversing tubes that connect $\ell$ sheets with one orientation to $\ell$ sheets with the other orientation. When $\ell>1$, the tube is twisted in such a way that the sheets are permuted when it is circumnavigated. We will call such twisted tubes ``branched'' orientation reversing tubes (see figure \ref{fig:branched ort}). All the tubes coming from a specific $\tilde \Omega$ factor sit at the same point in space-time. Because of the binomial pre-factor in \eqref{eq: zymnew}, the total number of these points is bounded by $2$ for $G=0$ and is equal to $0$ for $G=1$, while it can take any value for $G>1$.

Any term in \eqref{eq: zymnew} that comes with a factor of $N^{2-2g}$ is interpreted as the contribution of worldsheets (not necessarily connected) with genus $g$. The sum over all mappings is the exponent of the sum over connected mappings. It was argued in \cite{Gross:1993hu} that the sum in \eqref{eq: zymnew} includes all possible extremal-area mappings from any worldsheets to the space-time, with singularities of the types mentioned above.\footnote{Note that constant maps of the worldsheet to a point also extremize the area; their absence in \eqref{eq: zymnew} can be interpreted as due to a division by the volume of an infinite dimensional automorphism group acting on such maps.}  For $\lambda=0$, corresponding to the free (topological) gauge theory, the sum simplifies since only mappings with $s=t=0$ contribute, and it was argued in \cite{Cordes:1994sd,Horava:1995ic} that the coefficient of the term corresponding to mappings in a specific topological class is precisely the Euler number of the moduli space of extremal-area mappings in this class (up to a minus sign for every orientation-reversing tube). Some of these moduli spaces are singular, and this assumes a specific regularization of these singularities, described in \cite{Cordes:1994sd,Cordes:1994fc,Horava:1995ic}; we will discuss this in more detail in some examples below.

Our main goal in this paper will be to provide a worldsheet theory that reproduces \eqref{eq: zymnew} for $\lambda=0$. Note that for $\lambda=0$ the sum in \eqref{eq: zym} diverges for $G=0,1$ in any gauge theory (and for $U(N)$ theories it diverges for any $G$, because there is an infinite number of $U(N)$ representations with any dimension, that differ just by their $U(1)$ charge); and in fact for $G=0$ there is a large $N$ phase transition in \eqref{eq: zym} at a specific value of $\lambda$ \cite{Douglas:1993iia,Minahan:1993tp,Caselle:1993gc,Caselle:1993mq,Billo_caselle:1998fb,Gross:1994mr,Taylor:1994zm} such that the sum \eqref{eq: zymnew} only reproduces the correct partition function for $G=0$ above this critical value. Since our goal is to eventually use the worldsheet theories discussed in this paper for finite $\lambda$, and our discussion of the string theory derivation of the coefficients in \eqref{eq: zymnew} applies also in that case, we will ignore these issues here.

\subsection{Stringy interpretation of the \texorpdfstring{\ocmr-points}{omega-points}}\label{subsec: stringy interpretation of the ocmr points}

In section \ref{sec: various amplitudes} we will make extensive use of \eqref{eq: zymnew} and \eqref{eq: ocmr point expansion} for $\lambda=0$, so we need to elaborate a bit further on the various factors appearing there. The reader may consider skipping ahead and returning when needed. Here are a few comments:
%A few comments about the various factors appearing in \eqref{eq: zymnew} and \eqref{eq: ocmr point expansion} for $\lambda=0$:
\begin{itemize}
    \item As noted in \cite{Cordes:1994sd,Cordes:1994fc}, the factor:
    \be\label{eq: binom as Euler character}
    (-1)^k \binom{2G+k-3}{k} = \frac{1}{k!} \times (2-2G - (k-1))(2-2G-(k-2))\cdots(2-2G)
    \ee
    in \eqref{eq: zymnew} has the following interpretation: 
    \begin{itemize}
        \item $1/k!$ is a symmetry factor related to the exchange of \ocmr-points. In any particular contribution, $\tilde{\Omega}_{n,\tilde n}$ is replaced by some specific term in the sum over permutations \eqref{eq: ocmr point expansion}. When different factors of $\tilde\Omega$ are distinct, their permutations will be summed over and thereby partially cancel the symmetry factor, reflecting their non-equivalence.
        \item The factor:
        \bea \label{eulerk} & (2-2G - (k-1))(2-2G-(k-2))\cdots(2-2G)\\ &\qquad\qquad= \chi\left(\mathcal{M}_{G}\addslash \{p_1,\dots,p_{k-1}\}\right)\chi\left(\mathcal{M}_{G}\addslash \{p_1,\dots,p_{k-2}\}\right)\cdots \chi\left(\mathcal{M}_{G}\right) \eea
        is the Euler number of a configuration space of $k$ distinct points in the target space; first one chooses the first point to sit anywhere in the space-time $\mathcal{M}_G$, then the second point to sit anywhere except at the position of the first point, etc. Even when \eqref{eulerk} vanishes, it does not necessarily mean that the measure on the moduli space of such maps in the string theory vanishes -- only its Euler number.
        \item Some of the punctures appearing in \eqref{eq: ocmr point expansion} may be thought of as collisions of other punctures.
        When the punctures that collide form a puncture that also appears in \eqref{eq: ocmr point expansion}, the collision is called of type 1 or 2 (see \cite{Cordes:1994fc}). Collisions of type 1 occur when two branch points whose permutations have overlapping cycles combine into a new permutation (say, $(12)(23)=(123)$). Collisions of type 2 occur when non-overlapping cycles combine trivially (say $(12)\times(34)=(12)(34)$) -- in this case the colliding punctures simply happen to occupy the same point in the target space, but nothing really happens to them. In these cases, the contribution with $k$ \ocmr-points should combine with particular ``composite'' contributions with $(k-1)$ or fewer \ocmr-points including these collisions to, in essence, ``fill in'' the punctures. This is what is described in \cite{Cordes:1994fc} as ``a natural compactification of the moduli space''. Otherwise, the collision is of type 3 and should truly behave as a puncture in the moduli space (which should be removed). Collisions of type 3 happen, for instance when the permutations cancel out (as in $(12)(12)=1$ or $(123)(132)=1$), or when two tubes connecting overlapping sets of sheets collide.
    \end{itemize}
    \item The factor
    \be
  \ell^{v_{\ell}} v_{\ell}! \binom{\sigma_{(\ell)}}{v_{\ell}} \binom{\tau_{(\ell)}}{v_{\ell}}
    \ee
    in \eqref{eq: ocmr point expansion} can be interpreted as:
    \begin{itemize}
        \item $\ell^{v_{\ell}}$: There are $\ell$ ways of choosing which of the $\ell$ oriented sheets is connected through the tube to which of the $\ell$ anti-oriented sheets (the rest of the connections are then determined). This choice is repeated for each of the $v_{\ell}$ tubes.
        \item $v_{\ell}!$: This factor counts which of the $v_{\ell}$ stacks of $\ell$ cyclically-permuted sheets with one orientation is connected to which stack of the other orientation by a tube. Note that different ways of connecting the sheets may correspond to different worldsheet topologies (with the same total genus, but with different degrees of connectedness).
        \item $\binom{\sigma_{(\ell)}}{v_{\ell}}$: This counts which $v_{\ell}$ of the $\sigma_{(\ell)}$ cycles are connected branched tubes. Different choices correspond to composite \ocmr-points that have different decompositions into simpler ones. 
    \end{itemize}
\end{itemize}

%% file: horava_model/Horava_model.tex
\section{Review of \texorpdfstring{\horava}{Horava}'s topological rigid string}
\label{sec: horavasec}

In this section, we review \horava's proposal \cite{Horava:1995ic,Horava:1993aq} for a topological string theory meant to reproduce the Gross-Taylor expansion \cite{Gross:1993yt}.

A common feature of all the covering maps deduced from the expansion \eqref{eq: zymnew} is that they extremize the induced area of the worldsheet (although they do not necessarily minimize it), and thus they are solutions of the Nambu-Goto (NG) equation. This is consistent with the area-preserving diffeomorphism (``sDiff'') symmetry of the field theory. One should therefore seek a worldsheet theory that localizes to the space of such maps, and then, ostensibly, it is just a matter of obtaining the correct measure on that moduli space to reproduce the partition function.

For the special case of the terms with ${\tilde n}=0$ in the partition function \eqref{eq: zymnew} -- the \textbf{chiral} partition function -- these maps can all be chosen (in a conformal gauge for the worldsheet metric) to be holomorphic ($\bar{\partial}Z=0$ for $Z=x^1+i x^2$).

The objects that occur in those terms are various branch points, as well as orientation preserving tubes and handles, all of vanishing size. Branch points are naturally modeled by holomorphic maps which locally look like $Z=z^n$, where $Z$ and $z$ are, respectively, target space and worldsheet complex coordinates. Simple branch points correspond to $n=2$. An orientation preserving tube is the limit obtained as two simple branch points, whose associated permutations cancel out, approach one another. Zero-size handles may also be viewed as limits of purely holomorphic maps. Using the fact that topological string theories giving holomorphic maps are relatively well understood, several proposals have been suggested for such string theories that reproduce the chiral partition function \cite{Cordes:1994sd,Cordes:1994fc,Vafa:2004qa,KOMATSU_forthcoming,GABERDIEL_forthcoming,Rudd:1994ta}. For the special case of the torus, $G=1$, there is a simple relation between the full partition function and the chiral one, but this is not true on other manifolds.

Finding a string theory description of the full partition function is more complicated.
The issue is that in the complete partition function there are also (implicitly) orientation \textbf{reversing} tubes (``plumbing fixtures''), which connect sheets of opposite orientation, and are thus not holomorphic. They may be described as ``piecewise holomorphic - anti-holomorphic'', or \textbf{quasi-chiral}. In Wilson loop expectation values, when the loops intersect themselves or one another, there often appear, even at leading order in $N$, so-called ``twists'' \cite{Gross:1993yt}, which similarly reverse orientation; for instance this happens in the ``figure eight'' Wilson line. This requires going beyond holomorphic maps, and it seems that 
%These various ``non-chiral'' objects, as well as the sDiff symmetry of the theory, indicate that 
minimization (or, more generally, extremization) of the induced area is a more natural criterion than holomorphicity. The two are related, since in a conformal gauge, extremal area maps are quasi-chiral. 

It was argued in \cite{Cordes:1994sd,Cordes:1994fc,Horava:1995ic} that for zero Yang-Mills coupling, the weight of each class of maps in the Gross-Taylor series is precisely the Euler number of the moduli space of maps of that class, as we outlined at the end of section \ref{sec: 2d YM review}. Thus, to reproduce \eqref{eq: zymnew},
the integral over the moduli space of such maps should be weighed by a combination of \textbf{area two-forms} and \textbf{curvature two-forms},  with the area two-forms associated with the target space positions of some of the possible objects. In general, the positions of various special points in the mapping parametrize the moduli space. The moduli space integration of terms with \textbf{area} two-forms accounts for the polynomial-in-area factors in \eqref{eq: zymnew} (which vanish at zero coupling), while terms with \textbf{curvature} two-forms account for the Euler numbers of the moduli space that appear there, via the Chern-Gauss-Bonnet theorem.

Accordingly, Ho\^rava proposed a ``topological rigid string theory'' that localizes to extremal area maps. The name ``rigid string'' refers to the rigidity term $\Delta x \cdot \Delta x$, which in this formulation appears as the bosonic action for the coordinate field $x^\mu$ after integrating out an auxiliary field $B^\mu$. The effect of this term in effective string theory was studied by Polyakov and others \cite{Polyakov:1986cs,Kleinert:1986bk}, and in general creates a penalty for folds (which should be absent in the 2d Yang-Mills theory). However, the actual suppression of folds (and the restriction to the moduli space of extremal maps) in this theory is better understood as using supersymmetric localization to localize to the solutions of the NG equations $\Delta x^\mu=0$, and it is constructed using the Mathai-Quillen formalism \cite{Blau:1992pm}. 

The theory suggested by \horava\ is well-defined for any number of target space dimensions and for arbitrary geometry, and (for pedagogical reasons) we will review it first in that general case, but only for a 2d target space  it is argued to reproduce the Gross-Taylor expansion. The arguments in \cite{Horava:1995ic} are partly formal, and cover the computation of the YM partition function (no Wilson Loop or other insertions), primarily in the topological $\lambda=0$ case. As we shall see in section \ref{subsec: horava 2d target space}, there are certain pitfalls in attempting to take them beyond the level of formality, as well as certain pathologies associated specifically with the 2d case. In this section we will review \horava's suggestion and its problems, and our aim in the following sections is to remedy those problems.

Next, in section \ref{subsec: horava symmetries and fields} we will review the basic setup of the theory -- the symmetries, field content, superspace and some motivation. In section \ref{subsec: horava dynamics} we will discuss the dynamics in generality. We then turn to the specific properties of Nambu-Goto (NG) strings in $2d$ and what they imply for \horava's theory in section \ref{subsec: horava 2d target space}.

\input{horava_model/symmetries_and_fields}

\input{horava_model/2d_case}

%% file: horava_model/symmetries_and_fields.tex
\subsection{Symmetries and fields}\label{subsec: horava symmetries and fields}

We will take the target space to be Euclidean for simplicity (and since this is what we need for the 2d YM partition function).
By analogy with BRST gauge-fixing (and with twisted supersymmetric theories), \horava\, introduces a ghost $\psi^\mu$, and a nilpotent BRST operator $$\left[Q,x^\mu\right]=\psi^\mu,\, \left\{Q,\psi^\mu\right\}=0$$ to impose the ``gauge condition'' $\Delta x^\mu=0$ (the NG equation, equivalent to extremization of the induced area). An antighost multiplet $$\left\{Q,\chi^\mu\right\}=B^\mu,\, \left[Q,B^\mu\right]=0$$ is also added, and used to construct the main $Q$-exact term in the action:
\begin{align} \label{firstaction}
-it S_0 = it\{Q,\intop_\Sigma d^2 \sigma \sqrt{h} g_{\mu\nu}(x) \chi^\mu \Delta x^\nu  \},
\end{align}
where $g$ is the target space metric and $h_{ab}=g_{\mu\nu}\partial_a x^\mu \partial_b x^\nu$ is the induced metric on the worldsheet $\Sigma$. $\Delta$ which appears in the action and in the NG equation is the Laplace-Beltrami operator on the worldsheet, constructed using the induced metric. Carrying out the $Q$-transformation in \eqref{firstaction} produces a Lagrange multiplier term, linear in $B$, imposing the NG equation, and a Faddeev-Popov (FP) determinant term involving the ghost and antighost. This action is supposed to describe the zero-coupling limit of the Yang-Mills theory, namely the topological $BF$ theory, and we will briefly discuss deforming it to finite coupling in section \ref{paragraph: finte area}.

It is convenient to introduce also an anti-BRST operator $\bar Q$:
\begin{align}
    [\bar Q, x^\mu] & = \chi^\mu,\\
    \{\bar Q, \chi^\mu\} & = 0,\\
    \{ \bar Q,\psi^\mu\} & =-B^\mu, \\
    [ \bar Q,B^\mu] & = 0,
\end{align}
which anti-commutes with $Q$, thus combining all the fields into a single double-BRST multiplet. Now, after an integration by parts putting $S_0$ in a first-order form, one can write compactly:
\begin{align}
-it S_0 = -it\{Q,[\bar Q,\intop_\Sigma d^2 \sigma \sqrt{h}] \}.\label{eq:S0 no 1}
\end{align}

\subsubsection{Gauge symmetries}

In addition to the worldsheet diffeomorphism invariance required of any string theory, $S_0$ enjoys a number of additional gauge redundancies. Let us review them and explain their origin. 

There is something special about the Laplacian that is constructed specifically from the induced metric -- it satisfies \textbf{orthogonality}:
\begin{align}
    g_{\mu \nu}\partial_a x^\mu \Delta x^\nu=0 .
\end{align}
In other words, the NG equation is in the \textbf{normal bundle}. This property is the invariance of the action under infinitesimal target-space diffeomorphisms, and it will be true for any such action constructed only with the $x$'s. Indeed, one can write:
\begin{align}
    \Delta x^\mu = p^{\mu \nu} g_{\nu \rho} h^{ab}\Tilde{\nabla}_a\partial_b x ^\rho,
\end{align}
where $\Tilde{\nabla}_a$ is a derivative covariantized only with respect to $g_{\mu\nu}$ (not $h_{ab}$), and:
\begin{align}
    p^{\mu \nu} \equiv g^{\mu \nu} - h^{ab}\partial_a x^\mu \partial_b x^\nu \label{eq: normal projector definition}
\end{align}
is a projector onto the directions normal to the worldsheet\footnote{Obviously the behavior in two space-time dimensions, where there are no normal directions, will be very different, and we will discuss this specific case below.}. Antighosts and Lagrange multipliers always belong to the same bundle (have the same ``quantum numbers'') as the corresponding equation. Thus, $\chi$ and $B$ should also belong to the normal bundle -- their transverse components should be constrained to vanish, or be interpreted as gauge redundancies. Similarly, since $\psi$ represents a variation of $x$, whichever part of it is transverse to the worldsheet can be removed by a diffeomorphism, and so should be regarded similarly. This leads to an extension of the gauge symmetry to:
\begin{equation}
    \begin{split}
            \delta x & = \epsilon^a \partial_a x\\
            \delta \psi &= {\epsilon'}^a \partial_a x + \epsilon^a \partial_a \psi\\
            \vdots
    \end{split}
\end{equation}

parameterized by a double-BRST multiplet of four vectors $\epsilon,\epsilon',\dots$ (note that two of the vectors are fermionic). We will write it out in full, more compactly, using superfields below. The second bosonic gauge-symmetry (acting on $B^{\mu}$) is called Itoi-Kubota symmetry \cite{Itoi:1987de,Itoi:1988ji}.

Before describing the other global symmetries and other possible terms in the action, let us review the superspace formulation of this theory.

\subsubsection{Superspace formulation}\label{subsubsec: horava superspace formulation}

The superspace formulation of the theory is described in \cite{Horava:1998wf}. 
The various fields described above are combined into one superfield depending on Grassmann coordinates $\theta,\thetabar$:
\begin{equation}
    X^\mu = x^\mu + \theta \psi^\mu + \bar{\theta}\chi^\mu + \theta \bar{\theta} B^\mu.
\end{equation}
We then have:
\begin{equation}
    Q=\partial_\theta,\,\bar{Q} = \partial_\thetabar.
\end{equation}

In general, we will refer to any object that depends on the two supercoordinates as bi-graded. We review them in the appendix \ref{sec: bi graded algebra}. In particular, superfields are bi-graded fields. We will (mostly) designate such objects using uppercase letters and their components using lowercase letters:
\begin{equation}
    F(\theta,\thetabar) = f + \theta f_\theta + \thetabar f_\thetabar + \theta \thetabar f_{\theta \thetabar},
\end{equation}
so that, for example, $\psi \equiv x_\theta$. The induced metric is also extended to a superfield:
\begin{equation}
    H_{ab} = \partial_a X \cdot \partial_b X.
\end{equation}
Dot products denote contraction with the space-time metric $g_{\mu \nu}$.

\subsubsection{Gauge symmetries in Superspace formalism}

The diffeomorphisms, fermionic gauge symmetries and Itoi-Kubota transformations described above combine into bi-graded diffeomorphisms. These are transformations that map bosonic coordinates to functions of all the superspace coordinates. They are parameterized by bi-graded vectors
\begin{equation}
    \epsilon^a\left(\sigma^a ; \theta, \bar{\theta}\right)\label{eq: generalized diffs},
\end{equation}
and take the form
\begin{equation}
    \delta X = \epsilon^a \partial_a X.
\end{equation}

Any superfield constructed from $X$ and its \textbf{bosonic} derivatives transforms under bi-graded diff-s in the expected, tensorial way. For instance:
\begin{align}
    \delta H_{ab} & = \epsilon^c \partial_c H_{ab}+ \partial_{(a} \epsilon^c H_{b)c}\\
    \delta \sqrt{H} & = \epsilon^c \partial_c \sqrt{H}+ \partial_c \epsilon^c \sqrt{H}.
\end{align}

Superfields constructed with super-derivatives potentially transform non-tensorially:
\begin{align}
    \delta (\partial_\theta X) = \epsilon^c \partial_c\partial_\theta X+\underbrace{\partial_\theta\epsilon^c \partial_c X}_{\text{non-tensorial}}.
\end{align}
To construct a scalar $S$, we can contract this with the projector superfield:
\bea
P^{\mu \nu} & \equiv g^{\mu \nu}(X) - H^{ab}\partial_a X^\mu \partial_b X^\nu= p^{\mu \nu}+\dots
\eea
\bea\label{eq: Horava S invariant}
    S & \equiv  \partial_\theta X \cdot P \cdot \partial_\thetabar X = \psi \cdot p \cdot \chi + \dots
\eea

where $\cdots$ are the terms proportional to $\theta$ or/and $\bar\theta$.

\subsubsection{Global symmetries}

In this superspace formalism the action \eqref{eq:S0 no 1} can be written in more compact form:
\begin{equation}\label{eq: horava S_0 action in superspace}
    S_0 = \intop d^2 \sigma d^2\theta \sqrt{H},
\end{equation}
where $d^2\theta\equiv d\theta d\thetabar$. Clearly, this action is invariant under the Fermionic BRST and ``anti-BRST'' symmetries $Q=\partial_\theta,\,\bar Q = \partial_\thetabar$. In addition, there is a ghost number symmetry $\theta \partial_\theta -\thetabar \partial_\thetabar$, and a symmetry between ghosts and anti-ghosts, $\theta \partial_\thetabar,\thetabar \partial_\theta$. These three bosonic symmetry generators combine into an $SU(2)$ R-symmetry, under which $Q,\bar Q$ rotate in the fundamental representation.

In \cite{Horava:1998wf}, Ho\^rava calls the combination of global and gauge symmetries above ``semi-rigid'' Parisi-Sourlas supersymmetry, and shows how it can be obtained as the residual symmetry of an enlarged Parisi-Sourlas SUSY acting on an enlarged, but more redundant, field space. Although noteworthy, we will not have use of this symmetry in this paper.

We note that the theory is topological in the target space (as expected at zero Yang-Mills coupling) as a result of BRST symmetry. The action is not invariant under changes to the target space metric, but its change is a $Q$-exact term, so formally the theory is invariant.

\subsection{Dynamics}\label{subsec: horava dynamics}

%\subsubsection{The action}

Using the previously mentioned symmetries a more general $Q$-exact action can now be constructed using an arbitrary polynomial in the nilpotent scalar $S$:
\be 
S_{f} = \intop d^2 \sigma d^2\theta \sqrt{H}f(S).
\ee
The terms that depend on $S$, especially the linear term, are analogous to the $\xi$ term that appears when averaging over gauge-choices in ``$\xi$-gauge''. With them included, the action is no longer linear in the field $B$, and so it no longer acts as a Lagrange multiplier. Of course, one can still localize to solutions of the Lagrange constraint using super-symmetric localization, by making the coupling constant $t$ in $S_0$ \eqref{eq:S0 no 1} large. If $f$ is linear, then the action is quadratic in $B$ and it can be integrated out. The resulting action of $x$ is given by the rigidity term, which gives this theory its name in \cite{Horava:1995ic}. In addition to these terms, one can use arbitrary scalar functionals, just as in the ordinary NG string, but with the superfield $X$ replacing the field $x$.

Actions of this type are automatically $Q$ and $\bar Q$-exact, so the physics will not depend on the precise form of the function $f$. To construct terms that are neither $Q$ - exact nor $\bar Q$ - exact, we can include an explicit factor of $\theta \thetabar$ in the Lagrangian. Which such terms are $Q$-closed? If we include no super-derivatives, the expression will be constructed purely from $x$, and the action of $Q$ on it must give a total derivative. This simply means that any variation of it should be a total derivative. There are likely just two such terms in general:
\begin{enumerate}
    \item $\intop d^2 \sigma d^2\theta \sqrt{H} R \theta \thetabar$. This is the usual string-coupling term. In our case, we expect this term to appear with a coefficient related to $g_{\rm str} = 1/N$.
    \item The $B_{\mu \nu}$ term (not the same $B$ as in the multiplet of $x^\mu$) $\epsilon^{ab} B_{\mu \nu} \partial_a X^\mu \partial_b X^\nu$. We will ignore this term since in the 2d case it breaks space-time parity.
\end{enumerate}

\subsubsection{Quantization}\label{subsubsec: quantization (horava)}

To quantize the theory, one needs to gauge-fix and check for possible anomalies. Because both the fields and gauge symmetries are simply bi-graded versions of the standard NG string, all the same gauge conditions $G(x)$ can be used, including static gauge and orthogonal gauge, simply generalized to $G(X)$. In addition, \cite{Horava:1995ic} mentions Itoi-Kubota gauge, which fixes only the bi-graded extension of the diff-symmetry by setting $\partial x \cdot \psi= \partial x \cdot \chi = \partial x \cdot B=0$, however, it will be convenient for us to employ gauges that do not break the explicit $Q,\bar Q$ symmetries.

We will study the consistency of quantization in the Polyakov-like generalization of this theory in section \ref{subsec: polyakov central charge}; the same arguments for the bose-fermi cancellation of anomalies ought to apply here as well.

Upon gauge-fixing, we perform a localization to the solutions of the NG equation. If there is a moduli space of solutions, parametrized by some coordinates $a^i$ ($i=1,\cdots,n$), then these will all be accompanied by $2n$ fermi 0-modes and $n$ bosonic $B$ 0-modes, that can be identified with the tangent space to the moduli space at the point $a^i$ (see the appendix \ref{subsec: Bi-graded equations}). Concretely, if the solution space of the bosonic NG equation is given by:
\be
x(\sigma^a) = x_{\rm cl}(\sigma^a, a^i), 
\ee
then this generalizes to:
\be
X(\sigma^a,\theta,\thetabar) = x_{\rm cl}(\sigma^a, A^i(\theta,\thetabar)).
\ee
This implies:
\be
\partial_\theta X = \partial_\theta A^i \partial_i X.
\ee
If no term other than $S_0$ is included in the action, then the path integral has the undefined form of ``$0^{2n} /0^{n} \intop d^n a $''. So we need to add extra terms that will (at least) absorb the fermion zero modes.

If $Q$-exact terms are added which absorb the zero modes, then the theory will formally compute the Euler number of the moduli space, as explained in section \ref{subsubsec: polyakov moduli space} and in appendix \ref{subsec: bi-graded integration}.

%% file: horava_model/2d_case.tex
\subsection{\texorpdfstring{$2d$}{2d} target space}\label{subsec: horava 2d target space}

We will now specialize to the two-dimensional case, which is supposed to be related to $2d$ YM theory. We start by reviewing the standard bosonic NG equation in $2d$, and then return to our topological theory. Finally, we study some pathologies that arise in this case.

In $2d$, every vector in the target space is generically parallel to the worldsheet, and as a result almost all the information in the function $x^\mu(\sigma^a)$ (and its superpartners) is redundant (under the gauge transformations discussed above). In particular, the projector $p$ from \eqref{eq: normal projector definition} \textbf{vanishes at generic points}. This is not true at folds (where $\partial X$ vanishes in some direction). One might hope that it becomes a Dirac delta function on folds, or anywhere else where $\partial x$ is a singular matrix, but in fact it only equals $1$ there. This is to be expected from an object that squares to itself. $p$ can nevertheless be nontrivial, if it is contracted with expressions that diverge at such folds.

\subsubsection{Non-SUSY NG string in \texorpdfstring{$2d$}{2d}}\label{subsubsec: Non-SUSY NG string in 2d equation of motion}

The NG action is:
\be
S_{\rm NG}=\intop_\Sigma \sqrt{h},
\ee
for $h$ the induced metric
%. Note that if $g$ is the target space metric, then 
$h_{ab}=\partial_a x^{\mu} g_{\mu \nu} \partial_b  x^{\nu}$. For $d=2$, $\partial_a x^\mu$ is a square matrix and we can write:
\be
S_{\rm NG}=\intop_\Sigma \sqrt{g}|\det(\partial x)|.
\ee
This is invariant under target space area-preserving diffeomorphisms.
Since this action is just the absolute value of the topological $B$ term for $B_{\mu\nu}=\sqrt{g}\epsilon_{\mu\nu}$, and this term is a total derivative, it is clear that the equation of motion (EOM) should be trivial over patches where the induced orientation of the worldsheet (the sign of $\det(\partial x)$) is constant. The induced orientation changes only as one moves across a fold.
Indeed, the EOM is:
\be
\Delta x= 0,
\ee
but diffeomorphism invariance implies the orthogonality condition, $g_{\mu \nu} \partial_a x^\mu \Delta x^\nu=0$, even off-shell. At generic points on the worldsheet, every direction in the target space is parallel to the worldsheet, so for any vector $v^\mu$ we obtain $v\cdot\Delta x=0\Rightarrow\Delta x = 0$, and the EOM is seemingly trivial. 
This statement is not true only in codimension-1 regions traversed by folds. The NG equation is therefore potentially nontrivial only at points for which $\partial x$ (and therefore also the induced metric) is a \textbf{singular matrix} -- folds and branch points. In fact, the target space position of folds and branch points is a full characterization of the (local) gauge-invariant information in an embedding of the worldsheet. In practice, the NG equation will simply suppress (finite-sized) folds, as we shall now see. This is unsurprising, since folds always represent an unnecessary excess of area.

\paragraph{Folds:}
In the vicinity of a fold, we can simplify the NG equation by picking what we may call a ``fold gauge'' for the worldsheet diffeomorphisms, where the fold is at $\sigma=0$, and (in the case that the fold occupies a straight line in the target space):
\begin{equation}\label{eq: fold gauge}
    x^0 =\tau, \qquad
    x^1 = \frac{1}{2}\sigma^2.
\end{equation}
The induced metric $h$ becomes degenerate at the fold (located at $\sigma=0$), as can be diagnosed from its determinant:
\be
\sqrt{h}  = \sqrt{g} |\sigma|.
\ee
This non-smoothness results in a delta function:
\bea
\sqrt{h}\Delta x^\mu  & = \partial_a \left(\sqrt{h}h^{ab}\partial_b x^\mu \right)\\
    & = 0+\partial_\sigma \left(\sqrt{g} g^{1\mu} \sigma^{-1} |\sigma| \right)\\
    & = \sqrt{g} g^{1\mu} \left(2\delta(\sigma)\right).
\eea
So the EOM cannot be satisfied in the presence of a fold. There is an alternative derivation of this that can be generalized to the SUSY case. Note that for all values of $a,\mu=0,1$:
\begin{equation}
    \sqrt{h}h^{ab}\partial_b x^\mu  = \sqrt{g} |\sigma| g^{a\mu} (\sigma)^{-a}
\end{equation}
is finite at the fold in this gauge. Since the gauge transformation leading to this gauge choice is non-singular, this quantity is regular in any gauge. It appears in the variation of the NG action:
\begin{equation}
    \delta S_{\rm NG} = \intop_{\Sigma} \sqrt{h}h^{ab}\partial_b x^\mu \partial_a\delta x_\mu,
\end{equation}
which means we can safely evaluate this variation by splitting the worldsheet into fold-free patches $U_i$ and summing the contribution from each patch. In each patch the EOM is trivially satisfied, so we only get boundary terms:
\begin{equation}
    \delta S_{\rm NG} = \sum_i \pm \intop_{\partial U_i}  \star dx^\mu \delta x_\mu,
\end{equation}
where the sign depends on the orientation at which the string ends on the fold. The contributions of neighboring patches add up, representing the combined tension of the two strings that meet at the point of turnaround, and whose forces do not cancel, but instead add up, so
\begin{equation}
    \delta S_{\rm NG} = \sum_{folds} \pm 2 \intop \star dx^\mu \delta x_\mu,
\end{equation}
where the sign depends on the orientation of the fold. Thus, the variation of the action vanishes if and only if there are no finite-size folds.

We can rewrite the Laplacian in a more diff-invariant way as:
\be\label{eq: 2d induced laplacian w/ folds}
\Delta x^\mu = \pm 2\intop_{\rm folds}d\tau \sqrt{g}g^{\mu\nu}\epsilon_{\nu\rho}\dot{x}^{\rho}\frac{1}{\sqrt{h}}\delta^2\left(\sigma^a_{\rm fold}(\tau)-\sigma^a \right),
\ee
where $\sigma^a_{\rm fold}(\tau)$ is a parametrization of the worldsheet position of a fold and $\dot x=\dot{\sigma}^a \partial_a x$ is its target space velocity vector. This reveals that the equation can be satisfied when $\dot x=0$, which is to say, when the fold is of vanishing size, as in a degenerated tube.

\paragraph{Branch-points:}

In an appropriate gauge, a branch point of order $n$ locally looks like the holomorphic map $Z= z^{n+1}$, where $Z\equiv x^1+i x^2$ and $z=\sigma^1+i \sigma^2$. Evidently, the matrix $\partial x$ not only becomes singular, but vanishes entirely there. Nevertheless, it can easily be checked that $\Delta x=0$, since the induced metric is Weyl-equivalent to the flat metric, and so the Laplacian is simply $\partial \bar\partial Z=0$.

There is, however, a different natural geometric quantity that does measure such branch points. Note that branch points appear as conical excess defects, and so should appear in the curvature of the induced metric. Indeed, according to the Gauss equation, the Riemann tensor splits into a sum of two terms, one is the pullback of the target space Riemann tensor, and the other an extrinsic curvature term. In the Ricci scalar, the extrinsic contribution $K$ gives in the above case (see \eqref{eq: extrinsic curvature at branch point}):
\be\label{eq: 2d ext' curvature w/ branch points}
 K = p_{\mu\nu}h^{ab}h^{cd}\partial_{a}\partial_{[c}x^{\mu}\partial_{b]}\partial_{d}x^{\nu} = -\frac{1}{2}\pi n\frac{1}{\sqrt{h}}\delta^{2}\left(\sigma^a\right).
\ee
The extrinsic curvature is also sensitive to folds. In the presence of folds, it becomes:
\be\label{eq: 2d ext' curvature w/ folds}
K \propto \intop_{\rm folds}d\tau \frac{\sqrt{g}\epsilon_{\nu\rho}\dot{x}^{\rho}{\ddot{x}}^\nu}{\dot{x}^2}\frac{1}{\sqrt{h}}\delta^2\left(\sigma^a_{\rm fold}(\tau)-\sigma^a \right).
\ee
According to the Gauss-Bonnet theorem, the integral of the Ricci scalar computes the Euler number of the worldsheet. The split of the Ricci scalar into the target space Ricci scalar and the extrinsic contribution, together with the above formulae reproduce the Riemann-Hurwitz formula \cite{Cordes:1994fc}, which relates the Euler number of the worldsheet and the target space for branched covering maps.

\subsubsection{Supersymmetric version}\label{subsubsec: horava Supersymmetric version}

As we saw in \eqref{eq:S0 no 1}, the action is
\begin{align}\label{eq:S0 BRST inv}
-it S_0 = -it\{Q,[\bar Q,\intop_\Sigma d^2 \sigma \sqrt{h}] \}.
\end{align}
Just as in the purely bosonic case, all fields are purely gauge-redundant away from folds. Although the field content is enlarged, the amount of gauge symmetry is enlarged by the same amount. In this case, not only is the EOM localized to folds -- but so is the action itself. This is because the action \eqref{eq:S0 no 1} is given by a BRST variation of the NG action. We can readily find:
\begin{equation}
    S_{0} = \pm 2\intop_{\rm folds}d\tau \sqrt{g}\epsilon_{\mu\nu}\left(\dot{x}^{\mu}B^{\nu}+\dot{\psi}^{\mu}\chi^{\nu}\right).
\end{equation}
It is clear then that the $B$ EOM enforces $\dot{x}=0$, forcing the fold to map to a point in the target space as expected. In general, all the fields are restricted to constants at the fold. The only remaining degrees of freedom are then the target space position of the collapsed fold and its double-BRST partners.

Equations \eqref{eq: 2d induced laplacian w/ folds}, \eqref{eq: 2d ext' curvature w/ branch points} and \eqref{eq: 2d ext' curvature w/ folds}
generalize trivially to superfield versions of themselves.

\paragraph{Finite area}\label{paragraph: finte area}

\horava\ proposes \cite{Horava:1995ic} that to introduce area dependence, we should add terms to the action that are ``weakly BRST invariant''. Concretely, this means they are BRST closed when no folds are present (and thus on-shell). For instance, the ordinary NG action can be expressed as
\begin{equation}
    S_{NG} = \intop d^2 \sigma d^2\theta \sqrt{H}\theta\thetabar.
\end{equation}
Its BRST variation is proportional to the NG EOM, and thus vanishes at the saddles of $S_0$.

In the current work, we will focus on the zero coupling case where this term is not needed, and postpone its analysis to future work.

\paragraph{Localization}

Since $S_0$ \eqref{eq:S0 BRST inv} is $Q$-exact, changes in the coefficient $t$ do not affect the partition function. We can expand $X=X_{\rm cl}+t^{-1/2} \delta X$ around the saddles of $S_0$ and take $t\to\infty$. Formally, as described in appendix \ref{subsec: bi-graded integration}, we then need to evaluate the rest of the action on the moduli space of zero-modes of $S_0$, namely the moduli space of extremal area maps, and to evaluate the 1-loop determinant of the Hessian of $S_0$ on those extremal area maps.

One problem that arises when doing this is that, as noted in section \ref{subsubsec: quantization (horava)}, we need to introduce additional terms in the action to absorb the fermion zero-modes associated with this moduli space. For $d>2$, this role can naturally be played by the invariant $S$ of \eqref{eq: Horava S invariant}. However, for $d=2$, $S$ vanishes identically. The reason is that, due to the lack of normal directions away from folds, the projector $P$ vanishes away from folds and branch points. One might still hope that $P$ would become a delta function on folds / branch points, like the NG EOM is, but in fact, being a projector, it is only equal to $1$ there. Thus, unless the superfields $\partial_\theta x,\, \partial_\thetabar x$ are themselves for some reason divergent, $S$ will simply vanish.
One possible way to fix this issue is to generalize $S$ to a scalar $S_K$ that depends on some tensor $K_{\mu\nu}$:
\be
S_K \equiv \partial_\theta X \cdot P \cdot K \cdot P \cdot \partial_\thetabar X,
\ee
which reduces to \eqref{eq: Horava S invariant} for $K=g$. If $K$ is chosen to have a suitable divergence, the resulting action may do the job. We will see that similar problems will arise also in the Polaykov-like formalism that we will introduce in the next section, and that they can be resolved as discussed in
section \ref{subsubsec: polyakov moduli space}. 

% In practice, the 1-loop determinant here is ill-defined, since $S_0$ only couples to the fields on top of folds, but none are present at the saddles, except possibly degenerate ones. At the same time, maps with infinitesimal folds are present in the infinitesimal vicinity of the saddle, and the 1-loop fluctuations are supposed to be seen as living in that infinitesimal vicinity. In other words, the gauge-orbits of field space do not comprise a manifold, but a more pathological object, the points of which are not accompanied by tangent spaces that are truly vector spaces.

% Thus, the formal procedure fails. To remedy this, we will need to expand the field space further somehow, to embed the current field space in a larger, less pathological one. There are a number of possible ways to do this, but we will focus on reformulating the theory in a Polyakov-like formalism, which will introduce a new auxiliary variable -- the worldsheet bi-graded metric -- and thereby de-pathologize the space of maps.

A more general problem is that for $d=2$ the configuration space and the action on it are both singular. The gauge-invariant information in the mapping is just given by the branch points and folds and where they are mapped to in the target space. Whenever we have folds, these mappings are singular, and the configuration space is disconnected, with different components for different numbers of folds. Moreover, the action is non-vanishing just on the folds, so it is difficult to evaluate the one-loop determinant needed for localization. 

To remedy this, we would like to expand the field space and to embed it in a larger, less pathological one. There are a number of possible ways to do this, but we will focus on reformulating the theory in a Polyakov-like formalism, which will introduce a new auxiliary variable -- the world-sheet bi-graded metric. After introducing this variable the different components of the configuration space become connected, and we will see that we will be able to perform explicit computations using localization, and to regularize at least some of the singularities that are encountered. %

%% file: Polyakov_formalism.tex
\section{A Polyakov-like formalism}\label{sec: Polyakov formalism}

In this section we will reformulate Ho\^rava's worldsheet theory in a Polyakov-like formalism, with a dynamical metric $h$, and investigate its systematics, postponing specific computations to section \ref{sec: various amplitudes}. We will focus here just on the Yang-Mills theory at zero coupling, attempting to reproduce the topological BF theory (including the $\Omega$-points and twists of Gross and Taylor \cite{Gross:1993cw,Gross:1993hu,Gross:1993yt}). In particular, we should be able to see that the zero mode integrals compute Euler characters of the moduli space, and that maximal-area maps are accompanied by minus signs coming from fermion determinants (we expect one such sign for every orientation-reversing tube).

A reformulation of this type is possible because the NG equation and the Polyakov equations of motion are equivalent \cite{Polyakov:1981rd}. The superspace description of Ho\^rava's theory (see section \ref{subsubsec: horava superspace formulation}) gives a simple mechanism to write a theory that localizes to the space of solutions of the NG equation, but that mechanism can be instantly adapted to localize to any set of bosonic equations (see the appendix \ref{subsec: Bi-graded equations}), by replacing all dynamical objects with bi-graded versions of themselves. Thus, we will introduce a dynamical metric superfield $H_{ab}(\sigma^a,\theta,\thetabar)$. 

In this section, ($H$) $h$ will refer to the dynamical metric (super-)field on the worldsheet, and $\delta$ to the standard flat metric.

First, in subsections \ref{subsec: polyakov symmetries} and \ref{subsec: polyakov fields} we review the symmetries and the field content. In subsection \ref{subsec: polyakov action} we present the action and its relation to the NG formalism of \horava's theory \eqref{eq: horava S_0 action in superspace}. In subsection \ref{subsec: polyakov gauge fixing} we discuss gauge-fixing. In subsection \ref{subsec: polyakov central charge} we discuss the consistency of quantization. Finally, in subsection \ref{subsec: Polyakov path integral}, we discuss how to carry out the path integral.

\subsection{Symmetries}\label{subsec: polyakov symmetries}

As we pass to the Polyakov formalism and possibly attempt to introduce new ghost fields and terms to the action, we \textbf{need not} necessarily preserve all the symmetries of \horava's formalism (see section \ref{subsec: horava symmetries and fields}), at least not \textbf{manifestly}. Which should we hold on to?

\begin{enumerate}
    \item Worldsheet diffeomorphisms are non-negotiable.
    \item The target space diffeomorphism symmetry is reflected in the gauge symmetry $x\to x+ \delta x$, and the corresponding ghost $\psi$ should remain in the topological theory. This is related to field redefinitions on the worldsheet, so it is not a symmetry of the action, but rather of the path integral, and indeed the action changes by $Q$-exact terms. Thus, this symmetry manifests itself through the worldsheet global fermionic symmetry generated by $Q$.
    \item The Fermionic gauge symmetry $\delta \psi = \epsilon^a \partial_a x$  simply reflects that $\psi$ is a tangent vector to the space of maps.
    \item The additional Fermionic gauge symmetry and Itoi-Kubota symmetry should remain as long as $\chi,B$ are in the normal bundle, which they will be in the Polyakov formalism.
    \item The ``anti-BRST'' transformation $\delta x=  \chi,\, \delta\psi = B$ does not seem completely essential. %It also suffers from the fact that upon including the various extended diffeomorphisms above in the complete BRST operator, there's no clear way to write two BRST operators that anti-commute and keep the ``double-topological'' nature of the \horava string manifest.
    \item The three symmetries that include ghost number and ghost-antighost symmetry are also not clearly essential.
    \item To this we add, as usual in the Polyakov formalism, an extended Weyl symmetry acting on the double-BRST multiplet of the metric $H(\sigma, \theta,\thetabar)\to \Omega(\sigma, \theta,\thetabar) H(\sigma, \theta,\thetabar)$.
\end{enumerate}

In the present work, we will try to keep all the symmetries \textbf{intact} and see what we obtain.

In conformal gauge, described in subsection \ref{subsec: polyakov gauge fixing}, the bi-graded $\rm{diff}\rtimes \rm{Weyl}$ symmetry reduces to an bi-graded conformal symmetry.

\subsection{Fields}\label{subsec: polyakov fields}

In the Polyakov formalism, we'll have the following superfields:

\begin{itemize}
    \item As before we have the bi-graded target space coordinates $X^\mu = x^\mu + \theta \psi^\mu + \bar{\theta}\chi^\mu + \theta \bar{\theta} B^\mu.$
    \item The metric $H_{ab}= h_{ab} +\theta h_{\theta;ab} +\thetabar h_{\thetabar;ab} +\theta\thetabar h_{\theta\thetabar;ab}$.
\end{itemize}

If we gauge-fix to conformal gauge $H_{ab} = \delta_{ab}$ using the FP procedure (see subsection \ref{subsec: polyakov gauge fixing}), we need to add also:

\begin{itemize}
    \item An extended diffeomorphism ghost $C^a = c^a +\theta c_\theta ^a +\thetabar c_\thetabar^a + \theta\thetabar c_{\theta\thetabar}^a$. Note that $c_\theta ^a$ is denoted $\phi^a$ in \cite{Horava:1995ic}.
    \item An extended Weyl ghost $W = w + \theta w_\theta + \thetabar w_\thetabar + \theta\thetabar w_{\theta\thetabar}$.
    \item A symmetric-matrix Lagrange multiplier $\Lambda_{ab} = \lambda_{ab} + \cdots$.
    \item A symmetric-matrix antighost $B_{ab}=b_{ab} + \dots$ which will be made traceless by the integral over the Weyl ghost.
\end{itemize}

\subsection{The action}\label{subsec: polyakov action}

The action is:
\begin{equation}\label{eq: super-polyakov action}
    S=-it S_{0} + S_1 = -i\frac{t}{2} \intop d^2\sigma d^2\theta \sqrt{H}H^{ab}\partial_a X \cdot \partial_b X+S_1,
\end{equation}
where $S_1$ represents additional terms consistent with the symmetries that we'll discuss in section \ref{subsubsec: polyakov moduli space}. $S_0$ will be used in localization as $t$ is taken to $\infty$, and so it plays a distinguished role. We will therefore disregard $S_1$ in this section, until we discuss the integration over zero modes in section \ref{subsubsec: polyakov moduli space}. As in standard string theory, integrating out $H$ returns us to the NG-like formulation. In component language, integrating out $h_{ab;\theta\thetabar}$ sets $h_{ab}$ to be conformally equivalent to the induced metric. Then the Fermionic components of $H_{ab}$, upon being expanded around their classical values, fluctuate to cancel the 1-loop determinant of the bosons.  Overall, the entire effect is to replace $H$ by the induced metric $\partial X\cdot \partial X$. So naively, we get back the NG action. But (again as in standard string theory) since the determinant part of $H$ decouples (due to Weyl invariance), integration over it is ill-defined. A more well-defined procedure to deal with this mode is to gauge-fix the Weyl symmetry using the FP procedure, which would then lead back to the NG formalism. In this paper, this gauge-fixing will be part of the full gauge-fixing we will make to conformal gauge, that we will describe next.

The path integral is formally:
\be
Z=\intop \frac{DXDH}{{\rm Vol}\left({\rm diff}^{2|2}\ltimes {\rm Weyl}^{2|2}\right)}\exp\left(-S[X,H]\right).
\ee

\subsection{Gauge fixing}\label{subsec: polyakov gauge fixing}

To fix the full gauge symmetry, we pick \textbf{conformal gauge} $H_{ab} = \delta_{ab}$. In particular, all the component fields of $H$ are set to 0, besides the metric itself. The theory should be unchanged if we simply include in the action the double-BRST exact term:
\begin{equation}
    -i\intop d^2 \theta\Lambda^{ab} \left( H_{ab} - \delta_{ab}\right).
\end{equation}
Then we can simply integrate out $\Lambda$ and $H$, without worrying about 1-loop determinants, since those are cancelled between fermions and bosons\footnote{These simplifications are a corollary of the fact that the super-determinant of a diagonal operator acting on a super-vector space of equal bosonic and fermionic dimensions is 1, if the operator is the same in the bosonic and fermionic subspaces.} (see appendix \ref{subsec: bi-graded integration}). This indicates that the FP determinant should be trivial as well, but we should carefully include it to show this action is truly equivalent.

Note that any admissible gauge choice in bosonic string theory, including conformal gauge, can be straightforwardly extended to a bi-graded version, as can be seen in appendix \ref{subsec: Bi-graded equations}. Similarly, any residual gauge transformations, in this case conformal transformations, are automatically extended.

Following Polchinski's  analysis \cite{Polchinski_textbook:1998rq} of the bosonic string Polyakov path integral, we end up with a $b,c$  ghost system, where the ghosts are extended to superfields $B_{ab},C^a$, with $B$ traceless. If the worldsheet topology is such that the gauge-fixing involves $\mu$ complex structure moduli and $\kappa$  conformal killing vectors (CKVs), then, according to appendix \ref{subsec: Bi-graded equations}, $B$ ($C$)  will have a $2\mu\mid 2\mu$ ($2\kappa \mid 2\kappa$) dimensional vector-space of zero-modes, parameterized by bi-graded numbers:
\be
    \Xi^k(\theta,\thetabar)=\xi_k + \cdots,\, \qquad A^{j}(\theta,\thetabar)=a^{j}+\cdots, \, \qquad k\in \{1,\dots,\mu\}, \, \qquad j \in \{1,\dots,\kappa\}.
\ee 
We then need $\kappa$ additional gauge fixing conditions, which in standard bosonic string computations are provided by fixing the positions of $\kappa$ vertex operators. More abstractly, consider adding $\kappa$  bi-graded gauge conditions $G_j=0$, which may depend on either the fields of the theory or on singled-out coordinates, and which we'll assume are unaffected by Weyl transformations. We can define the FP determinant by:
\be
    1 = \Delta_{\rm FP}(H,G_j)\intop_{\left(\text{complex structures}\right)^{2|2}} d^{2\mu\mid 2\mu}\hat{T} \intop_{{\rm diff}^{2|2}\ltimes {\rm Weyl}^{2|2}}[DZ]\delta\left(H-{\hat H(\hat{T})}^Z\right)\prod_{j=1}^\kappa \delta^{2\mid2}\left(G_j^Z\right).
\ee
Hence:
\bea\label{eq: Fadeev Popov determinant}
     \Delta_{\rm FP}(\hat H,G_j) & = \frac{1}{n_R}\intop [DB][DC]d^{2\mu\mid 2\mu}
     \Xi \,d^{2\kappa\mid 2\kappa}A  \\
     & \times \exp\left(i\intop d^2\sigma d^2\theta \sqrt{\hat H} B^{ab}\left(2\left(P_1[\hat H (\hat{T})] C\right)_{ab}-\Xi^k \delta_k \hat H_{ab}\right)+A^j C^a \delta_a G_j\right),
\eea
where:
\begin{itemize}
    \item $\hat{T}^k$ parametrize the space of (bi-graded) complex structures and will be referred to as complex structure moduli.
    \item The integer $n_R$ may be needed to account for isolated ``Gribov copies''\footnote{$n_R$ should count solutions of the gauge-fixing condition ``up to sign'', in the sense that solutions around which the FP determinant is negative are counted with negative sign, much like in \cite{Eberhardt:2021ynh}.}, 
    \item $\delta_k {\hat H}$ is a holomorphic quadratic differential, representing an infinitesimal complex structure deformation,
    \item $\delta_a G_j= \partial_a X^\mu \frac{\delta}{\delta X^\mu}G$ is a gauge transformation of the gauge-fixing condition, and
    \item $P_1$ is, again, the differential operator from chapter 5 of \cite{Polchinski_textbook:1998rq}.
\end{itemize}

The integral \eqref{eq: Fadeev Popov determinant} is of the form \eqref{eq: general bi-graded gaussian}, where the vector ``$X$'' includes all the degrees of freedom $C,B,A,\Xi$, provided that the gauge conditions $G$ do not involve Grassmann derivatives, and so we obtain:
\be
\Delta_{\rm FP} = \frac{1}{n_R}.
\ee
The integral over $B,A$ produces a delta function (see appendix \ref{subsec: bi-graded integration}), setting $C=0$. If there is any other coupling to the ghost $C$ in the action, it can be ignored.

In summary, except for checking admissibility of the gauge choice, picking a gauge choice which does not involve Grassmann derivatives of the superfields, and including the factor $1/{n_R}$, the diff-ghosts can be ignored.

If there are complex structure moduli $\hat{T}$, they need to be integrated-over as usual. They appear through the metric $H[\hat{T}]$, but as in the bosonic string \cite{Polchinski_textbook:1998rq}, they can sometimes be ``pushed into'' the periodicity of the worldsheet. For instance, for a toric worldsheet with a bi-graded complex structure modulus $\tau(\theta,\thetabar)$, we can pick $H=\delta$ and coordinate identifications $z\sim z+ 2\pi w +2\pi n \tau(\theta,\thetabar)$ for integers $w,n$. This means that the superfields obey a bi-graded version of periodicity.

After gauge fixing, the partition function is:
\bea\label{eq: FP action}
& \frac{1}{n_R}\intop [DX][DB][DC] d^{2\mu|2\mu}\hat{T}d^{2\mu\mid 2\mu}
     \Xi \,d^{2\kappa\mid 2\kappa}A \exp\left(-S\right),\\
S & = -it S_0 + S_{\rm FP} + S_1,\\
S_{\rm FP} & = -i \intop d^2\sigma d^2\theta B^{ab}\left(2\left(P_1[\delta] C\right)_{ab}-\Xi^k \delta_k \hat H_{ab}\right)+A^j C^a \delta_a G_j.
\eea

\subsubsection{Discrete residual gauge symmetry}\label{subsubsec: discrete residual gauge symmetry}
Sometimes the above gauge fixing conditions, will leave us with a \textbf{discrete} residual gauge symmetry, as in section \ref{subsec: Torus on torus}, such as a discrete subgroup of translations, or for a toric worldsheet, an $SL(2,{\mathbb Z})$ modular group. We note that the bi-graded version of discrete groups is the same as the purely bosonic group (there are no tangent directions and therefore no fermions, see appendix \ref{sec: bi graded algebra}). This remaining gauge symmetry can be dealt with either by restricting some summation or integration to an appropriate \textbf{fundamental domain}, or by dividing by the cardinality of the group.

\subsection{Central charge and anomalies}\label{subsec: polyakov central charge}

The above only makes sense if there is no Weyl anomaly, and more broadly, no anomaly in any of the symmetries extending Weyl. Naively, this must be true, since each degree of freedom is accompanied by an identical but statistics-reversed BRST partner, which implements the notion of integration with differential forms, and makes the UV divergences cancel. To confirm this, we define the matter stress tensor superfield:
\begin{equation}
    T_{ab} = :\partial_a X \cdot \partial_b X: - \frac{1}{2} \delta_{ab} \delta^{cd} :\partial_c X \cdot \partial_d X :.
\end{equation}
The only non-vanishing components in complex coordinates are:
\begin{equation}
    T_{zz} = :\partial X \cdot \partial X:
\end{equation}
and its complex conjugate.  The two-point function is simply:
\begin{equation}
    \left\langle T_{zz}\left(0,\theta,\bar{\theta}\right)T_{zz}\left(z,\theta',\bar{\theta}'\right)\right\rangle =2\frac{1}{z^{2}}\cancelto{0}{\delta\left(\theta-\theta'\right)\delta\left(\theta-\theta'\right)}\cancelto{0}{\delta\left(\bar{\theta}-\bar{\theta}'\right)\delta\left(\bar{\theta}-\bar{\theta}'\right)}=0,
\end{equation}
so it vanishes identically. The ordinary matter stress tensor is given by the component proportional to $\theta \bar\theta$ :
\begin{equation}
    2:\partial x \cdot \partial B: + 2:\partial \psi \cdot  \partial \chi:.
\end{equation}
In this language, we see that the two-point function vanishes due to cancellation between Bosons and Fermions.

\subsection{Carrying out the path-integral}\label{subsec: Polyakov path integral}

To compute the partition function, we use supersymmetric localization. The relevant formulae to understand how this works are in appendix \ref{subsec: bi-graded integration}, and, more conceptually, in \cite{Blau:1992pm}. The first step is to find the saddles (see section \ref{subsubsec: Solving the Polyakov equation}), then, to compute the 1-loop determinant (see section \ref{subsubsec: polyakov 1-loop}) and finally to integrate over the moduli space of solutions (section \ref{subsubsec: polyakov moduli space}).

Note that we have written the action $S=-itS_0+\cdots$. The imaginary coefficient is because $B$ is a Lagrange multiplier (see equation \eqref{eq: euler character from Morse}). When we include also the term \eqref{eq: generlized S term} with a real coefficient, the relevant generalization is \eqref{eq: Mathai Quillen form in components} and the surrounding discussion. As we take $t\to \infty$ (without changing the path integral thanks to supersymmetry) this form of the action and \eqref{eq: Mathai Quillen form in components} guarantee that only saddle points of $S_0$ contribute.

\subsubsection{Solving the Polyakov equation}\label{subsubsec: Solving the Polyakov equation}

After gauge fixing $H=\delta$, the action \eqref{eq: super-polyakov action} becomes that of a non-linear sigma model:
\bea\label{eq: conf gauge super-polyakov action}
    S_{0} & =  S_{\rm NLSM}(\mathcal{M}^{2|2}),\\
    S_{\rm NLSM}(\mathcal{M}^{2|2})& =\intop d^2\sigma d^2\theta g_{\mu\nu}(X) \partial X^\mu \bar\partial X^\nu.
\eea
The equation of motion (EOM) in superspace is:
\bea\label{eq: NLSM superspace equation}
0 & = \delta^{ab}\partial_a(g_{\mu\nu}\partial_b X^\nu) - \frac{1}{2}\delta^{ab}g_{\rho\nu;\mu}(X) \partial_a X^\rho \partial_b X^\nu \\
& = \delta^{ab}g_{\mu\nu}\partial_a\partial_b X^\nu+\frac{1}{2}\delta^{ab}g_{\mu(\nu;\rho)}\partial_a X^\rho\partial_b X^\nu - \frac{1}{2}\delta^{ab}g_{\rho\nu;\mu}(X) \partial_a X^\rho \partial_b X^\nu \\
& = \delta^{ab}g_{\mu\nu}\left(\delta^\nu_\rho \partial_a+\partial_a X^\sigma \Gamma^\nu_{\sigma\rho}\right)\partial_b X^\rho\\
& = \delta^{ab}g_{\mu\nu}\nabla_a\partial_b X^\nu\\
& = g_{\mu\nu} (\nabla \bar \partial + \bar \nabla \partial)X^\nu,
\eea
where $\nabla_a$ is a worldsheet derivative covariantized with respect to the target space metric (a connection on the pullback of the tangent bundle).
As discussed in appendix \ref{subsec: Bi-graded equations}, we need only solve the bosonic version of the problem. 

In addition, the EOM of the metric is the Virasoro constraint; for the $(zz)$ component it is
\be\label{eq: Virasoro constraint}
T_{zz} = \frac{1}{2}g_{\mu\nu} \partial X^\mu \partial X^\nu = 0.
\ee
This means that the induced metric is Weyl-equivalent to the flat metric. Plugging it into the Polyakov action restores the NG action. When both the Virasoro constraint \eqref{eq: Virasoro constraint} and the NLSM EOM \eqref{eq: NLSM superspace equation} are satisfied, the NG equation is satisfied and we obtain an extremal area map. When only \eqref{eq: NLSM superspace equation} is satisfied, the map may or may not be of extremal area, but the induced metric is not necessarily Weyl-flat, and so the action is not equal to the induced area of the worldsheet.

Since the target space is $2d$, it is useful to pick target space coordinates for which $g$ is Weyl-flat. Namely, we pick complex coordinates $Z,\bar Z$ for which the only non-zero metric components are $g_{Z\bar Z}=g_{\bar Z Z}$. Then the EOM is:
\be\label{eq: NLSM superspace holomorphic}
0=  g_{Z\bar{Z}}\partial\bar{\partial}Z+g_{Z\bar{Z};Z}\partial Z\bar{\partial}Z,
\ee
and the Virasoro constraint is the complex equation
\be\label{eq: Virasoro constraint holmorphic}
T_{zz} = g_{Z \bar Z}\partial Z \partial \bar{Z} = 0.
\ee
\paragraph{Polyakov solutions vs. NLSM solutions:}
We call simultaneous solutions of \eqref{eq: Virasoro constraint holmorphic} and \eqref{eq: NLSM superspace holomorphic} ``Polyakov solutions''. \eqref{eq: Virasoro constraint holmorphic} implies that $Z$ is either holomorphic ($\bar\partial Z=0$), anti-holomorphic ($\partial Z=0$) or ``quasi-holomorphic'' (piece-wise holomorphic and anti-holomorphic). In the latter case, the induced orientation of the world-sheet, measured by the $B$-term:
\be
\epsilon^{ab}\sqrt{g}\epsilon_{\mu\nu} \partial_a X^\mu \partial_b X^\nu,
\ee
changes sign as one crosses an interface between patches, indicating that the string is folded there. Plugging such a map into \eqref{eq: NLSM superspace holomorphic} restores the NG equation and
%, as discussed in subsection \ref{subsec: polyakov action}, 
forces the size of the folds to vanish\footnote{Folds in two dimensional Nambu-Goto (NG) string theory were studied in \cite{Ganor:1994bq}.}, producing an extremal area map. Note that \eqref{eq: Virasoro constraint holmorphic} by itself does not force the folds to vanish, because it is first order in derivatives and so is insensitive to the jump-in-derivative that occurs there, unlike \eqref{eq: NLSM superspace holomorphic}. 

Quasi-holomorphic Polyakov solutions are therefore very ill-behaved. In each patch, the solution is (anti-)holomorphic and constant at the boundary -- but the only such functions are constant functions! Nevertheless, any extremal-area map with orientation reversing tubes -- the ``non-chiral'' maps with $n,{\tilde n} \geq 1$ discussed above -- has such vanishingly small tubes. It is a priori unclear how to expand around such solutions and compute the 1-loop fluctuations. 

In contrast, ``NLSM solutions'' solve only \eqref{eq: NLSM superspace holomorphic}. In carrying out the path integral, it is natural to first integrate out $X$ at fixed complex structure modulus $T$, essentially computing the NLSM partition function, and only then to integrate over $T$ to obtain the genus $g$ partition function (note that $\mu=0$ for $g=0$, $\mu =2$ for $g=1$ and $\mu=6(g-1)$ for $g>1$):
\be
Z_{g}=\intop d^{2\mu|2\mu} T Z_T=\intop d^{2\mu|2\mu} \hat{T} \intop_{\text{NLSM solutions $X_{\rm cl}[\hat{T}]$}} e^{-S[X_{\rm cl}[\hat{T}]]}\sign \left(\det\left(t\frac{\delta}{\delta X}\frac{\delta}{\delta X}S_0\right)\right),
\ee
where we have used the Gaussian integration formula \eqref{eq: general bi-graded gaussian} to carry out the 1-loop integral. The integral over $\hat{T}$ will then localize to saddles of $S_0$, where $\partial_{\hat{T}} S_0 =0$. However, extremization with respect to ${\hat{T}}$ is in essence extremization with respect to the dynamical metric $h$, \textbf{and so should enforce the Virasoro constraint!} More precisely, the moduli ${\hat{T}}$ represent the failure to fully gauge fix the metric, so variations of it represent variations of the metric that cannot be \textbf{traded off for variations of $X$} by a gauge transformation. Therefore, we expect the NLSM EOM by itself to ``nearly imply'' the Virasoro constraint, but not quite -- because of the ${\hat{T}}$'s. 

In summary, \textbf{NLSM solutions deviate from Polyakov solutions by precisely $\mu$ further conditions, and those are enforced by extremization with respect to the $\mu$ complex structure moduli ${\hat{T}}$}. To see how this comes about, note that while \eqref{eq: NLSM superspace holomorphic} does not in general imply \eqref{eq: Virasoro constraint holmorphic}, it does imply the nearly-as-powerful conservation equation:
\be
\bar \partial T_{zz}=0,
\ee
which means that the stress tensor is a \textbf{holomorphic quadratic differential}, and there is \cite{Polchinski_textbook:1998rq} precisely a $\mu$-dimensional vector space of such $b^i,\,i=1,\cdots,\mu$:
\be
T_{zz}= C_i (\{T^j\}) b^i.
\ee
This enables the CFT partition function to be computed without issues for generic $T$'s.
 
We will see examples of this procedure for toric worldsheets, that have $\mu=2g=2$, in sections \ref{subsec: Torus on torus} and \ref{subsec: torus on sphere}. In the latter, we will see how non-chiral Polyakov solutions with infinitesimal tubes are obtained as limits of well-behaved maps as the single complexified modulus (the Teichm\"uller parameter $\tau$) approaches its critical value (which will always be at the boundary of the moduli space).

A subtle but important case is that of the worldsheet sphere, for which $\mu=0$. In that case, NLSM solutions are automatically Polyakov solutions. This is good for (anti-)chiral maps, as we'll see in sections \ref{subsec: sphere on sphere once} and \ref{subsec: sphere on sphere twice}, but presents a complication for non-chiral maps, such as a mapping with a tube that we will study in section \ref{subsec: sphere on sphere w/ tube}. In those cases, something needs to be introduced ``by hand'' to play the role of ${\hat{T}}$ and regularize the mapping, as we will discuss in detail below.

For (anti-)chiral maps, the complicated nonlinear equation \eqref{eq: NLSM superspace holomorphic} can be replaced by the much simpler ($\partial Z=0$) $\bar \partial Z=0$. For toric target spaces, equation \eqref{eq: NLSM superspace holomorphic} becomes linear (a linear sigma model), as in subsection \ref{subsec: Torus on torus}. Otherwise, this equation may not be simple to solve analytically. We note, however, that for a spherical target space, it coincides with the \textbf{integrable} $O(3)$ model, which is equivalent to a sine-Gordon model \cite{Pohlmeyer:1976}.

Another subtlety concerns the moduli spaces of Polyakov and NLSM solutions. The moduli space of Polyakov solutions is parametrized by the positions of degenerate objects such as branch points and tubes, which can be moved independently without changing the action. In contrast, NLSM solutions also depend on the complex structure moduli. One might expect that NLSM solutions have simply an extended moduli space which includes both the Polyakov moduli and the complex structure moduli, and the Polyakov moduli space is simply a hypersurface embedded in it. In reality, ``turning on'' a complex structure modulus, deviating from its critical value, may partially or completely lift the Polyakov moduli. 

One sign that this is the case is that the space of (anti-)holomorphic maps is closed under \textbf{conformal transformations of the target space}, while general NLSM solutions are not. An example of this can be seen in subsection \ref{subsec: sphere on sphere twice}. Effectively, a larger symmetry group arises as the complex structure moduli approach their critical values. 

\subsubsection{1-loop determinant}\label{subsubsec: polyakov 1-loop}

Next in the localization procedure we must compute the 1-loop determinant for the Hessian of the localizing action $S_0$. According to \eqref{eq: general bi-graded gaussian}, this will give the sign of the determinant of the Hessian of the \textbf{purely bosonic version of} $S_0$, namely the Polyakov action $S_{\rm Pol}$, which is the lowest component of $S_0$. Upon finding a classical solution $X^\mu_{\rm cl}$, we expand:
\be
X = X_{\rm cl} + \delta X +O\left((\delta X)^2\right).
\ee
To ensure that $\delta X$ transforms like a vector, its relation to $X$ could be defined through the exponential map:
\be
X^{\mu}=\exp_{X_{\rm cl}}(\delta X) = X_{{\rm cl}}^{\mu}+\delta X^{\mu}-\frac{1}{2}\Gamma_{\nu\rho}^{\mu}\delta X^{\nu}\delta X^{\rho}+O\left(\left(\delta X\right)^{3}\right).
\ee
$S_0$ to second order is then:\footnote{If we chose simply $X=X_{\rm cl}+\delta X$ or some other relation, additional non-covariant terms would appear, though they would be total derivatives at this order anyway.}
\be\label{eq: linearized action 1}
\frac{1}{2}g_{\mu\nu}\nabla_{a}\delta X^{\mu}\nabla^{a}\delta X^{\nu}+\frac{1}{2}\partial_{a}X_{{\rm cl}}^{\mu}\partial^{a}X_{{\rm cl}}^{\nu}R_{\sigma\nu\rho\mu}\delta X^{\rho}\delta X^{\sigma},
\ee
with $R$ the Riemann tensor. For a 2d target space $R_{\sigma\nu\rho\mu}=\frac{1}{2}R\left(g_{\sigma\rho}g_{\nu\mu}-g_{\sigma\mu}g_{\nu\rho}\right)$, and we can write \eqref{eq: linearized action 1} as
\be \label{quadratic}
\frac{1}{2}\left(\nabla\delta X\right)^{2}+\frac{1}{4}R\left(\partial X_{{\rm cl}}\right)^{2}\left(\delta X\right)^{2}-\frac{1}{4}R\left(\partial X_{{\rm cl}}\cdot\delta X\right)^{2}.
\ee
The quadratic operator is therefore \textbf{self-adjoint}, and can be diagonalized. According to \eqref{eq: general bi-graded gaussian}, we only care about how many, if any, negative eigenvalues it possesses. 
\begin{itemize}
    \item If $R=0$ (as for a torus), the operator \eqref{quadratic} is non-negative.
    \item If $R<0$ everywhere, it's inconclusive.
    \item If $R>0$ everywhere, it's inconclusive for general NLSM solutions, but for Polyakov solutions, when $X_{\rm cl}$ is (anti-)holomorphic, we obtain:
    \be
    \frac{1}{4}R\left(\partial X_{{\rm cl}}\right)^{2}\left(\delta X\right)^{2}-\frac{1}{4}R\left(\partial X_{{\rm cl}}\cdot\delta X\right)^{2} = \frac{1}{8}R\left(\partial X_{{\rm cl}}\right)^{2}\left(\delta X\right)^{2},
    \ee
    so the operator is non-negative for Polyakov solutions and also for ``nearby'' NLSM solutions.
    \item Any other cases can likely be obtained by smoothly deforming the metric in the above special cases, and since such deformations are $Q$-exact, negative eigenvalues will probably (dis-)appear in pairs.
\end{itemize}

\subsubsection{Integrating over the moduli space (and 0-modes)}\label{subsubsec: polyakov moduli space}

According to \eqref{eq: bi graded solution}, \eqref{eq: generalized bi-graded solution} and the surrounding discussion, the collective coordinates of the moduli space of solutions are bi-graded coordinates $A^i(\theta,\thetabar)$, whose fermionic and ``auxiliary'' ($a_{\theta\thetabar}$) components represent tangent vectors to the space of bosonic solutions. They can be described as Fermi and auxiliary field zero-modes. The integration over the auxiliary zero-modes is ill-defined unless we add some $Q$-closed term $S_1$ to the action apart from $S_{0}$, as indicated in \eqref{eq: super-polyakov action}. Likewise, the fermi zero-modes need to be absorbed by the addition of such a term.

\horava's proposed additional term, which in the NG formalism is given by $\sqrt{h} S$ (with $S$ defined in \eqref{eq: Horava S invariant}), seems to vanish in 2d for smooth field configurations, as discussed in section \ref{subsubsec: horava Supersymmetric version}. We need to add a different, but similar term.

First, we'll discuss this in general, and then look at some concrete terms that can be added to the action. Consider a term of the form: 
\bea\label{eq: generlized S term}
S_1 =\frac{1}{2}\intop d^{2}\sigma d^{2}\theta\partial_{\theta}X^{\mu}K_{\mu\nu}\left[\sigma;X\right]\partial_{\bar{\theta}}X^{\nu}.
\eea
If the classical bosonic solutions can be parametrized using $n$ real moduli $a^{i}$, as $x_{{\rm cl}}\left(\sigma;\left\{ a^{i}\right\} \right)$, then the full solution in super-space is given (see \eqref{eq: generalized bi-graded solution}) by
\be
X_{\rm cl} = x_{{\rm cl}}\left(\sigma;\left\{ A^{i}\left(\theta,\bar{\theta}\right)\right\} \right),
\ee
with $A^{i}=a^{i}+\theta a_{\theta}^{i}+\bar{\theta}a_{\bar{\theta}}^{i}+\theta\bar{\theta}a_{\theta\bar{\theta}}^{i}$. Then:
\be\label{eq: on-shell S-term}
\intop d^{2}\sigma d^{2}\theta\partial_{\theta}X^{\mu}K_{\mu\nu}\partial_{\bar{\theta}}X^{\nu}\Big|_{X=X_{{\rm cl}}}=\intop d^{2}\theta\partial_{\theta}A^{i}M_{ij}\partial_{\bar{\theta}}A^{j},
\ee
with:
\be\label{eq: moduli metric from ws integral}
M_{ij}\equiv \intop d^{2}\sigma\partial_{i}X_{{\rm cl}}^{\mu}K_{\mu\nu}\left[\sigma;X_{{\rm cl}}\right]\partial_{j}X_{{\rm cl}}^{\nu}
\ee
the induced bi-graded metric on the moduli space. The action \eqref{eq: on-shell S-term} is of the form in equation \eqref{eq: euler character from Chern-Gauss-Bonnet}, so it computes the Euler character of the moduli space through the Chern-Gauss-Bonnet theorem \cite{chern:1945}. If we decompose $M_{ij}=m_{ij}+\cdots$ and $R$ is the Riemann tensor associated with $m$, we obtain from \eqref{eq: expansion of S-term into components} (after completing the square):
\be
\intop d^{2}\theta\partial_{\theta}A^{i}M_{ij}\partial_{\bar{\theta}}A^{j}=a_{\theta\bar{\theta}}^{i}m_{ij}a_{\theta\bar{\theta}}^{j}+\frac{1}{2}R_{ijkl}a_{\theta}^{i}a_{\theta}^{j}a_{\bar{\theta}}^{k}a_{\bar{\theta}}^{l}.
\ee
Now, from \eqref{eq: Euler density from 0-mode integration}, integration over the fermi and auxiliary 0-modes will give rise to:
\be\label{eq: Euler density from 0-mode integration 1}
 \intop_{\mathcal{M}} d^{2n}a \frac{1}{2^{2n}(2\pi)^n n! }{\det\left(m\right)}^{-1/2} \epsilon^{i_{1}j_{1}i_{2}j_{2}\dots i_{n}j_{n}}\epsilon^{k_{1}l_{1}k_{2}l_{2}\dots k_{n}l_{n}}R_{i_{1}j_{1}k_{1}l_{1}}\cdots R_{i_{n}j_{n}k_{n}l_{n}}
\ee
which is the Euler density on the moduli space. Thus, we obtain the Euler character of the moduli space.

Note that the moduli space is really the space of solutions \textbf{modulo} the conformal group. As discussed in subsection \ref{subsec: polyakov gauge fixing}, there are no measure factors from gauge-fixing.

The above discussion is relevant for the moduli space of Polyakov solutions, but the space of NLSM solutions includes moduli -- the complex structure parameters -- that \textbf{do} couple to $S_0$. These should be thought of as localizing to their critical values with respect to $S_0$.

Below are some concrete suggestions for terms of the form \eqref{eq: generlized S term}:
\begin{enumerate}
    \item 
    \begin{equation}
        K_{\mu\nu} = P_{\mu\nu}=g_{\mu\nu}-g_{\mu\rho}\partial_{a}X^{\rho}h_{\rm ind}^{ab}\partial_{b}X^{\sigma}g_{\sigma\nu}=0.
    \end{equation}
    Here, $h_{\rm ind}$  is the induced metric. In this case, we obtain \horava's $S$, which unfortunately vanishes.
\item The extrinsic curvature:
    \begin{equation}\label{eq: extrinsic S term}
        K_{\mu\nu} =  \sqrt{h_{\rm ind}}P_{\mu\rho}P_{\nu\sigma}h_{\rm ind}^{ab}h_{\rm ind}^{cd}\partial_{a}\partial_{[b}X^{\rho}\partial_{c]}\partial_{d}X^{\sigma} =\frac{1}{4}\pi ng_{\mu\nu}\delta^{2}\left(\sigma - \sigma_{\rm bp}\right),
    \end{equation}
which localizes to branch points, as indicated above for the case of a branch point of degree $n$ (see \eqref{eq: extrinsic curvature at branch point}). More generally, it also localizes to folds, so it will also couple to tubes. The main disadvantages of this term are its complexity and its singular nature, owing to its reliance on the induced metric. To properly compute it, one needs to regulate it, temporarily breaking its tensorial transformation properties -- a potential source of problems.
% \item Extra dimensions. In this case we extend $\mu=1,2$ to $I=1,2,\dots D$, but keep the strings from flying off in the extra directions by imposing a Dirichlet b.c. (for worldsheets with boundaries), and possibly something like a potential for closed strings.
% \begin{equation}
%     K_{IJ} = P_{IJ}.
% \end{equation}
% In this case, the projector doesn't vanish, instead giving $P_{IJ}=\delta_{Ii}\delta_{Jj}\delta^{ij}$ on-shell. Unfortunately, this still doesn't couple to the 0-modes, but if we keep the leading off-shell term which does:
% \begin{align}
% S & =\partial_{\theta}X^{i}\partial_{\bar{\theta}}X_{i}+\partial_{\theta}X_{{\rm cl}}^{(\mu}\partial_{\bar{\theta}}X^{i)}\delta_{ji}h^{ab}g_{\mu\nu}\partial_{b}X_{{\rm cl}}^{\nu}\partial_{a}X^{j}\\
%  & +\partial_{\theta}X_{{\rm cl}}^{\mu}\partial_{\bar{\theta}}X_{{\rm cl}}^{\nu}\delta_{kl}h^{ab}\partial_{a}X_{{\rm cl};\mu}\partial_{b}X^{k}h^{cd}\partial_{c}X_{{\rm cl};\nu}\partial_{d}X^{l},
% \end{align}
% And integrate out the perpendicular directions, we will get a subleading non-local $K_{\mu\nu}$ which can be used in principle. Although it's proportional to the $1/t$ and we take $t\to\infty$, since it is the only coupling to the 0-modes, it's effect is non-negligible. The dependence on $t$, however, is spurious, and will cancel out in the final answer due to bose-fermi cancellations. 
\item The simplest possibility would be:
\begin{equation}
    K_{\mu\nu}=g_{\mu\nu},
\end{equation}
but that gives rise to a term which is not diff-invariant. We can remedy this by using the combined BRST operator:
\be
\left[Q,X\right]=\partial_{\theta}X+C^{a}\partial_{a}X,
\ee
\be
\left\{ Q,C^{a}\right\} =\partial_{\theta}C^{a}+C^{b}\partial_{b}C^{a},
\ee
which is nilpotent. Concretely, we add to the action:
\begin{align}
\left\{ Q,-\partial_{\theta}X\cdot\partial_{\bar{\theta}}X\theta\right\}  & \sim\partial_{\theta}X\cdot\partial_{\bar{\theta}}X+\partial_{\theta}C^{a}\partial_{a}X\cdot\partial_{\bar{\theta}}X\theta-\partial_{\bar{\theta}}C^{a}\partial_{a}X\cdot\partial_{\theta}X\theta+\partial_{a}C^{a}\partial_{\theta}X\cdot\partial_{\bar{\theta}}X\theta.
 \end{align}
Thus, we obtain the desired term, in addition to some couplings to the diff-ghost superfield. Since all the terms in the action can be written as $Q$-exact, we can still perform the same localization. The integral over diff-ghost and diff-antighost nonzero modes sets $C=C_{\rm cl}$, which is just a sum over ghost 0-modes, representing the conformal group. Upon introducing residual gauge conditions, we will obtain $C_{\rm cl}=0$, and thus can ignore the additional terms. 
\item A related choice is:
\be
K_{\mu\nu}=(\partial X)^2 g_{\mu\nu}.
\ee
The same discussion applies here, but this term is at least invariant under conformal transformations, even if not under their bi-graded extension.
\end{enumerate}

In summary, the full action is given by:

\bea
S & = -it S_0 + S_{\rm FP} +S_1,
\eea
with $S_0$ given in \eqref{eq: super-polyakov action}, $S_{\rm FP}$ given in \eqref{eq: FP action} and $S_1$ given by \eqref{eq: generlized S term} for any choice of $K_{\mu\nu}$. The choice of $K_{\mu \nu}$ most suitable for generalization to the finite-area theory is that in \eqref{eq: extrinsic S term}, which we'll discuss in a specific example in section \ref{sec: various amplitudes}.

%% file: various_amplitudes.tex
\section{Various amplitudes}\label{sec: various amplitudes}

We now turn to some calculations. We will examine string amplitudes and relate them to the YM partition function \eqref{eq: zymnew} including the \ocmr-points \eqref{eq: ocmr point expansion}. 

The simplest cases are those of a torus covering the torus multiple times (section \ref{subsec: Torus on torus}), and the sphere covering the sphere once (section \ref{subsec: sphere on sphere once}), as they are holomorphic and they do not have any branch points or tubes. In these cases the formal arguments that our string theory gives the correct answer have no subtleties, but we still find it useful to see exactly how the correct answer is reproduced.

In section \ref{subsec: torus on sphere} we consider maps of the torus into the sphere with multiple tubes. This example showcases the capacity, discussed in section \ref{subsec: Polyakov path integral}, of the Polyakov formalism to obtain quasi-chiral maps as limits of smooth configurations. The tubes appear as sequences of sine-Gordon kinks and anti-kinks. On the other hand, certain problems arise with the 1-loop determinant, which require investigation.

In section \ref{subsec: sphere on sphere w/ tube} we consider similar maps from the sphere to the sphere with tubes. This time we need to employ a ``constrained instanton'' to regularize such maps, because the sphere has no complex structure moduli.

Finally, in section \ref{subsec: sphere on sphere twice} we consider double-covering the sphere by a sphere with two branch points. In this example, which is holomorphic, finding the solutions and performing the 1-loop fluctuation integral is simple, but a subtle problem arises due to collisions of type 3 (discussed in \cite{Cordes:1994fc} and in section \ref{subsec: stringy interpretation of the ocmr points}), which introduce a singularity into the moduli space.

\subsection{Sphere covering the sphere once}\label{subsec: sphere on sphere once}

Let's consider first a simple case where the target space is a sphere, and we are wrapping a world-sheet sphere on it once.
First we'll consider an orientation-preserving map. 

In conformal gauge, the worldsheet sphere has no moduli, and as explained in section \ref{subsec: Polyakov path integral}, the stress tensor must vanish on-shell, implying that the map is quasi-chiral (see section \ref{sec: horavasec}). In the orientation-preserving case it is holomorphic.

The map of a sphere wrapping the sphere once, in an orientation-preserving way, in the conformal gauge, is (using the $z$ complex coordinate for the worldsheet sphere, and the $Z$ complex coordinate for the space-time sphere, with metric $g_{Z\bar Z} = \frac{4}{(1+|Z|^2)^2}$)
\begin{equation}
    Z = \frac{a z + b}{c z + d},
\end{equation}
where $ad-bc=1$. This is simply a M\"obius transformation with bi-graded parameters. We can fix the remaining $SL^{2|2}(2,\mathbb{C})$ conformal symmetry on the worldsheet sphere by taking $Z = z$. This can be achieved, for instance, by gauge-fixing the images of $z=0,1,\infty$ to be, respectively $Z=0,1,\infty$, a permissible gauge choice in the topological sector of these maps. This gauge choice is of the form discussed in section \ref{subsec: polyakov gauge fixing}, free of Grassmann derivatives as required. Thus, there is no moduli space of such maps, as expected since there are no tubes / branch points.
% and $Z=-z$ for orientation reversing case. Note, since the action is quadratic in $X^i$ we will get same answer for both the cases.

For this map we have:
\begin{equation}
    \frac{1}{2} (\partial x)^2 = g_{Z\bar Z}  \partial Z \bar\partial \bar Z= g_{Z\bar Z} = \frac{4}{(1+|z|^2)^2}.
\end{equation}
The on-shell Polyakov action\footnote{Note that this is the ordinary Polyakov action, not its bi-graded version $S_0$. Of course, in general, extremizing a bi-graded function automatically extremizes its bosonic version, which is simply the lowest component of it (see appendix \ref{subsec: Bi-graded equations}).} is then simply given by the area of the target space:
\begin{equation}
%\begin{split}
    S_{\rm Pol} = \int_{S^2} d^2 z \frac{4}{(1+|z|^2)^2}
    = 4\pi.
%\end{split}
\end{equation}
The 1-loop determinant will be $1$ according to section \ref{subsubsec: polyakov 1-loop}.

The orientation-reversing case is simply the complex conjugate of the one above.

\subsubsection{Disconnected spheres wrapping the sphere}

To reproduce the disconnected mapping to the sphere with $n$ orientation-preserving and $\tilde n$ orientation-reversing sheets, we need to consider a world-sheet $S^2\otimes S^2 \otimes \cdots S^2 \otimes \bar S^2\otimes \bar S^2 \otimes \cdots \bar S^2$, where $S^2$ wraps the target space sphere once in an orientation-preserving way and $\bar S^2$ wraps the same target space in an orientation-reversing way. Note that here we have to divide by $n!\times \tilde n!$ to account for the over-counting of the integration region in the path integral.

The contribution of each disconnected worldsheet is exactly as in the single-wrapping case. We will use $Z = z$ for the orientation-preserving case and $Z=\bar z$ for the orientation-reversing case. So the only novelty will be the inclusion of the factor $\frac{1}{n!\tilde n!}$.

\subsubsection{Comparison to YM theory}
The Gross-Taylor formula \eqref{eq: zymnew} says that at leading order in $N$ the partition function gives:
\begin{equation}
    Z(G=0,\lambda A\rightarrow 0, N) = \sum_{n,\tilde n} \frac{N^{2(n+\tilde n)}}{n!\tilde n!} + O\left(N^{2(n+\tilde n)-2}\right),
\end{equation}
in agreement with the above.

\subsection{Torus covering the torus multiple times}\label{subsec: Torus on torus}

Next we discuss a worldsheet torus mapping to a space-time torus. The worldsheet is given by $z=x+it$ with the identifications (for integer $k,k'$)
\be
z \sim z+ 2\pi k + 2\pi k' \tau,\,\qquad {\rm Im}(\tau)>0.
\ee
We have a residual translation symmetry, as well as $SL(2,\mathbb{Z})$ modular transformations which act as:
\be
z'=\frac{z}{c\tau+d}, \, \tau'=\frac{a\tau+b}{c\tau+d},\,\qquad ad-bc=1,
\ee
and which we'll need to fix. We'll take the target space to be a ``square'' torus $X^i \sim X^i + 2\pi$ for simplicity. 

The solutions to the NLSM equation are linear functions:
\be
x^i(z) = x_{0}^{i}+\frac{1}{2i{\rm Im}(\tau)}z\left(N^{i}-W^{i}\tau^{*}\right)+{\rm c.c.},\qquad
N^i,W^i\in \mathbb{Z},\, \qquad i=1,2.
\ee
The worldsheet translation symmetry is broken to the discrete group
\be \label{eq: discretetrans}
z\to z+2\pi\frac{\epsilon_{ij}q^{i}N^{j}+\epsilon_{ij}W^{i}q^{j}\tau}{\epsilon_{ij}W^{i}N^{j}}.
\ee
Upon gauge fixing $x_{0}^i=0$ by choosing $G^i=x^i(0)$ in \eqref{eq: Fadeev Popov determinant}, the remaining translation symmetry is a discrete group parametrized by the integer-component vector $\vec{q}$, modulo $\vec{q}\sim \vec{q}+\vec{W}\sim \vec{q}+\vec{N}$. This symmetry exchanges the sheets of the cover, and so it is the group of automorphisms of the covering space. We must \textbf{divide by the cardinality} of this group, which is given by $|\epsilon_{ij}W^i N^j|$ (this is derived below upon choosing a gauge for $\vec{N},\vec{W}$). This is also the \textbf{order of the cover}, since each translation in \eqref{eq: discretetrans} takes us between distinct pre-images of a given target space point.

The modular group of the worldsheet torus acts in its fundamental representation on the two-component vectors $(N^i,W^i)$ $(i=1,2)$: 
\be
\left(\begin{array}{c}
{N^{i}}'\\
{W^{i}}'
\end{array}\right)=\left(\begin{array}{cc}
d & -b\\
-c & a
\end{array}\right)\left(\begin{array}{c}
N^{i}\\
W^{i}
\end{array}\right).
\ee
This leaves invariant the determinant $\epsilon_{ij}W^{i}N^{j}$.

The solutions give
\begin{align}
\frac{1}{2}\left(\partial x\right)^{2} & =\frac{1}{2\left({\rm Im}(\tau)\right)^{2}}\sum_{i}\left(N^{i}-W^{i}\tau^{*}\right)\left(N^{i}-W^{i}\tau\right)\\
 & =\frac{1}{2\left({\rm Im}(\tau)\right)^{2}}\left(\vec{N}^{2}+\vec{W}^{2}\tau\tau^{*}-\vec{N}\cdot\vec{W}\left(\tau+\tau^{*}\right)\right),
\end{align}
and:
\be
\intop_{0}^{2\pi{\rm Im}(\tau)}dt\intop_{2\pi\Re(\tau)}^{2\pi+2\pi\Re(\tau)}d\sigma=\left(2\pi\right)^{2}{\rm Im}(\tau).
\ee
The on-shell action is thus:
\be
S_0=\intop d^2\theta \frac{\left(2\pi\right)^{2}}{2{\rm Im}(\tau)}\left(\vec{N}^{2}+\vec{W}^{2}\tau\tau^{*}-\vec{N}\cdot \vec{W}\left(\tau+\tau^{*}\right)\right).
\ee
The extremum (with respect to $\tau$), which is in fact a minimum, occurs at:
\be
\tau=\frac{N^{1}\pm iN^{2}}{W^{1}\pm iW^{2}}=\frac{\vec{N}\cdot\vec{W}+i\left|\epsilon_{ij}N^{i}W^{j}\right|}{\vec{W}^{2}},
\ee
which is precisely the value for which the map is (anti-)holomorphic. The extremal value of the action is:
\be
S_{\rm Pol}= \left(2\pi\right)^{2}\left|\epsilon_{ij}N^{i}W^{j}\right|,
\ee
indicating the number of times the target space is covered.

Next, we gauge-fix the residual $SL(2,\mathbb{Z})$. We can either fix $\tau$ to the fundamental domain, in which case the $\tau$-extremum will only be attained for certain values of $N,W$ (the rest will give a vanishing contribution in the $t\to\infty$ limit), or we can fix $N,W$, and allow $\tau$ to vary over the entire upper-half plane. Here we choose the latter route.

 Without loss of generality (up to a modular transformation) we can assume $N^2\neq0$ (we'll ignore the degenerate case when $W^2=N^2=0$). We can then set ${W^2}'=0$ by choosing:
\be
a=\frac{N^2}{\gcd(W^2,N^2)},\,c=\frac{W^2}{\gcd(W^2,N^2)}.
\ee
Because $a,c$ are co-prime, we can choose $b$ and $d$ to satisfy the constraint $ad-bc=1$. We can further set $N^2>0$ using minus the identity matrix. The remaining symmetry is given by the matrices:
\be
\left(\begin{array}{cc}
1 & a\\
0 & 1
\end{array}\right),
\ee
and can be used to fix $0\leq N^1<|W^1|$. We will ignore the degenerate case $W^1=0$. Thus, after gauge-fixing we have
\be
W^2=0,\,N^2>0,\,0\leq N^1 < |W^1|,\,W^1\in \mathbb{Z}\backslash \{0\}.
\ee
The invariant becomes:
\be
n\equiv\left|\epsilon_{ij}N^{i}W^{j}\right|=|W^1|N^2.
\ee
Assume a holomorphic mapping ($W^1>0$). For fixed $n$, $W^1$ can be any divisor of $n$, and so the number of total allowed mappings is:
\be
\sum_{d|n}d
%= \sigma_{1}(A),
\ee
where $d|n$ means $d$ divides $n$.
%and $\sigma$ is the divisor function.

\paragraph{The order of the cover:} In the present gauge, it is easy to see that the number of distinct $\vec{q}$ values in \eqref{eq: discretetrans} is equal to $n$. That is because the vector $\vec{W}$ has 0 in its second component. So $\vec{q}$ is restricted to the lattice $q^1=0,1,\dots,|W^1|-1,\,q^2=0,1,\dots,N^2-1$. Hence, dividing by the remaining discrete symmetry group we obtain a factor of $1/n$. 

\subsubsection{Comparison to YM theory} Since we are interested only in connected maps at order $N^{2-2g}=N^0$ in the YM partition function, we can limit attention to chiral maps without punctures, which cover the target space $n$ times. The contribution of such maps in the Gross Taylor formula is given by $1/n!$ times a sum over assignments of holonomies $\sigma_1,\sigma_2\in S_n$ to the two cycles of the target space, which satisfy:
\begin{itemize}
    \item The holonomy constraint: $\sigma_1 \sigma_2 \sigma_1^{-1}\sigma_2^{-1}=1$.
    \item Connectedness: Each sheet of the cover can be reached from any other sheet by a suitable combination of the holonomies, or: ${\rm orbit}({\rm span}(\sigma_1,\sigma_2))=\{1,\dots,n\}$.
\end{itemize}

Let us describe the various solutions of this pair of constraints:
\begin{itemize}
    \item $\sigma_1=1$ -- In this case, the holonomy constraint is trivial, and connectedness implies that $\sigma_2$ is a cycle of length $n$. There are $(n-1)!$ such cycles, so we get $\frac{(n-1)!}{n!}=1/n$.
    \item $\sigma_1$ is a cycle of length $n$. There are $(n-1)!$ such cycles. Upon choosing one, connectedness is satisfied. To satisfy the holonomy constraint, we must pick $\sigma_2= \sigma_1^k$ for $k=0,\dots,n-1$, which yields $n$ options. In total, we get $\frac{(n-1)!n}{n!}=n/n=1$.
    \item Using notation suggestive of the worldsheet analysis -- $\sigma_1$ decomposes into $N^2$ cycles of length $W^1=n/N^2$, for $N^2$ any divisor of $n$. There are 
    \be
   \frac{n!}{\left(W^{1}\right)^{N^{2}}N^{2}!}=\underbrace{\frac{n!}{\left(W^{1}!\right)^{N^{2}}}}_{\text{Ways of selecting \ensuremath{N^{2}} subsets}}\left(\underbrace{\left(W^{1}-1\right)!}_{\text{distinct cycles of length \ensuremath{W_{1}}}}\right)^{N_{2}}\underbrace{\frac{1}{N^{2}!}}_{\text{subsets indistinguishable}}
    \ee
    such permutations. The two constraints are invariant under composition of $\sigma_2$ with $\sigma_1$, so the possible choices of $\sigma_2$ decompose into orbits of length $W^1$ under this operation -- in correspondence with the possible choices of $N^1=0,\dots,W^1-1$ in the worldsheet analysis. To satisfy connectedness, $\sigma_2$ must connect between all the cycles of $\sigma_1$, but to satisfy the holonomy constraint, it must do so by permuting the cycles of $\sigma_1$. It mustn't separate positions that share a cycle, or disturb their cyclic ordering within a cycle. There are:
    \be
    \underbrace{\left(N^{2}-1\right)!}_{\begin{array}{c}
\text{cyclic permutations of}\\
\text{the cycles of \ensuremath{\sigma_{1}}}
\end{array}}\underbrace{\left(W^{1}\right)^{N^{2}}}_{\begin{array}{c}
\text{Choice of \ensuremath{\sigma_{2}} image for}\\
\text{the first element of each}\\
\sigma_{1}\text{ cycle}
\end{array}},
    \ee
    possible choices for $\sigma_2$. Putting it all together, we get a contribution of:
    \be
    \frac{1}{n!}\frac{n!}{\left(W^{1}\right)^{N_{2}}N^{2}!}\left(N^{2}-1\right)!\left(W^{1}\right)^{N^{2}}=\frac{1}{N^{2}}=\frac{W^{1}}{n},
    \ee
    as expected from the worldsheet analysis.
    \item More general cycle structures for $\sigma_1$ are disallowed. The combination of connectedness and the holonomy constraint means $\sigma_2$ must permute \textbf{all} the cycles of $\sigma_1$ while preserving their structure, but this is not possible if any two of the cycles have different lengths.
\end{itemize}

Thus, the worldsheet analysis precisely matches the field theory result.

\subsection{Torus covering the sphere with (branched) orientation-reversing tubes}\label{subsec: torus on sphere}

As before, we take the worldsheet to be a torus, but now the target space is a sphere, with metric $g_{Z\bar{Z}}=4(1+Z\bar{Z})^{-2}$. In this section, we consider non-holomorphic extremal-area mappings that wrap the sphere with both orientations, with zero-size orientation-reversing tubes.

\begin{figure}[p]
    \centering
    \begin{minipage}{0.45\columnwidth}
    \includegraphics[width=\linewidth]{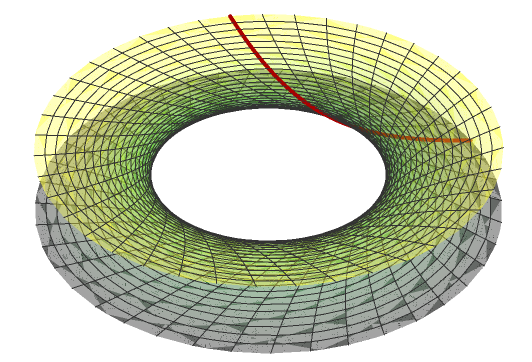}
    \caption{An orientation-reversing tube for $p\neq0$ (The fictitious height dimension is for visualization). The red line signifies the time evolution of a bit of string.}
    \label{fig:twisted ort}
    \end{minipage}\hfill
    \begin{minipage}{0.45\columnwidth}
    \includegraphics[width=\linewidth]{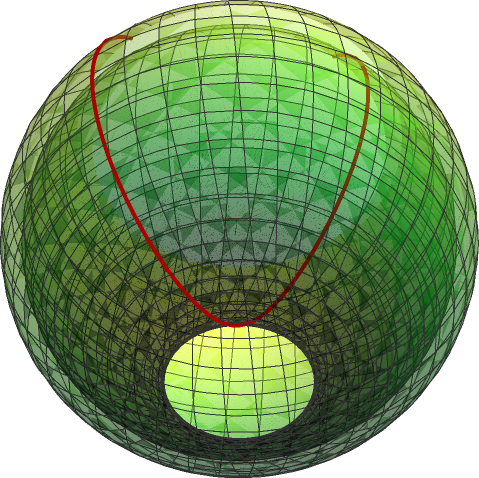}
    \caption{A torus covering the sphere with two orientation-reversing tubes at $l=n=1$ and $p\neq0$ (The fictitious radial dimension is for visualization). The red line signifies the time evolution of a bit of string.}
    \label{fig:torus on sphere w tubes}
    \end{minipage}
    \vspace{4cm}
    \begin{minipage}{0.45\columnwidth}
    \includegraphics[width=0.9\linewidth]{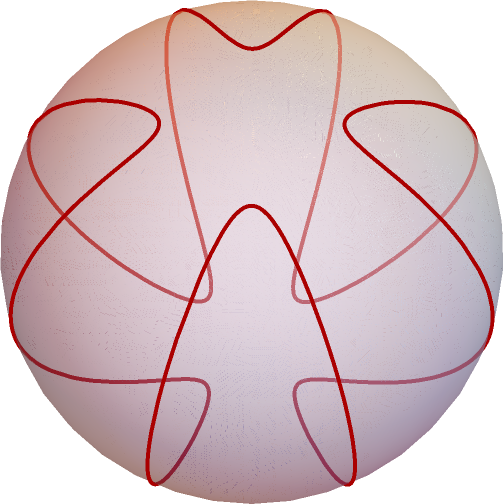} 
    \caption{The trajectory of a bit of string in the case $p\neq0$ and $l=4$.}
    \label{fig:string bit trajectory for l equals 4}
    \vspace{0 cm}
    \end{minipage}
    
\end{figure}

The NLSM equation becomes:
\be\label{eq: O(3) equation for torus}
g_{Z\bar{Z}}\partial\bar{\partial}Z+g_{Z\bar{Z};Z}\partial Z\bar{\partial}Z=0,
\ee
with periodicity (for any integer $k,k'$):
\be
Z(z+2\pi k+2\pi \tau k') = Z(z). 
\ee
This is essentially the $O(3)$ model, which is integrable and can be recast in terms of the Sine-Gordon model \cite{Pohlmeyer:1976}. We will here focus on ``single frequency'' solutions of the form:
\be
Z(x+it) = e^{in x}e^{i \phi(t)}\tan(\theta(t)/2),
\ee
using standard polar coordinates on $S^2$, where $\theta(t)$ will oscillate $l$ times between two peak values (figure \ref{fig:torus on sphere w tubes} shows the case $l=n=1$). In the limit where the peak values approach $\theta=0$ and $\theta=\pi$, we will obtain extremal area maps that cover the sphere $n l$ times with one orientation and $n l$ times with the other, connected by $2l$ \textbf{(branched) orientation-reversing tubes} (see discussion in section \ref{subsec: stringy interpretation of the ocmr points}), half of which are located at the North Pole, and half at the South Pole. These will come from \ocmr-points in the expansion of the YM partition function. The case $n=1$ corresponds to ordinary orientation-reversing tubes, while for $n>1$ the tubes have a non-trivial holonomy surrounding them which permutes the sheets. We will discuss the full moduli space in section \ref{subsubsec: torus on sphere moduli space}.

% \begin{wrapfigure}{r}{0.5\textwidth}
% \centering 
% \includegraphics[width=0.9\linewidth]{various_amplitudes/string_bit_trajectory_for_l_equals_4.png} 
%     \caption{The trajectory of a bit of string in the case $p\neq0$ and $l=4$.}
%     \label{fig:string bit trajectory for l equals 4}
%     \vspace{0 cm}
% \end{wrapfigure}

In the $\theta,\phi$ variables, the action becomes:
\be
\frac{1}{2}\dot{\theta}^{2}+\frac{1}{2}\left(n^{2}+\dot{\phi}^{2}\right)\sin^{2}(\theta).
\ee
Periodicity becomes $\phi(t+2\pi{\rm Im}(\tau))=\phi(t)-2\pi n \Re(\tau) \mod 2\pi$, and $\theta(t+2\pi{\rm Im}(\tau))=\theta(t)$. Time translation symmetry leads to conservation of the energy:
\be
2E=\dot{\theta}^{2}+\left(\dot{\phi}^{2}-n^{2}\right)\sin^{2}(\theta),
\ee
while spatial translations, which for single-frequency maps coincide with the shift symmetry of $\phi$, lead to conservation of the momentum:
\be
p=\dot{\phi}\sin^{2}(\theta).
\ee
We can solve for $\phi$ in terms of $p$ and $\theta$ and plug into the energy, creating a fictitious potential for $\theta$:
\be
2E = \dot{\theta}^{2}+p^{2}\csc^{2}\left(\theta\right)-n^{2}\sin^{2}\left(\theta\right).
\ee
We can characterize the amplitude of oscillation by $\theta_{\rm min}\in (0,\pi/2)$, and then:
\be
2E = p^{2}\csc^{2}\left(\theta_{\rm min}\right)-n^{2}\sin^{2}\left(\theta_{\rm min}\right).
\ee
From this we can compute the period of oscillation in the well:
\be
T = 4\intop_{\theta_{{\rm min}}}^{\pi/2}d\theta\left(2E-\left(p^{2}\csc^{2}(\theta)-n^{2}\sin^{2}(\theta)\right)\right)^{-1/2},
\ee
and the phase shift over one oscillation:
\be
\Delta \phi = 4p\intop_{\theta_{{\rm min}}}^{\pi/2}d\theta\left(2E-\left(p^{2}\csc^{2}(\theta)-n^{2}\sin^{2}(\theta)\right)\right)^{-1/2}\csc^{2}(\theta).
\ee
We must therefore tune $E,p$ so that $T=2\pi{\rm Im}(\tau)/l$ and $\Delta \phi = -\frac{2\pi}{l} n \Re(\tau) \mod \frac{2\pi}{l}$, to ensure periodicity after $l$ oscillations (see figure \ref{fig:string bit trajectory for l equals 4}). We can similarly compute the on-shell action, which takes a suggestively thermodynamic form:
\bea
S_{\rm Pol}& =2\pi\left(F-2\pi p\left(n \Re(\tau)+k\right)-E{\rm Im}(\tau)\right),\\
F&\equiv 4l\intop_{\theta_{{\rm min}}}^{\pi/2}d\theta\sqrt{2E-p^{2}\csc^{2}(\theta)+n^{2}\sin^{2}(\theta)},
\eea
where $\Delta\phi=-2\pi (n \Re(\tau)+k)/l$ with $k$ an integer. The quantity $F$ satisfies the equation $\partial_\tau F= \partial_\tau E {\rm Im} (\tau) -\partial_\tau p l\Delta \phi$, so we obtain:
\be \label{eq: tau_extrema}
\partial_{{\rm Im}(\tau)} S_{\rm Pol} = -2\pi E,\, \qquad \partial_{\Re (\tau)}S_{\rm Pol}=-2\pi p.
\ee
Thus, the saddle is at $E=p=0$ (which is also what we get from the Virasoro constraint), and there it is easy to see that $F=2nl\Rightarrow S_{\rm Pol}=4\pi n l=nl {\rm Area}(S^2)$. Since these conditions occur at ${\rm Im}(\tau)\to\infty$, one might worry whether this is truly a saddle point. The \textbf{crucial point} is that $S_{\rm Pol}$ decays to its extremal value \textbf{faster than linearly}, so a simple ${\rm Im}(\tau)\to\frac{1}{{\rm Im} (\tau)}$ transformation does not remove the saddle. 

We pause to make a few comments:
\begin{enumerate}
    \item Although at this stage we keep $\tau$ fixed and arbitrary, we note that extremization with respect to it will ultimately impose the Virasoro constraint (see the general discussion in section \ref{subsec: Polyakov path integral}), which in this case leads to $E=p=0$. At $p=0$, $\theta$ becomes a particle in a Sine-Gordon potential, and as $E\to0$ we get $T\to\infty\Rightarrow{\rm Im}(\tau)\to\infty$ and the solution approaches an alternating sequence of $l$ Sine-Gordon kinks and $l$ anti-kinks, infinitely removed from one another. $E=p=0$ is equivalent to $p=\theta_{\rm min}=0$.
    %\item The states at $p=0$ and $E>0$ are supposedly propagating states that move ``past'' the limits $\theta=0,\pi$. To interpret this as a well-defined mapping into the sphere, we must think of this as a reflective b.c., thus trapping the particle in an infinite well. This is consistent with taking the limit $p\to0$ at fixed $E>0$. These maps have tubes of vanishing size, and so have extremal area, even though $E\neq0$ -- because they are not holomorphic. 
    \item The solutions have two additional translational modes $\phi_0\in[0,2\pi),t_0\in [0,T)$. These can be fixed to 0 using the residual worldsheet translation symmetry. Spatial translations are then broken down to $\mathbb{Z}_n$, and translations in the $\tau$ direction to $\mathbb{Z}_l$, that leave invariant the solution, so we must divide by a symmetry factor of $nl$, equal to the number of sheets of any given orientation.
    \item In addition, we need to fix the modular group. We can do so by fixing $\tau$ to the fundamental domain $\mathcal{F}$, but then we still are left with the $\pi$-rotation $z\to-z$. This map takes us from frequency $n$ to $-n$ and reverses time, but this can be undone through the target space transformation $Z\to1/Z$ (or equivalently, $\phi\to-\phi,\,\theta\to\pi-\theta$), so this just exchanges the kink and anti-kink. Effectively this is the Bose symmetry of the tubes, and we can ignore it by including a symmetry factor of $1/2$, and then considering the two tube positions as independent.
\end{enumerate}

\begin{figure}[p]
    \centering
    \begin{minipage}{0.45\columnwidth}
    \includegraphics[width=\linewidth]{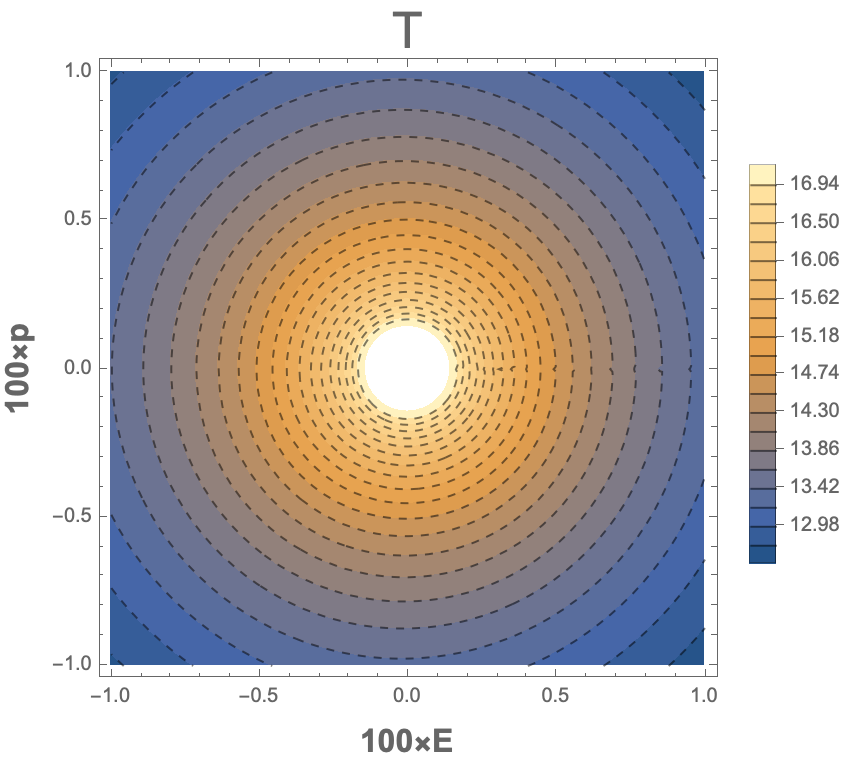}
    \caption{The period $T(E,p)$ in a small range of values around the extremum of $S_0$. The width of the contours grows linearly with distance, showing the logarithmic divergence of $T$ at the origin.}
    \label{fig:period contour small}
    \end{minipage}
    \hfill
    \begin{minipage}{0.45\columnwidth}
    \includegraphics[width=\linewidth]{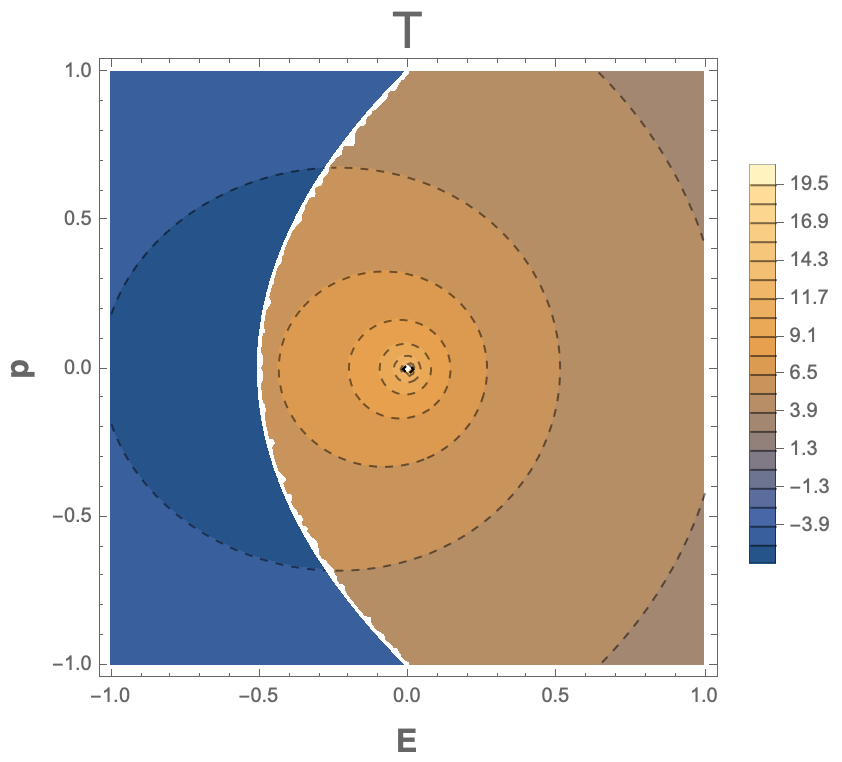}
    \caption{The period $T(E,p)$ in a large range of values around the extremum of $S_0$. The region to the left of the parabola $2E=p^2-n^2$ is unphysical, with impossible energies.}
    \label{fig:period contour large}
    \end{minipage}
    \vfill
    \begin{minipage}{0.45\columnwidth}
    \includegraphics[width=\linewidth]{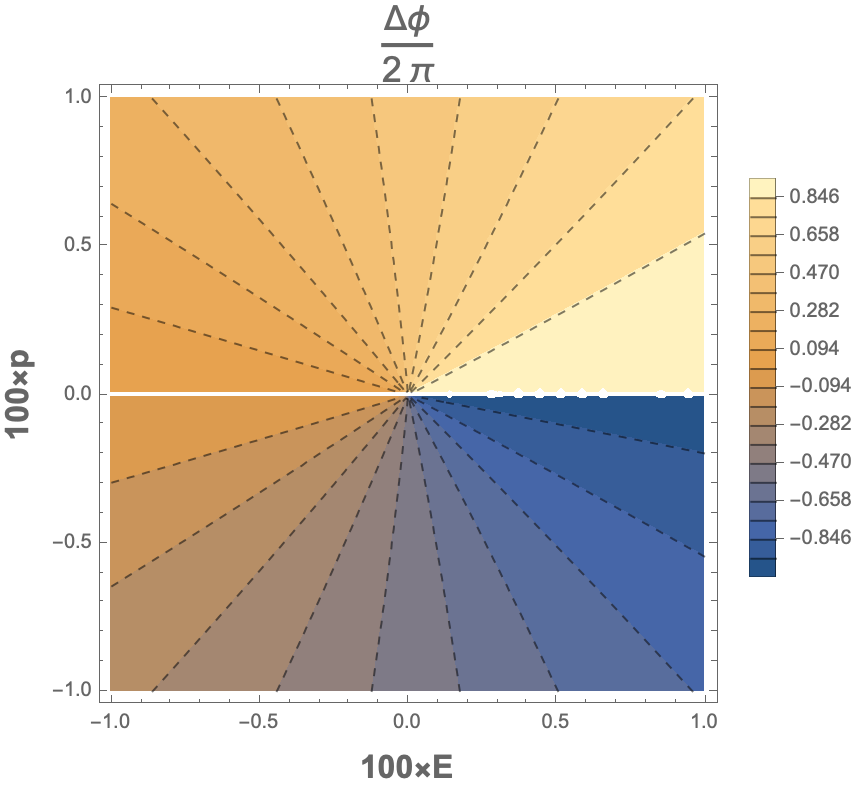}
    \caption{The phase $\frac{\Delta\phi(E,p)}{2\pi}$ in a small range of values around the extremum of $S_0$. Evidently, the phase approaches a linear function of the phase of $E+ip$ in this region.}
    \label{fig:phase contour small}
    \end{minipage}\hfill
    \begin{minipage}{0.45\columnwidth}
    \includegraphics[width=\linewidth]{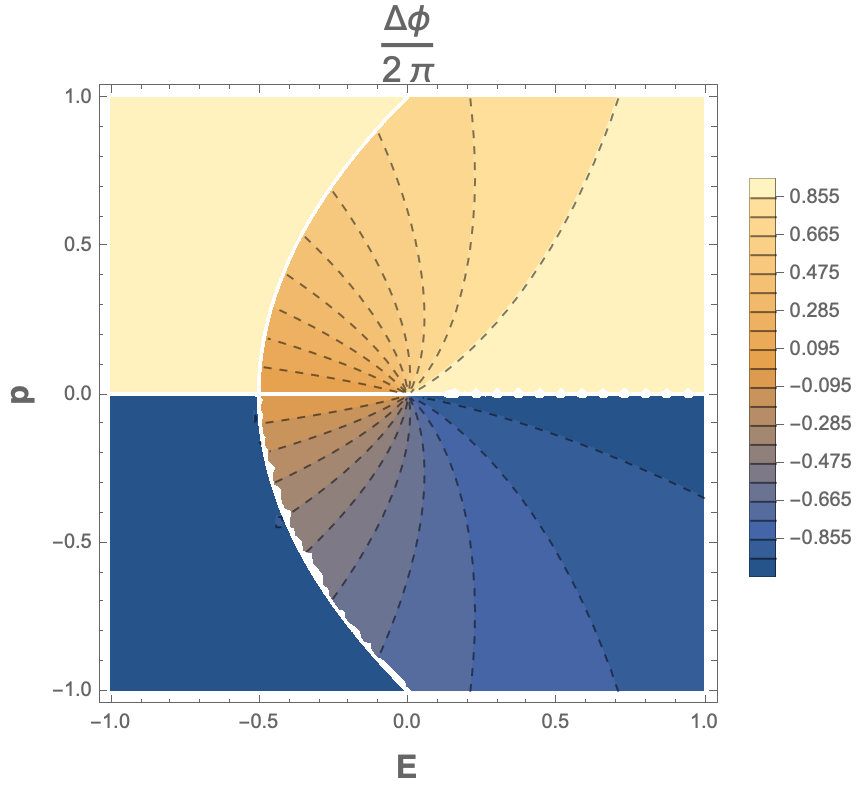}
    \caption{The phase $\frac{\Delta\phi(E,p)}{2\pi}$ in a large range of values around the extremum of $S_0$. The region to the left of the parabola $2E=p^2-n^2$ is unphysical, with impossible energies.}
    \label{fig:phase contour large}
    \end{minipage}
\end{figure}
% \begin{figure}[ht]
%     \centering
%     \begin{minipage}{0.45\columnwidth}
%     \includegraphics[width=\linewidth]{various_amplitudes/phase_contour_small_EPplane.png}
%     \caption{The phase $\frac{\Delta\phi(E,p)}{2\pi}$ in a small range of values around the extremum of $S_0$. Evidently, the phase approaches a linear function of the phase of $E+ip$ in this region.}
%     \label{fig:phase contour small}
%     \end{minipage}\hfill
%     \begin{minipage}{0.45\columnwidth}
%     \includegraphics[width=\linewidth]{various_amplitudes/phase_contour_large_EPplane.png}
%     \caption{The phase $\frac{\Delta\phi(E,p)}{2\pi}$ in a large range of values around the extremum of $S_0$. The region to the left of the parabola $2E=p^2-n^2$ is unphysical, with impossible energies.}
%     \label{fig:phase contour large}
%     \end{minipage}
% \end{figure}

We now must ask how many solutions exist near $E=p=0$. In figures \ref{fig:period contour small},\ref{fig:phase contour small}, we plot $T,\Delta\phi$ as functions of $E,p$, near the $E=p=0$ locus. At sufficiently large $T$, the fixed $T$ contours intersect once all the possible values of $\Delta\phi$. The phase $\Delta\phi$ varies from $-2\pi$ to $2\pi$, and those two values should be identified with one another.\footnote{Any given bit of string passes near the South / North Pole, and when it passes from the right the phase is $\pi$ compared with when it passes on the left.} In fact, near $E=p=0$ we have $E+ip\propto \exp(\frac{i}{2}(2\pi-\Delta\phi + i T))$.  The limits occur at the origin in $p,\theta_{\rm min}$ space. This means that for given $\tau$, at large ${\rm Im}(\tau)$, there are $2l$ solutions to the equations of motion, with $\Delta\phi=-2\pi (n\Re(\tau)+k)/l$ where $k$ is an integer taking $2l$ values consistent with the range of $\Delta\phi$. If we continuously increment $\tau\to \tau+1/n$, the solutions flow until they are cyclically identified $k\to k+1$. 

The amplitude associated to maps with tubes in the north and south poles is therefore:
\be\label{eq: path integral with matter integrated - torus on sphere w tube}
\mathcal{A}_{n,l} = \frac{1}{2} \frac{1}{nl} \intop_{\mathcal{F}^{2|2}}d^{2|2}\tau \sum_{k=0}^{2l-1}\exp(i t S_{n,l,k}(\tau)+\dots),
\ee
where $S_{n,l,k}(\tau)$ is $S_0$ evaluated at the solution and the ellipsis indicates terms like \eqref{eq: generlized S term} that would give rise to the moduli space measure. 

It is subtle to count how many of these solutions contribute near $E=p=0$ (where the integral over $\tau$ is localized). $E$ and $p$ both decay exponentially with ${\rm Im}(\tau)\to\infty$, and all the second derivatives of the action vanish there, leading to an ill-defined 1-loop determinant, as discussed below. 

Let's begin with the case of $n=1$. We can first perform the integral in \eqref{eq: path integral with matter integrated - torus on sphere w tube} over the real part of $\tau$, and only then over the imaginary part of $\tau$. For every large enough value of ${\rm Im}(\tau)$, equation \eqref{eq: tau_extrema} implies that only the two solutions at $p=0$ extremize the integral over ${\rm Re}(\tau)$ (which goes over a circle in the $(E,p)$-plane, as discussed above); if we choose the fundamental domain to have $-1/2 \leq {\rm Re}(\tau) \leq 1/2$ then these are the solutions with $k=0$ and with $k=l$ (at ${\rm Re}(\tau)=0$). 
Moreover, if we follow these solutions to ${\rm Im}(\tau)=\infty$ (where the integral over the imaginary part of $\tau$ is extremized) then it seems that they both go into the same zero-size-tube solution there. Thus, we believe that \eqref{eq: path integral with matter integrated - torus on sphere w tube} gives
\be\label{eq: A_1l ampltidue in terms of A_11}
\mathcal{A}_{1,l}=
%\frac{1}{l}\mathcal{A}_{1,1}=
\frac{1}{2l} \sign(\det(\partial\partial S_{{\rm Pol}}))\omega,
\ee
where $\omega$ is the moduli space measure.

The general case of $n>1$ is almost the same, since the space of solutions at a given $\tau$ is mapped to itself as $\tau\to\tau+1/n$, and it is easy to see that all the $2 n$ extrema inside the fundamental domain give the same contribution. So summing over the solutions cancels the factor of $n$ discussed above, and we have
\be\label{eq: A_nl ampltidue in terms of A_11}
\mathcal{A}_{n,l}=
%\frac{1}{l}\mathcal{A}_{1,1}=
\frac{1}{2l} \sign(\det(\partial\partial S_{{\rm Pol}}))\omega.
\ee

The precise analysis of the contribution of these solutions is subtle because of the vanishing of the second derivatives at $E=p=0$.
This is related to the appearance of additional zero-modes at $E=p=0$. The zero-modes related to overall $SO(3)$ transformations are already present away from $E=p=0$, but additional ones may appear there. In particular, there should be zero-modes related to \textbf{independent} displacements of the two tubes, possibly expressible as target-space non-rigid conformal transformations. In addition, note that independent worldsheet-time translations of each kink and anti-kink should become zero-modes in this limit, as the kinks become infinitely separated in time. The study of these subtleties will be postponed to future work.

\subsubsection{The moduli space}\label{subsubsec: torus on sphere moduli space}

For general $l,n$, the moduli space of extremal area maps should be $4l$ dimensional. The target space position of each of the $2l$ tubes should be able to independently move. A detailed study of its manifestation in the present model is beyond the scope of this work, but let us briefly outline what is expected:
\begin{itemize}
    \item At general values of $\tau$ all the moduli besides those related to rigid rotations are lifted. 
    \item As $\tau$ approaches $\tau_{\rm crit} = i \infty$, these moduli will begin to manifest as vanishing eigenvalues of the fluctuation determinant.
\end{itemize}

\subsubsection{Comparison to YM theory}

The contribution in the Gross-Taylor formula \eqref{eq: zymnew} relevant to us is one with $nl$ orientation-preserving sheets, the same number of orientation-reversing sheets, and a number of \ocmr-points ranging from $2$ to $2l$, with a total of $2l$ branched orientation-reversing tubes. Each orientation-reversing tube includes an order $n$ cyclic permutation, which is chosen in such a way as to produce a connected worldsheet. 

The simplest case is $l=1$. Then there are $(n-1)!$ choices for the cyclic permutation $\sigma$ of one of the \ocmr-points, and that fixes the permutation of the other to be the inverse. Similarly, the choice of permutation $\tau$ of the orientation \textbf{reversed} sheets contributes another $(n-1)!$. The combinatoric factors in the \ocmr-point \eqref{eq: ocmr point expansion} at $v_n=1$ are just $(-n)$ from each \ocmr-point, related to the choice of which orientation reversed sheet is connected to which orientation-preserving one through the tube. In total, we get:
\be
\frac{1}{n!}\frac{1}{n!} \underbrace{(n-1)!}_{\text{choice of perm' $\sigma$}}\underbrace{(n-1)!}_{\text{choice of perm' $\tau$}}(- n)^2  =1,
\ee
in addition to the factor from \eqref{eq: binom as Euler character}:
\be
\binom{2}{2} = \underbrace{\frac{1}{2!}}_{\text{bose symmetry}}\underbrace{2\cdot 1}_{\text{Euler number of conf' space of 2 points}}.
\ee
Since the above factor of $2\cdot1$ comes from integration over the moduli space, we expect the contribution of a given point in the moduli space to be $1/2$ times the Euler density, in agreement with \eqref{eq: A_nl ampltidue in terms of A_11}, provided $\sign(\det(\partial\partial S))=1$ and is well-defined, despite the potential zero modes discussed above.

In the case $l>1$, let's first consider $2$ \ocmr-points, each with $l$ cyclic permutations of order $n$. One \ocmr-point will carry disjoint cyclic permutations $\sigma_k,\,\tau_k$ and the other will carry the inverse cycles, to satisfy the holonomy constraint. The cycles must be disjoint to satisfy connectivity, so we arrive at all the sheets of the worldsheet. The combinatoric factor we get is:
\bea
& \underbrace{\frac{1}{l!}\binom{nl}{n,n,\dots,n}}_{\text{division into $l$ cycles for the $\sigma$-s}}\times
\underbrace{\frac{1}{l!}\binom{nl}{n,n,\dots,n}}_{\text{division into $l$ cycles for the $\tau$-s}}\times\underbrace{((n-1)!)^l}_{\text{Choice of cycle orders for the $\sigma$-s}}\\
\times & \underbrace{((n-1)!)^l}_{\text{Choice of cycle orders for the $\tau$-s}}\times  \underbrace{(-n)^{l} l!}_{\begin{array}{cc}
     \text{combinatoric factor}  \\
     \text{from first \ocmr-point}
\end{array}}\underbrace{(-n)^{l} (l-1)!}_{\begin{array}{cc}
     \text{combinatoric factor}  \\
     \text{from second \ocmr-point}
\end{array}}\\
= & \, \frac{1}{l} (nl)! (nl)!.
\eea
Note that the first \ocmr-point contributed $l!$ whereas the second contributed $(l-1)!$. This is because the choice of ways to connect cycles across tubes is constrained by the requirement of producing a connected worldsheet. After making such a choice for the ``first'' \ocmr-point, the choices for the second must induce a cyclic permutation \textbf{of the $l$ cycles}, and there are $(l-1)!$ such permutations. Together with the factor of $\frac{1}{(nl)!}\frac{1}{(nl) !}$ in \eqref{eq: zymnew} and the binomial coefficient from \eqref{eq: binom as Euler character} we get:
\be
\frac{1}{2l} 2\cdot 1.
\ee
Hence, the coefficient at a given point in the moduli space is $1/2l$, again, in agreement with \eqref{eq: A_nl ampltidue in terms of A_11}. 

The contributions with more than $2$ \ocmr-points vanish due to the binomial coefficient in \eqref{eq: binom as Euler character}.

\subsection{Sphere covering the sphere with an orientation-reversing tube}\label{subsec: sphere on sphere w/ tube}

We consider maps from the sphere into the unit sphere, with metric $g_{Z\bar{Z}}=4(1+Z\bar{Z})^{-2}$. The NLSM equation is:
\begin{equation}\label{eq: NLSM eq' sphere on sphere w/ tube}
g_{Z\bar{Z}}\partial\bar{\partial}Z+g_{Z\bar{Z};Z}\partial Z\bar{\partial}Z=0,
\end{equation}
or, equivalently:
\begin{equation}
\partial\bar{\partial}Z-\frac{2\bar{Z}}{1+Z\bar{Z}}\partial Z\bar{\partial}Z=0.
\end{equation}
Because the sphere has no holomorphic quadratic differentials, there are no complex structure moduli and correspondingly the Virasoro constraint is automatically satisfied on-shell:
\begin{equation}
g_{Z\bar{Z}} \partial Z \partial \bar{Z}=0.
\end{equation}
This completely determines the solutions of \eqref{eq: NLSM eq' sphere on sphere w/ tube} to be piecewise (anti-)holomorphic maps, ostensibly freeing us from dealing with the nonlinearity of the equation. More precisely, meromorphic maps are also allowed, since the factor $g_{Z \bar{Z}}$ vanishes at poles and $Z=\infty$ is just a regular point on the sphere. Furthermore, maps which blow up at infinity are permitted, since the target space metric ensures that the action is finite. Thus, permitted embeddings in the sphere include maps such as:
\begin{equation}
Z=z^n,\,Z=1/z^n,\,Z=\bar{z}^n,\, Z=1/\bar{z}^n, \cdots.
\end{equation}
These describe various wrappings of the target space. 

Our current interest is in maps that cover the target space with two sheets, one holomorphic and one anti-holomorphic, connected by a single orientation-reversing tube. These are maximal area maps in the topological sector of connected \textbf{contractible} maps (since the tube can be widened until the covering unwraps itself from the target space). Schematically, they have the form $Z=az+a/\bar{z}$, in the singular limit $a\to0$ which puts most of the holomorphic covering near $z=\infty$ and the anti-holomorphic one near $z=0$. Here, a difficulty arises which is common to all such cases of maps that change orientation, in the absence of complex structure moduli. These maps are singular limits of well-behaved maps like the one above, none of which actually solve the NLSM equation. They are, in a sense, saddle points at infinity in field space. By ``infinity'' we are referring to all the non-smooth limits of smooth maps -- the limits that establish the noncompactness of field space.

To deal with this problem, we will use the method of ``constrained instantons'' \cite{AFFLECK1981429}. We choose to perform the path integral in a particular order, first keeping fixed a certain conveniently chosen degree of freedom so that the integral over the rest will enjoy a well-behaved saddle. These saddles now make up a one-parameter family of maps, parametrized by the ``frozen'' degree of freedom, that asymptote to the true extremal area map of interest at one end, and to another (in this case the constant map) at the other.

Concretely, consider a one-parameter family of maps $Z_{A}$ with $A(\theta,\bar{\theta})$ the parameter-multiplet. If we expand the fields in fluctuations around it, $Z=Z_A+\delta Z/\sqrt{t}$, the action will include a tadpole term for the fluctuations:
\begin{equation}
\sqrt{t} \intop d^2z d^2 \theta \left( g_{Z\bar{Z}}(Z_A,\bar{Z}_A)\partial\bar{\partial}Z_A+g_{Z\bar{Z};Z}(Z_A,\bar{Z}_A)\partial Z_A\bar{\partial}Z_A \right)\delta \bar{Z}.
\end{equation}
The growth of this term in the large $t$ limit spoils localization to $Z_A$ -- unsurprisingly, since it is not a saddle of the localizing term. To cancel this term, we must add a ``constrained instanton'' term.  But first, let us fix the residual gauge symmetries.

\paragraph{Gauge fixing:}
Let's assume that the map covers the North Pole $Z=\infty$ at least twice (once with each orientation), so we can pick the gauge $Z(0)=Z(\infty)=\infty$.  To impose the gauge in the Faddeev Popov procedure, we must insert a resolution of 1 as an integral over a delta function. We can normalize the delta function with $-1/2$, to account for the two possibilities of which inverse image of infinity is mapped to 0 and which to $\infty$, and for the negative sign of the Fermi-determinant of the matrix:
\begin{equation}
\left(\begin{array}{cc}
\partial Z & \partial\bar{Z}\\
\bar{\partial}Z & \bar{\partial}\bar{Z}
\end{array}\right),
\end{equation}
at the point which covers infinity with a negative orientation. Alternatively, we could include just a factor of $-1$ and require that $z=0$ is on the negative orientation sheet of the cover. This gauge-fixing breaks the conformal group of the worldsheet sphere to just dilatations and rotations. It is now natural to decompose the mapping into Fourier modes with respect to to the phase $\alpha$ of $z=re^{i\alpha}$:
\bea
    Z(z) & =\sum_{k=-\infty}^\infty e^{i k \alpha} Z_k (r)\\
    Z_k(r) & = \intop \frac{d\alpha}{2\pi}e^{-ik\alpha}Z(r,\alpha).
\eea
For the maps of interest to us $Z_1(r)$ behaves like $ c_\infty r$ at large $r$ and like $c_0 /r$ at small $r$. We can fix our remaining gauge symmetry by setting $|c_\infty|=|c_0|,\,Z_1(1) \in \mathbb{R}^+$.

\paragraph{Constrained instanton:}
Next, to obtain a ``constrained instanton'' with a real parameter $A$, we insert into the path integral:
\bea\label{eq: Constrained instanton Z1(1)}
    1& =\intop d^{2|2} A \delta^{2|2}(A - Z_1(1))\\
    & =\intop d^{2|2} A  d^{2|2}\Lambda  \exp \left(i \intop d^2\theta {\Lambda}(A - Z_1(1))\right).
\eea
Now $\Lambda$ acts as a source term in the $Z$ EOM, allowing $\partial_r Z_1(r)$ to jump (by a real number) at $r=1$. We therefore find a solution to the combined $Z$ and $\Lambda$ equations, at fixed $A$:
\begin{equation} \label{eq: constsols}
    Z_A=\begin{cases}
Az & \qquad |z|>1\\
A/\bar{z} & \qquad |z|<1
\end{cases},\, \qquad \Lambda = \frac{4}{(1+A^2)^2} (2A).
\end{equation}

Integration over the fluctuations $\delta Z$ and $\delta\Lambda$ can now be carried out using localization, and the 1-loop integral gives 1 on general grounds. We are left only with integration over $A$, for which we need to study the on-shell action $S_0[Z_A]$. Since $Z$ is piecewise holomorphic / anti-holomorphic, the action becomes equal to the induced area, which since the worldsheet covers the $|Z|>A$ region twice is equal to:
\begin{equation}
%    -iS_0[Z_A] = 2 \times \frac{4\pi}{1+A^2}.
        S_{\rm Pol}[Z_A] = 2 \times \frac{4\pi}{1+A^2}.
\end{equation}
The two extrema, as expected, are at $A=0$ (maximum) and $A=\infty$ (minimum).

To analyze whether there are other solutions, we would have to study the full non-linear equation more carefully. It is likely that additional branches exist that are not quasi-holomorphic for generic $A$, and where the value of $|Z_1|$  attains multiple extrema as a function of $r$. These maps will correspond, upon extremizing with respect to $A$, to multiple coverings of the target space that are connected by multiple tubes, some centered at $Z=0$, and others at the $Z=\infty$. These should, for critical $A$, be embedded in a larger moduli space of maps in which the positions of the tubes are independently varied.

We expect that different choices of the ``constrained instanton'' terms would give different families of solutions, that would agree with \eqref{eq: constsols} in the extreme limits but not for finite $A$. For instance,
a smoother one-parameter family of configurations is given by ${\tilde Z}_A=A(z+1/\bar{z})$. 
%Presumably, a suitable choice of constrained instanton ``insertion'' will give rise to it. 
If we can find a ``constrained instanton'' that would give rise to it, then after integrating over the rest of the modes, the on-shell action would be:
\be
S_{\rm Pol}[{\tilde Z}_A] = 4\pi \left(\frac{4A^2+2}{4A^2+1}+\frac{4(A^2)^2\log\left(\frac{2A^2-\sqrt{4A^2+1}+1}{2A^2+\sqrt{4A^2+1}+1}\right)}{\left(4A^2+1\right)^{3/2}}\right).
\ee
This attains its maximal value of $8\pi$ -- twice the target space area -- at $A=0$ as expected. At large $A$ it decays to $0$ as $\frac{8\pi}{3A^2}$, approaching the degenerate case where the worldsheet is mapped to a point, in this case the North Pole.

\paragraph{The moduli space of the problem:} So far, we considered only the case where the tube is centered around the origin $Z=0$ in the target space (the South Pole). However, the action is invariant under bi-graded $SO(3)$ rotations of the target space $Z\to e^{i \alpha (\theta,\thetabar)} \frac{Z-Z_0 (\theta,\thetabar)}{Z Z_0^* (\theta,\thetabar)+1}$, and those can be used to bring the center of the tube to any other position in the target space. Rotations that fix the origin can be undone by worldsheet bi-graded phase rotations, but the rest of the rotation group parameterizes the possible positions to which the origin can be mapped. The moduli space of our solutions is thus expected to also be a sphere, and no point on it is distinguished. According to \eqref{eq: Euler density from 0-mode integration 1} and the surrounding discussion, the measure on the moduli space will be the Euler density, and the integral will produce the Euler number, which is $\chi(S^2)=2$.

A subtle problem, however, is that this moduli space is lifted by the constrained instanton we introduced, which breaks the rotational symmetry when the dynamical variable $A$ \textbf{is kept fixed}\footnote{The rotational symmetry is preserved if $A,\Lambda$, or any such dynamical object is endowed with a suitable transformation property.} (it is also partly lifted by our gauge-fixing). The full moduli space will likely make itself felt again only as $A$ reaches its critical value. To keep the moduli space for general $A$ we add to the path integral (without modifying it) another degree of freedom $Z_0$, and then for any given $Z_0$ we replace \eqref{eq: Constrained instanton Z1(1)} by a suitably $SO(3)^{2|2}$ transformed version of itself. For every $Z_0$ we can then regularize the specific mapping with a tube at that $Z_0$, and integrating over it reproduces the full moduli space.

Concretely, consider the rotated map:
\be\label{eq: rotated map of sphere into wphere w tube}
Z=\frac{Z_{A}-Z_{0}}{Z_{A}\bar{Z}_{0}+1}.
\ee
It is easy to verify using the covariance of the EOM \eqref{eq: NLSM eq' sphere on sphere w/ tube} that upon replacing $Z_A$ with its transformed version we get:
\be\label{eq: rotated NLSM equation for ZA}
g_{Z\bar{Z}}\partial\bar{\partial}Z_{A}+g_{Z\bar{Z};Z}\partial Z_{A}\bar{\partial}Z_{A}\to\frac{\left(\bar{Z}_{A}Z_{0}+1\right)^{2}}{1+Z_{0}\bar{Z}_{0}}\left(g_{Z\bar{Z}}\partial\bar{\partial}Z_{A}+g_{Z\bar{Z};Z}\partial Z_{A}\bar{\partial}Z_{A}\right).
\ee
Therefore, if we generalize the constrained instanton \eqref{eq: Constrained instanton Z1(1)} to depend on $Z_0$ through the inverse transformation \footnote{We must also modify the residual gauge condition $Z_1(1)\in {\mathbb R}^+$ to depend on $Z_0$, so that the rotated expression is the one in ${\mathbb R}^+$. This can be done because world-sheet rotations act as a phase on both $Z$ and $Z_0$.}:
\be
\intop d^{2}\theta{\Lambda}\left(A-\intop d^{2}z\frac{e^{-i\alpha}}{2\pi}\delta\left(r-1\right)\frac{Z-\left(-Z_{0}\right)}{Z\left(-\bar{Z}_{0}\right)+1}\right)+({\rm c.c.}),
\ee
which coincides with \eqref{eq: Constrained instanton Z1(1)} for $Z_0=0$, we ensure that \eqref{eq: rotated map of sphere into wphere w tube} is a solution. To see this, we compute the $Z$ variation of this term:
\be
\delta\frac{Z-\left(-Z_{0}\right)}{Z\left(-\bar{Z}_{0}\right)+1}=\frac{1+Z_{0}\bar{Z}_{0}}{\left(-Z\bar{Z}_{0}+1\right)^{2}}\delta Z,
\ee
and for \eqref{eq: rotated map of sphere into wphere w tube}:
\be
\frac{1+Z_{0}\bar{Z}_{0}}{\left(-Z\bar{Z}_{0}+1\right)^{2}}=\frac{1+Z_{0}\bar{Z}_{0}}{\left(-\frac{Z_{A}-Z_{0}}{Z_{A}\bar{Z}_{0}+1}\bar{Z}_{0}+1\right)^{2}}=\frac{\left(Z_{A}\bar{Z}_{0}+1\right)^{2}}{1+Z_{0}\bar{Z}_{0}},
\ee
giving the same pre-factor as in \eqref{eq: rotated NLSM equation for ZA}.\footnote{Up to complex conjugation, but \eqref{eq: rotated NLSM equation for ZA} is the $\bar Z$ EOM so it's consistent.} To introduce $Z_0$ into the path integral we can insert (see \eqref{eq: euler character from Chern-Gauss-Bonnet}):
\be\label{eq: introduction of Z_0 into path integral}
1=\frac{\chi\left(S^{2}\right)}{2}=\frac{1}{2}\intop d^{2|2}Z_{0}d^{2|2}\bar{Z}_{0}\exp\left(-c\intop d^{2}\theta\left(\partial_{\theta}Z_{0}g_{Z_{0}\bar{Z}_{0}}\partial_{\bar{\theta}}\bar{Z}_{0}+\partial_{\theta}\bar{Z}_{0}g_{Z_{0}\bar{Z}_{0}}\partial_{\bar{\theta}}Z_{0}\right)\right).
\ee
The factor of $1/2$ above will eventually cancel because for each value of $Z_0$ there will be two solutions related by the antipodal transformation. Our gauge conditions $Z(0)=Z(\infty)=\infty$ can now be modified to $Z(0)=Z(\infty)=1/{{\bar Z}_0}$ and the gauge condition on the asymptotic behaviors near these points can similarly be adapted by defining:
\be
c_0 = \intop d^2 z e^{-i\alpha} \delta(r) \frac{Z-\left(-Z_{0}\right)}{Z\left(-\bar{Z}_{0}\right)+1},\,\qquad c_\infty = \intop d^2 z e^{-i\alpha} r^{-4}\delta(1/r) \frac{Z-\left(-Z_{0}\right)}{Z\left(-\bar{Z}_{0}\right)+1}.
\ee
Finally, after localizing to the critical values of $A$ we'll be left with an integral over the modulus $Z_0$. In addition to the term \eqref{eq: on-shell S-term} we will now have also the measure from \eqref{eq: introduction of Z_0 into path integral}. This should not change the final answer (by $Q$-exactness), but in any case we can take the coefficient $c$ to be arbitrarily small.

\subsubsection{Comparison to Yang-Mills theory} 

The $n=\bar{n}=1$ contribution to the partition function on the sphere ($G=0$) comes just from the adjoint representation, which gives:
\be
(\dim ({\rm Adjoint}))^2= N^4 -2 N^2 +1.
\ee
The first term is the contribution from two disconnected spheres. In this section, we correctly reproduce the middle contribution (from a single genus zero sphere). The factor of $2$ comes from the Euler number of the moduli space, while the factor of $(-1)$ is due to the maximality of the saddle.

\subsubsection{Generalization to multi-covers} 

We can generalize our discussion to maps schematically of the form $Z_A=A (z^n +1/\bar{z}^n)$. One needs to change the constrained instanton \eqref{eq: Constrained instanton Z1(1)} (and its generalizations) to $A=Z_{n}(1)$. The on-shell action is then $n$ times the previous result, corresponding to $n$ covers with each orientation. We expect the extremal area solutions to have a branched tube with a cyclic permutation of order $n$ at the South Pole, and two branch points of order $n-1$ (one for each orientation) at the North Pole. A symmetry factor of $1/n$ should be included due to worldsheet rotations that leave invariant the map. 

There should now be additional moduli corresponding to the possibility of independently moving around the various objects, as well as splitting the branch points into simpler ones, each with its own position modulus. The collision that forms the higher order branch point is a collision of mixed types $1,2$ (see discussion in section \ref{subsec: stringy interpretation of the ocmr points}) and should not be associated with a singularity in the moduli space. However, because these moduli do not correspond to the action of a symmetry, they are likely lifted as long as $A$ is not at its critical value.

Much like in the case of the torus covering the sphere (section \ref{subsec: torus on sphere}), we can consider additional mappings with another index $l$, representing a map with $(2l-1)$ (possibly branched) tubes that alternate in position between the North and South Poles. However, unlike in that case, it is not simple to find solutions of the form above that have a simple single frequency ($Z_k=0$ for $k\neq1$), using the ``constrained instanton'' chosen above. %If we include \textbf{two} constrained instantons, say by ``fixing'' $Z_1(r)$ at \textbf{two} values of $r$, then such solutions may appear. In the language of section \ref{subsec: torus on sphere}, $\theta = 2\tan(|Z_1|)$ lives in a sine-Gordon potential. If we have two constrained instantons, its motion is interrupted at two times, at which its energy can change. The particle's energy can be 0 in the first interval and the last, while its energy in the intermediate time interval can be tuned to obtain the necessary $l$ oscillations in the well.

\subsection{Sphere covering the sphere twice}\label{subsec: sphere on sphere twice}

In this section we consider maps from the sphere to itself which cover it twice with the same orientation; this is a representative example of mappings that are regular, but where the moduli space of mappings has a singularity (which is like an orbifold). 

For this case,
as in section \ref{subsec: sphere on sphere once}, the stress tensor vanishes, and we consider holomorphic maps  $Z(z)$. We require that each point in the target space has two pre-images. This restricts to rational functions $p(z)/q(z)$ with $p,q$ quadratic. This is a 5 complex parameter family of solutions, but we also have the 3 complex parameter M\"obius group as a redundancy. We can use that to bring the solutions into the form of a \textbf{target space M\"obius transformation} applied to the trivial solution $Z=z^2$:
\begin{equation}\label{eq: sphere wrapping twice gauge choice 1}
   Z = \frac{Z_{2}z^{2}+Z_{1}}{z^{2}+1},
\end{equation}
which has only two parameters $Z_{1,2}$.  These correspond to the positions of the branch points in the target space, whereas their worldsheet positions have been gauge-fixed to $z=0$ and $z=\infty$.

Other useful gauges, related to this one by M\"obius transformations, are possible. We should note that this gauge can be characterized as setting the worldsheet positions of the branch points to $z=0,\infty$, and also setting $Z(i)=\infty$. This gauge can be chosen generically, but not when either of the branch points is taken to infinity in the target space, and we would like to describe such configurations as well. Gauge choices that avoid this problem are possible, but complicate the computation. One way to include configurations where branch points are at infinity is to divide the moduli space into patches and to use a different mapping in each patch, carefully taking into account any contributions from the boundaries between the patches. For instance, a different choice which is not problematic when branch points go to infinity is
\be
Z = \frac{z^2+1}{z^2 Z_1 +Z_2}Z_1 Z_2,
\ee
which sets $Z(i)=0$ and is problematic if and only if a branch point goes to $Z=0$. Similarly:
\be
Z = \frac{z^2+Z_1}{z^2 +Z_2} Z_2,
\ee
can be used in the patch of moduli space where $Z_1$ and $Z_2$ ``threaten'' to approach, respectively, $0$ and $\infty$.

The only remaining conformal transformations for the gauge choice \eqref{eq: sphere wrapping twice gauge choice 1} are $z\to-z$ (which permutes the two sheets)  and $z\to 1/z$ (which exchanges $Z_1 \leftrightarrow Z_2$), so we can simply divide by 4 and take $Z_{1,2}$ to be independent parameters, albeit symmetric under exchange. After extracting this factor of $1/4$, the Euler number, based on the YM theory, \textbf{should come out to be $2$} (see below).

The moduli space is then given by the positions $Z_{1,2}$, but the map becomes singular when the two branch points collide, leading to a corresponding singularity in the moduli space metric. This is a type 3 collision (see the discussion in section \ref{subsec: stringy interpretation of the ocmr points}). Evidently, the moduli space is either not a Riemannian manifold (if the collision is included) or it is noncompact (if it is excluded). In either case, standard topological arguments cannot be appealed to in determining the contribution of the singular region to the integral of the Euler density. 

From the field theory side, we know that the contribution at order $N^2$ from representations with $n=2$ is $N^2/2$. On the string theory side, the corresponding maps are either the sphere double-covering the sphere, or the disconnected union of a sphere covering the sphere once and a torus covering the sphere once with a degenerated handle. The latter contribution should be 0, since the torus covering the sphere once doesn't contribute at $n=1$. Taking into account the division by 4, we expect the regulated Euler number of the moduli space presently under consideration to come out to be 2, if we are to reproduce the field theory result.

The field theory result can also simply be written using \eqref{eq: zymnew} as:
\be
\frac{1}{2!} {\binom{2}{2}} = \underbrace{\frac{1}{2!}}_{\text{sheet automorphisms}}\underbrace{\frac{1}{2!}}_{\text{Bose symmetry}} 2\cdot 1.
\ee
The combinatorics match the stringy analysis above, but, as discussed in section \ref{subsec: stringy interpretation of the ocmr points}, the Euler number indicates that the collision should behave like a puncture which is removed.

In order to reproduce this, we must compute the metric on the moduli space and then derive the Euler density from it.  Note that the holomorphic dependence  of $Z$ on the moduli gives the metric a complex structure. There is no guarantee that every choice of metric on the moduli space will give the same, or even a finite, Euler number, due to the noncompactness / singularity. We will therefore obtain an ``Euler character'' which is not completely topologically invariant. It will depend on the nature of the singularity of the metric, but not on the rest of its details.

To see this, consider that the moduli space can be split into a small neighborhood of the singularity, and the rest, and the Euler character can be written as a sum by adding and subtracting the boundary term from the Chern-Gauss-Bonnet theorem \cite{GILKEY1975334} along the interface of the two regions. The contribution from the regular part (including its boundary term) will be determined by the Chern-Gauss-Bonnet theorem to be $2$ (the Euler number of $[(S^2\times S^2) - S^2]$), while the contribution from the singular neighborhood will depend on the asymptotic behavior of the metric.

Below, we will consider three possible actions (leading to three different metrics for the moduli space). The first choice (section \ref{subsubsec: K = extrinsic curvature}) gives an answer that formally reproduces the expected result. The other 2 options exhibit some interesting properties, but fail to give the expected result, presumably due to contributions near the singularity which are not correctly taken into account.

\subsubsection{\texorpdfstring{$K_{\mu\nu}=$}{K =} extrinsic curvature}\label{subsubsec: K = extrinsic curvature}

We can start by adding to the action \eqref{eq: generlized S term} with \eqref{eq: extrinsic S term}. Since the maps \eqref{eq: sphere wrapping twice gauge choice 1} with $Z_1 \neq Z_2$ have only branch points, the worldsheet integral \eqref{eq: moduli metric from ws integral} defining the metric on the moduli space will simply localize to the branch points. This means that the metric on the moduli space will factorize into two 2d metrics in the $Z_1$ and $Z_2$ directions. In each block, the metric will be simply the \textbf{pullback} of the target space metric\footnote{We assume here that we describe the regions of moduli space where $Z_i\to \infty$ by dividing the moduli space into patches, and using mappings which are non-singular in this limit for that part of the moduli space. Since the different patches give us precisely the same moduli space metric in this case, we do not get any contributions from the boundaries between the patches.}. However, the points with $Z_1=Z_2$ do not appear in the moduli space,
%the metric on the moduli space vanishes, 
since in that case the mapping \eqref{eq: sphere wrapping twice gauge choice 1} in our gauge choice degenerates to a constant map. Thus, our moduli space is simply $[(S^2 \times S^2) - S^2]$, and integrating the Euler density on it gives
\be
\chi(S^2\times S^2)-\chi(S^2)=4-2=2
\ee
as expected.

We expect that this will generalize to arbitrary type 3 collisions of singular points, and that these will always be removed from the moduli space, as anticipated in the discussion of section 5.3 of \cite{Horava:1995ic}.

\subsubsection{ \texorpdfstring{$K_{\mu\nu}=\left(\partial X\right)^2g_{\mu\nu}$}{K = (dX)(dX) g}} 

In this case, conformal invariance is maintained, and the metric is:
\begin{equation}
    m_{i\bar{j}}=\intop d^{2}z\frac{1}{\left(1+z\bar{z}\right)^{2}}\frac{\left(1+z\bar{z}\right)^{2}}{\left(1+Z\bar{Z}\right)^{2}}\partial Z\bar{\partial}\bar{Z}\frac{1}{\left(1+Z\bar{Z}\right)^{2}}\partial_{i}Z\partial_{\bar{j}}\bar{Z}.
\end{equation}
The worldsheet metric decouples as expected from conformal invariance. Using conformal invariance, we can take $z\to\sqrt{z}$, followed by a M\"obius transformation to bring $Z$  to the form $Z=z$. The integrals giving each component of the metric are then simple:
\begin{gather}
    m_{i\bar{j}} = \left(\begin{array}{cc}
\frac{2\pi}{3} & -\frac{2\pi\bar{Z}_{+}}{3\bar{Z}_{-}}\\
-\frac{2\pi Z_{+}}{3Z_{-}} & \frac{2\pi\left|Z_{+}\right|^{2}+\pi}{3\left|Z_{-}\right|^{2}}
\end{array}\right),
\end{gather}
where $Z_+ = Z_1 + Z_2,\, Z_- = Z_1 -Z_2$. This evaluation method can be generalized for arbitrary rational mappings.

The Euler density for this metric evaluates to exactly 0.  This remains true also if we modify $K_{\mu\nu}=\left(\partial X\right)^2g_{\mu\nu} \sqrt{g}^n$ for arbitrary $n\geq0$, which also enjoys conformal invariance. However, for different gauges we get different, nonzero results, signaling the pathological nature of the singularity in the moduli space, as well as the inadmissibility of the gauge choice at $Z_{1,2}\to \infty$ (which needs to be dealt with separately).

\subsubsection{ \texorpdfstring{$K_{\mu\nu}=g_{\mu\nu}$}{K = g}} 

The metric on the moduli space induced by the term $\partial_{\theta}X\cdot\partial_{\bar{\theta}}X$  is:
\begin{equation}
    m_{i\bar{j}}=\intop d^{2}z\frac{1}{\left(1+z\bar{z}\right)^{2}}\frac{1}{\left(1+Z\bar{Z}\right)^{2}}\partial_{i}Z\partial_{\bar{j}}\bar{Z}.
\end{equation}
Interestingly, this metric is K\"ahler, with:
\begin{align}
    m_{i\bar{j}} & =\partial_{i}\partial_{\bar{j}}K\\
    K & =\intop d^{2}z\frac{1}{\left(1+z\bar{z}\right)^{2}}\log\left(1+Z\bar{Z}\right).
\end{align}
The K\"ahler potential can be approximated near the singularities, and the Euler number evaluates numerically to roughly $-\frac{7}{2}$, again indicating that the singularity is pathological.

%% file: appendices.tex
\section{Bi-graded algebra}\label{sec: bi graded algebra}

The superspace used in this paper is taken from \cite{Horava:1998wf}. Here we develop some formulae using it.

We define a bi-graded number to be an element of an algebra generated by two Grassmann numbers, $X(\theta,\thetabar)=x+\theta x_\theta + \thetabar x_\thetabar + \theta\thetabar x_{\theta \thetabar}$. 

More generally, given any sort of object taking values in a manifold $\mathcal{M}$, we can bi-grade it, extending it to a function of the Grassmann numbers. In that case, the new components should always be thought of as existing in the tangent space to the manifold at $y$, $T_y\mathcal{M}$, and representing infinitesimal variations of $y$. 
For instance, for some Lie group element $y\in G$ we can write $Y\in G^{2|2}$ and then:\footnote{An alternative definition: $Y=y+\theta y^*y_\theta + \dots$ with $y_\theta$ a tangent vector at the identity element might be more natural in the case of groups.}
\be
Y=y+\theta y_\theta + \dots
\ee
is a formal combination of a group element $y$ with \textbf{Lie-algebra} elements $y_\theta,y_\thetabar,y_{\theta \thetabar}$. The inverse is $Y^{-1}= y^{-1} - y^{-1} \theta y_\theta y^{-1} - \dots$, automatically 
translating the vector $y_\theta$ from $T_{y}G$ to $T_{y^{-1}}G$.
% In particular, consider conjugation of  $Y$ by some element $H$:
% \be
% H Y H^{-1}= h Y h^{-1} + \theta
% \ee

\subsection{Coordinate transformations}
Given some reparameterization ${X'}^i (X^j,\theta,\thetabar)$, the Jacobian matrix is an operator acting on a super-vector space, with both bosonic and fermionic directions. If we write $A=1,2,3,4$ and $x^i=x^i_1,\,x^i_{\theta}=x^i_2,\,x^i_\thetabar=x^i_3,\,x^i_{\theta\thetabar}=x^i_4$, then the matrix is upper-diagonal in the $A$ index. The diagonal elements are all the same, and equal to the ordinary Jacobian elements of ${x'}^i(x^j)$. The super-determinant, or Berezinian, is unaffected by the above-diagonal components and so becomes:
\be
{\rm sdet}(\partial X') = \frac{\det\left|{\begin{array}{cc}
    \partial x &  \\
     & \partial x
\end{array}}\right|}{\det\left({\begin{array}{cc}
    \partial x &  \\
     & \partial x
\end{array}}\right)} = 1
\ee
for orientation preserving reparameterizations. Thus, the fermionic components, acting as differential forms for their bosonic counterparts, enable an extensive coordinate-invariance.

\subsection{Bi-graded equations}\label{subsec: Bi-graded equations}
Every equation in bosonic variables can be automatically extended to a bi-graded one:
\be
f(x)=0\to f(X)=0.
\ee
One should first examine the bottom component, which gives back the original equation:
\be
f(X)\mid_{\theta=\thetabar=0}=f(x)=0.
\ee
Then:
\bea
f(X)\mid_{\theta,',\thetabar=0} & = \partial_i f(x)x^i_\theta =0,\\
f(X)\mid_{\theta=0,\,\thetabar} & = \partial_i f(x)x^i_\thetabar =0,\\
f(X)\mid_{\theta\thetabar} & = \partial_i f(x)x^i_{\theta\thetabar} - \partial_i \partial_j f(x)x^i_\theta x^j_\thetabar = 0.
\eea
If $df(x_0)$ is non-degenerate at a solution $x_0$ to the Bosonic equation, as in the case of an isolated solution, then all the extended components are set to $0$, and we simply get $X=x_0$. 
If there is a moduli space of solutions, parametrized by some coordinates $a^\alpha$:
\be
x=x_0(\{a^\alpha\}),
\ee
then some of the derivatives and higher derivatives of $f$ will vanish, allowing for nontrivial components. This can be summarized as simply:
\be\label{eq: bi graded solution}
X=x_0(\{A^\alpha (\theta,\thetabar)\}).
\ee
In summary, if there is a moduli space of solutions to the bosonic equation, then the bi-graded version of the solution space is obtained by bi-grading the moduli / collective coordinates.

If $f$ is extended to a superfield, $F(X,\theta,\thetabar)=0$, the only thing that changes is the appearance of ``source terms'' in the equations for $x_\theta,x_\thetabar,x_{\theta\thetabar}$. In short, if $f$ includes some parameters $b^q$, on which the solutions depend: $x_0(\{a^\alpha\},\{b^q)\}$ and $F$ is defined by bi-grading the parameters, then the solutions are:
\be\label{eq: generalized bi-graded solution}
X=x_0(\{A^\alpha (\theta,\thetabar)\},\{B^q (\theta,\thetabar)\}).
\ee

\subsection{Integration}\label{subsec: bi-graded integration}
Note that for any function $f(x)$:
\bea
f(X) & = f(x)+ \theta f'(x) x_\theta + \thetabar f'(x)x_\thetabar +\theta\thetabar ( f'(x) x_{\theta \thetabar} - f''(x) x_\theta x_\thetabar),
\eea
and its integral is given by
\bea
i \intop d^2 \theta f(X) & = i f'(x) x_{\theta \thetabar} - i f''(x) x_\theta x_\thetabar, \\
d^2\theta &\equiv d\thetabar d\theta.
\eea
If $f$ plays the role of an action, evidently the ``auxiliary'' component $x_{\theta \thetabar}$ is a Lagrange multiplier. Hence, the integral over $X$ will only make sense when the factor of $i$ above is included. Concretely, we define the integration measure:
\be
d^{2|2}X \equiv \frac{i}{2\pi}dx dx_\thetabar dx_\theta dx_{\theta \thetabar},
\ee
which is invariant under arbitrary reparametrizations of $X$. We find that the integral localizes:
\be
\intop d^{2|2}X \exp \left(i \intop d^2 \theta f(X)\right) = \intop dx \delta(f'(x))f''(x) = \sum_{x|f'(x)=0} \sign(f''(x)).
\ee
This can be generalized for $x$ an element of an $n$-dimensional manifold $\mathcal{M}_n$. The integral gives the Euler number of the space, according to the Poincar\'e-Hopf theorem, with $f$ acting as a Morse function:
\be\label{eq: euler character from Morse}
\intop_{\mathcal{M}^{2n|2n}} d^{2n|2n}X \exp \left(i \intop d^2 \theta f(X)\right) =  \sum_{x|df(x)=0} \sign(\det(\partial_i\partial_j f(x)))=\chi(\mathcal{M}_n).
\ee
We can further generalize this for $F(X)=f(X)+\theta f_\theta(X)+\dots$ a bi-graded function. Since $x_{\theta\thetabar}$ still appears as a Lagrange multiplier, we obtain:
\be
\intop_{\mathcal{M}^{2|2}} d^{2n|2n}X \exp \left(i \intop d^2 \theta F(X)\right) =  \sum_{x|df(x)=0} e^{if_{\theta\thetabar}(x)+i(df_\theta(x))^{\rm T} (\partial \partial f(x))^{-1} df_\thetabar(x)}\sign(\det(\partial_i\partial_j f(x))).
\ee
\paragraph{Gaussian integration:}
\be
\intop d^{2n|2n}X \exp\left(i \intop d^2\theta X^{\rm T} a X\right) = \sign(\det(a)),
\ee
where $a$ is an $n$ by $n$ matrix. This is ill-defined for singular $a$. In fact, the Gaussian formula holds also for a bi-graded matrix $A=a+\theta a_\theta +\cdots$:
\be\label{eq: general bi-graded gaussian}
\intop d^{2n|2n}X \exp\left(i \intop d^2\theta X^{\rm T} A X\right) = \sign(\det(a)).
\ee
\paragraph{Delta function:}
\be
\delta^{2|2}(X) = -2\pi i\delta(x)\delta(x_\theta)\delta(x_\thetabar)\delta(x_{\theta\thetabar}) = \intop d^{2|2}\Lambda \exp\left( -i \intop d^2\theta \Lambda X\right).
\ee
It follows that:
\be
\delta^{2|2}(f(X)) =  \sum_{x_0|f'(x_0)=0} \sign(f'(x_0))\delta^{2|2} (X-x_0).
\ee
\paragraph{Grassmann derivatives:}
Upon introducing Grassmann derivatives into the action, $x_{\theta\thetabar}$ is no longer a Lagrange multiplier. However, \textbf{as long as} we don't break the invariance of the action under $Q=\partial_\theta$, by introducing explicitly $\theta$-dependent terms, we can perform localization with respect to $Q$-exact terms. 

A particularly useful action is of the form:
\be
S= \intop d^2\theta \partial_\theta X^i m_{ij}(X) \partial_\thetabar X^j,
\ee
where $m$ is a metric on $\mathcal{M}$. In components:
\bea \label{eq: expansion of S-term into components}
S & = x_{\theta\bar{\theta}}^{i}m_{ij}x_{\theta\bar{\theta}}^{j}-x_{\theta\bar{\theta}}^{i}\partial_{k}m_{ij}x_{\theta}^{k}x_{\bar{\theta}}^{j}-x_{\theta}^{i}\partial_{k}m_{ij}x_{\bar{\theta}}^{k}x_{\theta\bar{\theta}}^{j}+x_{\theta}^{i}x_{\theta\bar{\theta}}^{k}\partial_{k}m_{ij}x_{\bar{\theta}}^{j}\\
&+ x_{\theta}^{i}x_{\bar{\theta}}^{l}x_{\theta}^{k}\partial_{l}\partial_{k}m_{ij}x_{\bar{\theta}}^{j}\\
 & =x_{\theta\bar{\theta}}^{i}m_{ij}x_{\theta\bar{\theta}}^{j}-x_{\theta\bar{\theta}}^{i}\left(\partial_{(k}m_{j)i}-\partial_{i}m_{kj}\right)x_{\theta}^{k}x_{\bar{\theta}}^{j}+x_{\theta}^{i}x_{\bar{\theta}}^{l}x_{\theta}^{k}\partial_{l}\partial_{k}m_{ij}x_{\bar{\theta}}^{j}\\
& \downarrow \,(\text{shift}\,\, x_{\theta\bar{\theta}}^{i}\to x_{\theta\bar{\theta}}^{i}+\Gamma_{kl}^{i}x_{\theta}^{k}x_{\bar{\theta}}^{l}) \\
&=x_{\theta\bar{\theta}}^{i}m_{ij}x_{\theta\bar{\theta}}^{j}+\left(-\Gamma_{kj}^{n}m_{nm}\Gamma_{il}^{m}+\partial_{l}\partial_{k}m_{ij}\right)x_{\theta}^{i}x_{\bar{\theta}}^{l}x_{\theta}^{k}x_{\bar{\theta}}^{j}\\
 & =x_{\theta\bar{\theta}}^{i}m_{ij}x_{\theta\bar{\theta}}^{j}+\left(-\Gamma_{kj}^{n}m_{nm}\Gamma_{il}^{m}+\partial_{l}\left(m_{ni}\Gamma_{jk}^{n}+m_{nj}\Gamma_{ik}^{n}\right)\right)x_{\theta}^{i}x_{\bar{\theta}}^{l}x_{\theta}^{k}x_{\bar{\theta}}^{j}\\
 & =x_{\theta\bar{\theta}}^{i}m_{ij}x_{\theta\bar{\theta}}^{j}+\left(\cancel{-\Gamma_{kj}^{n}m_{nm}\Gamma_{il}^{m}+m_{mn}\Gamma_{il}^{m}\Gamma_{jk}^{n}}+m_{mi}\Gamma_{nl}^{m}\Gamma_{jk}^{n}+m_{ni}\Gamma_{jk;l}^{n}\right)x_{\theta}^{i}x_{\bar{\theta}}^{l}x_{\theta}^{k}x_{\bar{\theta}}^{j}\\
 & =x_{\theta\bar{\theta}}^{i}m_{ij}x_{\theta\bar{\theta}}^{j}+m_{ni}\left(\Gamma_{ml}^{n}\Gamma_{jk}^{m}+\Gamma_{jk;l}^{n}\right)x_{\theta}^{i}x_{\bar{\theta}}^{l}x_{\theta}^{k}x_{\bar{\theta}}^{j}\\
 & =x_{\theta\bar{\theta}}^{i}m_{ij}x_{\theta\bar{\theta}}^{j}+\frac{1}{2}m_{ni}R_{ljk}^{n}x_{\theta}^{i}x_{\bar{\theta}}^{l}x_{\theta}^{k}x_{\bar{\theta}}^{j}\\
 & =x_{\theta\bar{\theta}}^{i}m_{ij}x_{\theta\bar{\theta}}^{j}+\frac{1}{2}R_{ijkl}x_{\theta}^{i}x_{\theta}^{j}x_{\bar{\theta}}^{k}x_{\bar{\theta}}^{l}.
\eea
Thus, when the dimension of $\mathcal{M}$ is even $d=2n$:
\bea \label{eq: Euler density from 0-mode integration}
& \intop_{\mathcal{M}^{2|2}} d^{2(2n)|2(2n)}X \exp \left(-\frac{1}{2}\intop d^2\theta \partial_\theta X^i M_{ij}(X) \partial_\thetabar X^j\right)\\
= & \intop_{\mathcal{M}^{2|2}} d^{2(2n)|2(2n)}X \exp \left(-\frac{1}{2}x_{\theta\bar{\theta}}^{i}m_{ij}x_{\theta\bar{\theta}}^{j}\right)\frac{1}{n!}\left(-\frac{1}{4}R_{ijkl}x_{\theta}^{i}x_{\theta}^{j}x_{\bar{\theta}}^{k}x_{\bar{\theta}}^{l}\right)^n\\
= & \intop_{\mathcal{M}} d^{2n}x \left(\frac{i}{2\pi}\right)^{2n}(2\pi)^{n}{\det\left(m\right)}^{-1/2}\frac{1}{n!}\left(-\frac{1}{4}\right)^n\\
\times & \epsilon^{i_{1}j_{1}i_{2}j_{2}\dots i_{n}j_{n}}\epsilon^{k_{1}l_{1}k_{2}l_{2}\dots k_{n}l_{n}}R_{k_{1}l_{1}i_{1}j_{1}}\cdots R_{k_{n}l_{n}i_{n}j_{n}}\\
= & \intop_{\mathcal{M}} d^{2n}x \frac{1}{2^{2n}(2\pi)^n n! }{\det\left(m\right)}^{-1/2}\epsilon^{i_{1}j_{1}i_{2}j_{2}\dots i_{n}j_{n}}\epsilon^{k_{1}l_{1}k_{2}l_{2}\dots k_{n}l_{n}}R_{i_{1}j_{1}k_{1}l_{1}}\cdots R_{i_{n}j_{n}k_{n}l_{n}}.
\eea
The integrand in the last expression is the \textbf{Euler density} \cite{GILKEY1975334}, as can be seen by expressing the Riemann tensor in terms of the curvature two-form $\Omega$:
\be
\Omega_{\rho\sigma}=\frac{1}{2}R_{\mu\nu\rho\sigma}dx^{\mu}dx^{\nu}
\ee
\be
\frac{1}{2^{2n}(2\pi)^{n}n!}m^{-1/2}\epsilon^{i_{1}j_{1}i_{2}j_{2}\dots i_{n}j_{n}}\epsilon^{k_{1}l_{1}k_{2}l_{2}\dots k_{n}l_{n}}R_{i_{1}j_{1}k_{1}l_{1}}\cdots R_{i_{n}j_{n}k_{n}l_{n}}=\frac{1}{(2\pi)^{n}}\star{\rm Pf}\left(\Omega\right)=\star e_{\nabla}\left(T\mathcal{M}\right),
\ee
where $\star$ is the Hodge star, $\rm Pf$ is the Pfaffian and $e_\nabla$ is a closed form representing the Euler class. In summary, we find:
\be\label{eq: euler character from Chern-Gauss-Bonnet}
\intop_{\mathcal{M}^{2|2}} d^{2(2n)|2(2n)}X \exp \left(-\intop d^2\theta \partial_\theta X^i M_{ij}(X) \partial_\thetabar X^j\right)=\chi(\mathcal{M}).
\ee
\eqref{eq: euler character from Morse} and \eqref{eq: euler character from Chern-Gauss-Bonnet} both compute the Euler number, the former through Morse theory and the latter through the Chern-Gauss-Bonnet theorem. As explained in \cite{Blau:1992pm}, the Mathai-Quillen formalism \cite{Mathai:1986tc} interpolates between these two representations of the Euler number. We obtain the Mathai-Quillen form by combining the two ``actions'' from these two formulae and completing the square for $x_{\theta\thetabar}$:
\bea\label{eq: Mathai Quillen form in components}
&it_{0}\intop d^{2}\theta f(X)-\frac{t_{1}}{2}\intop d^{2}\theta\partial_{\theta}X^{i}M_{ij}(X)\partial_{\bar{\theta}}X^{j}\\=&it_{0}\partial_{i}fx_{\theta\bar{\theta}}^{i}-it_{0}\partial_{i}\partial_{j}fx_{\theta}^{i}x_{\bar{\theta}}^{j}-t_{1}\left(\frac{1}{2}x_{\theta\bar{\theta}}^{i}m_{ij}x_{\theta\bar{\theta}}^{j}-x_{\theta\bar{\theta}}^{i}m_{ij}\Gamma_{kl}^{j}x_{\theta}^{k}x_{\bar{\theta}}^{l}+\frac{1}{2}x_{\theta}^{i}x_{\bar{\theta}}^{l}x_{\theta}^{k}\partial_{l}\partial_{k}m_{ij}x_{\bar{\theta}}^{j}\right)\\=&-it_{0}\partial_{i}\partial_{j}fx_{\theta}^{i}x_{\bar{\theta}}^{j}-t_{1}\left(\frac{1}{2}x_{\theta\bar{\theta}}^{i}m_{ij}x_{\theta\bar{\theta}}^{j}-x_{\theta\bar{\theta}}^{i}\left(-i\frac{t_{0}}{t_{1}}\partial_{i}f+m_{ij}\Gamma_{kl}^{j}x_{\theta}^{k}x_{\bar{\theta}}^{l}\right)+\frac{1}{2}x_{\theta}^{i}x_{\bar{\theta}}^{l}x_{\theta}^{k}\partial_{l}\partial_{k}m_{ij}x_{\bar{\theta}}^{j}\right)\\\to&-it_{0}\partial_{i}\partial_{j}fx_{\theta}^{i}x_{\bar{\theta}}^{j}-t_{1}\left(\begin{array}{c}
\frac{1}{2}x_{\theta\bar{\theta}}^{i}m_{ij}x_{\theta\bar{\theta}}^{j}+\frac{1}{2}x_{\theta}^{i}x_{\bar{\theta}}^{l}x_{\theta}^{k}\partial_{l}\partial_{k}m_{ij}x_{\bar{\theta}}^{j}\\
-\frac{1}{2}\left(-i\frac{t_{0}}{t_{1}}\partial_{i_{1}}f+m_{i_{1}j}\Gamma_{kl}^{j}x_{\theta}^{k}x_{\bar{\theta}}^{l}\right)m^{i_{1}i_{2}}\left(-i\frac{t_{0}}{t_{1}}\partial_{i_{2}}f+m_{i_{2}j}\Gamma_{kl}^{j}x_{\theta}^{k}x_{\bar{\theta}}^{l}\right)
\end{array}\right)\\=&-it_{0}\partial_{i}\partial_{j}fx_{\theta}^{i}x_{\bar{\theta}}^{j}-t_{1}\left(\begin{array}{c}
\frac{1}{2}x_{\theta\bar{\theta}}^{i}m_{ij}x_{\theta\bar{\theta}}^{j}-\frac{1}{2}\left(-i\frac{t_{0}}{t_{1}}\partial_{i_{1}}f\right)m^{i_{1}i_{2}}\left(-i\frac{t_{0}}{t_{1}}\partial_{i_{2}}f\right)\\
-\frac{1}{2}\left(-i\frac{t_{0}}{t_{1}}\partial_{i_{1}}f\right)m^{i_{1}i_{2}}\left(m_{i_{2}j}\Gamma_{kl}^{j}x_{\theta}^{k}x_{\bar{\theta}}^{l}\right)\\
-\frac{1}{2}\left(m_{i_{1}j}\Gamma_{kl}^{j}x_{\theta}^{k}x_{\bar{\theta}}^{l}\right)m^{i_{1}i_{2}}\left(-i\frac{t_{0}}{t_{1}}\partial_{i_{2}}f\right)+\frac{1}{4}R_{ijkl}x_{\theta}^{i}x_{\theta}^{j}x_{\bar{\theta}}^{k}x_{\bar{\theta}}^{l}
\end{array}\right)\\=&-\frac{1}{2}\frac{t_{0}^{2}}{t_{1}}\partial_{i}fm^{ij}\partial_{j}f-\frac{t_{1}}{2}x_{\theta\bar{\theta}}^{i}m_{ij}x_{\theta\bar{\theta}}^{j}-it_{0}\left(\partial_{i}\partial_{j}f+\partial_{k}f\Gamma_{ij}^{k}\right)x_{\theta}^{i}x_{\bar{\theta}}^{j}-\frac{t_{1}}{4}R_{ijkl}x_{\theta}^{i}x_{\theta}^{j}x_{\bar{\theta}}^{k}x_{\bar{\theta}}^{l}\\
=&-\frac{1}{2}\frac{t_{0}^{2}}{t_{1}}\partial_{i}fm^{ij}\partial_{j}f-\frac{t_{1}}{2}x_{\theta\bar{\theta}}^{i}m_{ij}x_{\theta\bar{\theta}}^{j}-it_{0}\nabla_{i}\partial_{j}fx_{\theta}^{i}x_{\bar{\theta}}^{j}-\frac{t_{1}}{4}R_{ijkl}x_{\theta}^{i}x_{\theta}^{j}x_{\bar{\theta}}^{k}x_{\bar{\theta}}^{l}.
\eea
If we take $t_0\to\infty$, which by BRST-exactness doesn't affect the answer, we recover \eqref{eq: euler character from Morse}. A particular hybrid case of interest is when we take $t_0\to\infty$ but $f$ has flat directions. Then the vanishing locus of $df$ is a submanifold of $\mathcal{M}$, which in the current context we'll call the moduli space. Upon localizing to this submanifold, the integration measure will again be the Euler density, this time constructed from the induced metric on the moduli space from its embedding in $\mathcal{M}$.

\section{The extrinsic curvature tensor}\label{sec: extrinsic curvature tensor}

In this section, we'll study the invariant:
\be
p_{\mu\nu}h^{ab}h^{cd}\nabla_{a}\partial_{[b}x^{\mu}\nabla_{c]}\partial_{d}x^{\nu},
\ee
where $h$ is the induced metric and $x$ is a map from the 2d worldsheet to a 2d target space. $p$ is the projector onto the normal bundle. As discussed in subsection \ref{subsec: horava 2d target space}, the projector vanishes generically. We will show it becomes a delta function at branch points:
for a branch point at the origin where $(n+1)$ sheets are cyclically permuted,
\be\label{eq: extrinsic curvature at branch point}
p_{\mu\nu}h^{ab}h^{cd}\partial_{a}\partial_{[c}x^{\mu}\partial_{b]}\partial_{d}x^{\nu}=-\frac{1}{2}\pi n\frac{1}{\sqrt{h}}\delta^{2}\left(\sigma^{a}\right).
\ee
Since $dx$ vanishes at branch points, so does the induced metric. Consequently, the inverse metric is singular and we must regulate it wherever it appears. We can take $h^{-1}\to\frac{\det(h) h^{-1}}{\det(h) + \epsilon}$. In particular, we must regulate the projection operator, which vanishes at generic points for a 2d target space:
\be
p^{\mu\nu}=g^{\mu\nu} - \partial_a x^\mu h^{ab}\partial_b x^\nu\to p_\epsilon^{\mu\nu}=g^{\mu\nu} - \partial_a x^\mu \frac{\det(h) h^{ab}}{\det(h) + \epsilon}\partial_b x^\nu = \frac{\epsilon}{\det(h)+\epsilon}g^{\mu\nu}.
\ee
Now:
\be
p_{\mu\nu}h^{ab}h^{cd}\nabla_{a}\partial_{[b}x^{\mu}\nabla_{c]}\partial_{d}x^{\nu}\to\frac{\epsilon}{\left(\det\left(h\right)+\epsilon\right)^{3}}\left(\det\left(h\right)h^{ab}\right)\left(\det\left(h\right)h^{cd}\right)g_{\mu\nu}\partial_{a}\partial_{[b}x^{\mu}\partial_{c]}\partial_{d}x^{\nu}.
\ee

\subsection{Branch points}
Consider the map with a single order $n$ branch point at the origin: in complex coordinates
\be
Z=\frac{1}{n+1}z^{n+1}.
\ee
For simplicity, assume the target space is flat. The only non-zero induced-metric components are:
\be
h_{z\bar{z}}=\left(z\bar{z}\right)^{n}=r^{2n},\,h^{z\bar{z}}=\frac{1}{\left(z\bar{z}\right)^{n}}=r^{-2n}\Rightarrow\det(h)=r^{4n}.
\ee
Now:
\bea
&\frac{\epsilon}{\left(\det\left(h\right)+\epsilon\right)^{3}}\left(\det\left(h\right)h^{ab}\right)\left(\det\left(h\right)h^{cd}\right)g_{\mu\nu}\partial_{a}\partial_{[b}x^{\mu}\partial_{c]}\partial_{d}x^{\nu}\\=&-\frac{2\epsilon}{\left(r^{4n}+\epsilon\right)^{3}}\left(r^{2n}\right)\left(r^{2n}\right)\left(nz^{n-1}\right)\left(n\bar{z}^{n-1}\right)\\=&-2n^{2}r^{2n-2}\frac{\epsilon r^{4n}}{\left(r^{4n}+\epsilon\right)^{3}}.
\eea
Evidently this vanishes as $\epsilon\to0$ away from the origin, but its integral in the vicinity of the origin is:
\bea
%&
\intop d^{2}\sigma\sqrt{h}p_{\mu\nu}h^{ab}h^{cd}\partial_{a}\partial_{[c}x^{\mu}\partial_{b]}\partial_{d}x^{\nu}
%\\=&
=&
-2\pi\intop rdr\left(r^{2n}\right)2n^{2}r^{2n-2}\frac{\epsilon r^{4n}}{\left(r^{4n}+\epsilon\right)^{3}}\\x=r^{4n}\Rightarrow=&-\pi n\intop_{0}^{\infty}dx\frac{\epsilon x}{\left(x+\epsilon\right)^{3}}\\y=x/\epsilon \Rightarrow=&-\pi n\intop_{0}^{\infty}dy\frac{y}{\left(y+1\right)^{3}}=-\frac{1}{2}\pi n,
\eea
thus proving \eqref{eq: extrinsic curvature at branch point}.

%% file: refs.bib
@article{Migdal:1975zg,
    author = "Migdal, Alexander A.",
    editor = "Khalatnikov, I. M. and Mineev, V. P.",
    title = "{Recursion Equations in Gauge Theories}",
    reportNumber = "PRINT-75-1043 (LANDAU-INST)",
    journal = "Sov. Phys. JETP",
    volume = "42",
    pages = "413",
    year = "1975"
}

@article{Rusakov:1990rs,
    author = "Rusakov, B. E.",
    title = "{Loop averages and partition functions in U(N) gauge theory on two-dimensional manifolds}",
    doi = "10.1142/S0217732390000780",
    journal = "Mod. Phys. Lett. A",
    volume = "5",
    pages = "693--703",
    year = "1990"
}

@article{Witten:1992xu,
    author = "Witten, Edward",
    title = "{Two-dimensional gauge theories revisited}",
    eprint = "hep-th/9204083",
    archivePrefix = "arXiv",
    doi = "10.1016/0393-0440(92)90034-X",
    journal = "J. Geom. Phys.",
    volume = "9",
    pages = "303--368",
    year = "1992"
}

@article{Horava:1995ic,
    author = "Horava, Petr",
    title = "{Topological rigid string theory and two-dimensional QCD}",
    eprint = "hep-th/9507060",
    archivePrefix = "arXiv",
    reportNumber = "PUPT-1547",
    doi = "10.1016/0550-3213(96)00036-3",
    journal = "Nucl. Phys. B",
    volume = "463",
    pages = "238--286",
    year = "1996"
}

@inproceedings{Horava:1993aq,
    author = "Horava, Petr",
    title = "{Topological strings and QCD in two-dimensions}",
    booktitle = "{NATO Advanced Research Workshop on New Developments in String Theory, Conformal Models and Topological Field Theory}",
    eprint = "hep-th/9311156",
    archivePrefix = "arXiv",
    reportNumber = "EFI-93-66",
    month = "11",
    year = "1993"
}

@article{Horava:1998wf,
    author = "Horava, Petr",
    title = "{On QCD string theory and AdS dynamics}",
    eprint = "hep-th/9811028",
    archivePrefix = "arXiv",
    reportNumber = "CALT-68-2199",
    doi = "10.1088/1126-6708/1999/01/016",
    journal = "JHEP",
    volume = "01",
    pages = "016",
    year = "1999"
}

@article{Vafa:2004qa,
    author = "Vafa, Cumrun",
    title = "{Two dimensional Yang-Mills, black holes and topological strings}",
    eprint = "hep-th/0406058",
    archivePrefix = "arXiv",
    reportNumber = "HUTP-04-A026",
    month = "6",
    year = "2004"
}

@article{Strassler:1992zr,
    author = "Strassler, Matthew J.",
    title = "{Field theory without Feynman diagrams: One loop effective actions}",
    eprint = "hep-ph/9205205",
    archivePrefix = "arXiv",
    reportNumber = "SLAC-PUB-5757",
    doi = "10.1016/0550-3213(92)90098-V",
    journal = "Nucl. Phys. B",
    volume = "385",
    pages = "145--184",
    year = "1992"
}

@article{tHooft:1973alw,
    author = "'t Hooft, Gerard",
    editor = "Taylor, J. C.",
    title = "{A Planar Diagram Theory for Strong Interactions}",
    reportNumber = "CERN-TH-1786",
    doi = "10.1016/0550-3213(74)90154-0",
    journal = "Nucl. Phys. B",
    volume = "72",
    pages = "461",
    year = "1974"
}

@article{Ganor:1994bq,
    author = "Ganor, O. and Sonnenschein, J. and Yankielowicz, S.",
    title = "{The String theory approach to generalized 2-D Yang-Mills theory}",
    eprint = "hep-th/9407114",
    archivePrefix = "arXiv",
    reportNumber = "TAUP-2182-94",
    doi = "10.1016/0550-3213(94)00397-W",
    journal = "Nucl. Phys. B",
    volume = "434",
    pages = "139--178",
    year = "1995"
}

@article{Ganor:1994rm,
    author = "Ganor, O. and Sonnenschein, J. and Yankielowicz, S.",
    title = "{Folds in 2-D string theories}",
    eprint = "hep-th/9404149",
    archivePrefix = "arXiv",
    reportNumber = "TAUP-2152-94",
    doi = "10.1016/0550-3213(94)90275-5",
    journal = "Nucl. Phys. B",
    volume = "427",
    pages = "203--244",
    year = "1994"
}

@article{Rudd:1994ta,
    author = "Rudd, Robert E.",
    title = "{The String partition function for QCD on the torus}",
    eprint = "hep-th/9407176",
    archivePrefix = "arXiv",
    reportNumber = "RU-94-58",
    month = "7",
    year = "1994"
}

@article{Douglas:1993iia,
    author = "Douglas, Michael R. and Kazakov, Vladimir A.",
    title = "{Large N phase transition in continuum QCD in two-dimensions}",
    eprint = "hep-th/9305047",
    archivePrefix = "arXiv",
    reportNumber = "LPTENS-93-20, RU-93-17",
    doi = "10.1016/0370-2693(93)90806-S",
    journal = "Phys. Lett. B",
    volume = "319",
    pages = "219--230",
    year = "1993"
}

@article{Minahan:1993tp,
    author = "Minahan, Joseph A. and Polychronakos, Alexios P.",
    title = "{Classical solutions for two-dimensional QCD on the sphere}",
    eprint = "hep-th/9309119",
    archivePrefix = "arXiv",
    reportNumber = "CERN-TH-7016-93, UVA-HET-93-08",
    doi = "10.1016/0550-3213(94)00153-7",
    journal = "Nucl. Phys. B",
    volume = "422",
    pages = "172--194",
    year = "1994"
}

@article{Gross:1994mr,
    author = "Gross, David J. and Matytsin, Andrei",
    title = "{Instanton induced large N phase transitions in two-dimensional and four-dimensional QCD}",
    eprint = "hep-th/9404004",
    archivePrefix = "arXiv",
    reportNumber = "PUPT-1459",
    doi = "10.1016/S0550-3213(94)80041-3",
    journal = "Nucl. Phys. B",
    volume = "429",
    pages = "50--74",
    year = "1994"
}

@article{Maldacena:1997re,
    author = "Maldacena, Juan Martin",
    title = "{The Large N limit of superconformal field theories and supergravity}",
    eprint = "hep-th/9711200",
    archivePrefix = "arXiv",
    reportNumber = "HUTP-97-A097, HUTP-98-A097",
    doi = "10.4310/ATMP.1998.v2.n2.a1",
    journal = "Adv. Theor. Math. Phys.",
    volume = "2",
    pages = "231--252",
    year = "1998"
}

@article{Ooguri:2002gx,
    author = "Ooguri, Hirosi and Vafa, Cumrun",
    title = "{World sheet derivation of a large N duality}",
    eprint = "hep-th/0205297",
    archivePrefix = "arXiv",
    reportNumber = "CALT-68-2386, CITUSC-02-019, HUTP-02-A018",
    doi = "10.1016/S0550-3213(02)00620-X",
    journal = "Nucl. Phys. B",
    volume = "641",
    pages = "3--34",
    year = "2002"
}

@article{Fine:1990zz,
    author = "Fine, D. S.",
    title = "{Quantum Yang-Mills on the two-sphere}",
    doi = "10.1007/BF02097703",
    journal = "Commun. Math. Phys.",
    volume = "134",
    pages = "273--292",
    year = "1990"
}

@article{Gopakumar:1998ki,
    author = "Gopakumar, Rajesh and Vafa, Cumrun",
    editor = "Vafa, Cumrun and Yau, S. -T.",
    title = "{On the gauge theory / geometry correspondence}",
    eprint = "hep-th/9811131",
    archivePrefix = "arXiv",
    reportNumber = "HUTP-98-A078",
    doi = "10.4310/ATMP.1999.v3.n5.a5",
    journal = "Adv. Theor. Math. Phys.",
    volume = "3",
    pages = "1415--1443",
    year = "1999"
}

@article{tHooft:1974pnl,
    author = "'t Hooft, Gerard",
    title = "{A Two-Dimensional Model for Mesons}",
    reportNumber = "CERN-TH-1820",
    doi = "10.1016/0550-3213(74)90088-1",
    journal = "Nucl. Phys. B",
    volume = "75",
    pages = "461--470",
    year = "1974"
}

@article{Taylor:1994zm,
    author = "Taylor, Washington",
    title = "{Counting strings and phase transitions in 2-D QCD}",
    eprint = "hep-th/9404175",
    archivePrefix = "arXiv",
    reportNumber = "MIT-CTP-2297",
    month = "4",
    year = "1994"
}

@article{Blau:1992pm,
    author = "Blau, Matthias",
    editor = "Gielerak, R. and Borowiec, Andrzej",
    title = "{The Mathai-Quillen formalism and topological field theory}",
    eprint = "hep-th/9203026",
    archivePrefix = "arXiv",
    reportNumber = "NIKHEF-H-92-07",
    doi = "10.1016/0393-0440(93)90049-K",
    journal = "J. Geom. Phys.",
    volume = "11",
    pages = "95--127",
    year = "1993"
}

@article{Mathai:1986tc,
    author = "Mathai, Varghese and Quillen, Daniel G.",
    title = "{Superconnections, Thom classes and equivariant differential forms}",
    doi = "10.1016/0040-9383(86)90007-8",
    journal = "Topology",
    volume = "25",
    pages = "85--110",
    year = "1986"
}

@article{Blau:1991mp,
    author = "Blau, Matthias and Thompson, George",
    title = "{Quantum Yang-Mills theory on arbitrary surfaces}",
    reportNumber = "NIKHEF-H-91-09, MZ-TH-91-17",
    doi = "10.1142/S0217751X9200168X",
    journal = "Int. J. Mod. Phys. A",
    volume = "7",
    pages = "3781--3806",
    year = "1992"
}

@inproceedings{Gross:1993cw,
    author = "Gross, David J. and Taylor, Washington",
    title = "{Two-dimensional QCD and strings}",
    booktitle = "{International Conference on Strings 93}",
    eprint = "hep-th/9311072",
    archivePrefix = "arXiv",
    reportNumber = "MIT-CTP-2250, PUPT-1431",
    month = "10",
    year = "1993"
}

@article{Gross:1993hu,
    author = "Gross, David J. and Taylor, Washington",
    title = "{Two-dimensional QCD is a string theory}",
    eprint = "hep-th/9301068",
    archivePrefix = "arXiv",
    reportNumber = "LBL-33458, PUPT-1376, UCB-PTH-93-02",
    doi = "10.1016/0550-3213(93)90403-C",
    journal = "Nucl. Phys. B",
    volume = "400",
    pages = "181--208",
    year = "1993"
}

@article{Gross:1993yt,
    author = "Gross, David J. and Taylor, Washington",
    title = "{Twists and Wilson loops in the string theory of two-dimensional QCD}",
    eprint = "hep-th/9303046",
    archivePrefix = "arXiv",
    reportNumber = "CERN-TH-6827-93, PUPT-1382, LBL-33767, UCB-PTH-93-09",
    doi = "10.1016/0550-3213(93)90042-N",
    journal = "Nucl. Phys. B",
    volume = "403",
    pages = "395--452",
    year = "1993"
}

@article{Cordes:1994sd,
    author = "Cordes, Stefan and Moore, Gregory W. and Ramgoolam, Sanjaye",
    title = "{Large N 2-D Yang-Mills theory and topological string theory}",
    eprint = "hep-th/9402107",
    archivePrefix = "arXiv",
    reportNumber = "YCTP-P23-93, RU-94-20",
    doi = "10.1007/s002200050102",
    journal = "Commun. Math. Phys.",
    volume = "185",
    pages = "543--619",
    year = "1997"
}

@article{Cordes:1994fc,
    author = "Cordes, Stefan and Moore, Gregory W. and Ramgoolam, Sanjaye",
    title = "{Lectures on 2-d Yang-Mills theory, equivariant cohomology and topological field theories}",
    eprint = "hep-th/9411210",
    archivePrefix = "arXiv",
    reportNumber = "YCTP-P11-94",
    doi = "10.1016/0920-5632(95)00434-B",
    journal = "Nucl. Phys. B Proc. Suppl.",
    volume = "41",
    pages = "184--244",
    year = "1995"
}

@article{Polyakov:1981rd,
    author = "Polyakov, Alexander M.",
    editor = "Khalatnikov, I. M. and Mineev, V. P.",
    title = "{Quantum Geometry of Bosonic Strings}",
    reportNumber = "Print-81-0351 (LANDAU INST)",
    doi = "10.1016/0370-2693(81)90743-7",
    journal = "Phys. Lett. B",
    volume = "103",
    pages = "207--210",
    year = "1981"
}

@book{Polchinski_textbook:1998rq,
    author = "Polchinski, J.",
    title = "{String theory. Vol. 1: An introduction to the bosonic string}",
    doi = "10.1017/CBO9780511816079",
    isbn = "9780511816079",
    publisher = "Cambridge University Press",
    series = "Cambridge Monographs on Mathematical Physics",
    month = "12",
    year = "2007"
}

@article{GILKEY1975334,
	author = {Peter B Gilkey},
	doi = {https://doi.org/10.1016/0001-8708(75)90141-3},
	issn = {0001-8708},
	journal = {Advances in Mathematics},
	number = {3},
	pages = {334-360},
	title = {The boundary integrand in the formula for the signature and Euler characteristic of a Riemannian manifold with boundary},
	url = {https://www.sciencedirect.com/science/article/pii/0001870875901413},
	volume = {15},
	year = {1975}
}

@unpublished{KOMATSU_forthcoming,
    author = "Komatsu, Shota and Maity, Pronobesh",
    title = "{String Dual of Two-Dimensional Yang-Mills and Symmetric Product Orbifiolds}",
    year = {2024},
    note = "To appear.",
    url = {https://online.kitp.ucsb.edu/online/bootstrap-c23/}
}

@unpublished{GABERDIEL_forthcoming,
    author = "Gaberdiel, M. R. and Galvagno, F. and Knighton, B. and Naderi, K.",
    year = {2024},
    note = "To appear."
}

@article{Polyakov:1986cs,
    author = "Polyakov, Alexander M.",
    title = "{Fine Structure of Strings}",
    doi = "10.1016/0550-3213(86)90162-8",
    journal = "Nucl. Phys. B",
    volume = "268",
    pages = "406--412",
    year = "1986"
}

@article{Kleinert:1986bk,
    author = "Kleinert, H.",
    title = "{The Membrane Properties of Condensing Strings}",
    doi = "10.1016/0370-2693(86)91111-1",
    journal = "Phys. Lett. B",
    volume = "174",
    pages = "335--338",
    year = "1986"
}

@article{Itoi:1988ji,
    author = "Itoi, Chigak and Kubota, Hiroshi",
    title = "{Gauge Invariance Based on the Extrinsic Geometry in the Rigid String}",
    reportNumber = "TIT/HEP-127",
    doi = "10.1007/BF01557339",
    journal = "Z. Phys. C",
    volume = "44",
    pages = "337",
    year = "1989"
}

@article{Itoi:1987de,
    author = "Itoi, Chigak and Kubota, Hiroshi",
    title = "{{BRST} Quantization of the String Model With Extrinsic Curvature}",
    reportNumber = "TIT/HEP-122",
    doi = "10.1016/0370-2693(88)90489-3",
    journal = "Phys. Lett. B",
    volume = "202",
    pages = "381--384",
    year = "1988"
}

@article{Eberhardt:2021ynh,
    author = "Eberhardt, Lorenz and Pal, Sridip",
    title = "{The disk partition function in string theory}",
    eprint = "2105.08726",
    archivePrefix = "arXiv",
    primaryClass = "hep-th",
    doi = "10.1007/JHEP08(2021)026",
    journal = "JHEP",
    volume = "08",
    pages = "026",
    year = "2021"
}

@article{Pohlmeyer:1976,
	author = {Pohlmeyer, K. },
	doi = {10.1007/BF01609119},
	id = {Pohlmeyer1976},
        isbn = "9780521822671",
	journal = {Communications in Mathematical Physics},
	number = {3},
	pages = {207--221},
	title = {Integrable Hamiltonian systems and interactions through quadratic constraints},
	url = {https://doi.org/10.1007/BF01609119},
	volume = {46},
	year = {1976}}

@article{chern:1945,
 ISSN = {0003486X},
 URL = {http://www.jstor.org/stable/1969203},
 author = {Shiing-shen Chern},
 journal = {Annals of Mathematics},
 number = {4},
 pages = {674--684},
 publisher = {Annals of Mathematics},
 title = {On the Curvatura Integra in a Riemannian Manifold},
 urldate = {2023-12-10},
 volume = {46},
 year = {1945}
}

@article{AFFLECK1981429,
	author = {Ian Affleck},
	doi = {https://doi.org/10.1016/0550-3213(81)90307-2},
	issn = {0550-3213},
	journal = {Nuclear Physics B},
	number = {2},
	pages = {429-444},
	title = {On constrained instantons},
	url = {https://www.sciencedirect.com/science/article/pii/0550321381903072},
	volume = {191},
	year = {1981}}

@article{Strominger:1980xa,
    author = "Strominger, Andrew",
    title = "{Loop Space Solution of Two-dimensional {QCD}}",
    reportNumber = "MIT-CTP-901",
    doi = "10.1016/0370-2693(81)90311-7",
    journal = "Phys. Lett. B",
    volume = "101",
    pages = "271--276",
    year = "1981"
}

@article{Billo_caselle:1998fb,
    author = "Billo, M. and Caselle, M. and D'Adda, A. and Provero, P.",
    title = "{Matrix string states in pure 2-D Yang-Mills theories}",
    eprint = "hep-th/9809095",
    archivePrefix = "arXiv",
    reportNumber = "DFTT-56-98",
    doi = "10.1016/S0550-3213(98)00865-7",
    journal = "Nucl. Phys. B",
    volume = "543",
    pages = "141--169",
    year = "1999"
}

@inproceedings{Caselle:1993mq,
    author = "Caselle, M. and D'Adda, A. and Magnea, Lorenzo and Panzeri, S.",
    title = "{Two-dimensional QCD on the sphere and on the cylinder}",
    booktitle = "{Trieste Summer School on High-energy Physics and Cosmology (Part 1 (15 Jun - 3 Jul) includes Workshop on Superstrings and Related Topics, 2-3 Jul 1992) Part 2 will be held 6-31 Jul (Note: change of dates from 15 Jun-14 Aug)}",
    eprint = "hep-th/9309107",
    archivePrefix = "arXiv",
    reportNumber = "DFTT-50-93",
    month = "9",
    year = "1993"
}

@article{Caselle:1993gc,
    author = "Caselle, M. and D'Adda, A. and Magnea, Lorenzo and Panzeri, S.",
    title = "{Two-dimensional QCD is a one-dimensional Kazakov-Migdal model}",
    eprint = "hep-th/9304015",
    archivePrefix = "arXiv",
    reportNumber = "DFTT-15-93",
    doi = "10.1016/0550-3213(94)90553-3",
    journal = "Nucl. Phys. B",
    volume = "416",
    pages = "751--770",
    year = "1994"
}

@article{Nguyen_unsal:2021naa,
    author = {Nguyen, Mendel and Tanizaki, Yuya and \"Unsal, Mithat},
    title = "{Noninvertible 1-form symmetry and Casimir scaling in 2D Yang-Mills theory}",
    eprint = "2104.01824",
    archivePrefix = "arXiv",
    primaryClass = "hep-th",
    reportNumber = "YITP-21-29",
    doi = "10.1103/PhysRevD.104.065003",
    journal = "Phys. Rev. D",
    volume = "104",
    number = "6",
    pages = "065003",
    year = "2021"
}

@article{Pantev_Sharpe:2023dim,
    author = "Pantev, T. and Sharpe, E.",
    title = "{Decomposition and the Gross-Taylor string theory}",
    eprint = "2307.08729",
    archivePrefix = "arXiv",
    primaryClass = "hep-th",
    month = "7",
    year = "2023"
}

@article{Kimura:2008gs,
    author = "Kimura, Yusuke and Ramgoolam, Sanjaye",
    title = "{Holomorphic maps and the complete 1/N expansion of 2D SU(N) Yang-Mills}",
    eprint = "0802.3662",
    archivePrefix = "arXiv",
    primaryClass = "hep-th",
    reportNumber = "QMUL-PH-08-03",
    doi = "10.1088/1126-6708/2008/06/015",
    journal = "JHEP",
    volume = "06",
    pages = "015",
    year = "2008"
}
